\documentclass[letterpape,11pt]{article}
\pdfoutput=1
\usepackage{physics}    
\usepackage{jheppub}
\usepackage{graphicx}
\usepackage{amsmath}
\usepackage{slashed}
\allowdisplaybreaks
\usepackage[utf8]{inputenc}
\usepackage{color}
\usepackage[usenames,dvipsnames]{xcolor}
\usepackage[normalem]{ulem}
\usepackage{soul}
\usepackage{units}
\usepackage{rotating}
\usepackage{hhline,multirow,tabularx}
\usepackage{bm}
\usepackage{hyperref}
\usepackage{comment}
\newcommand*{\FigPath}{./}%

\def\lqcd{\L_{\rm QCD}}

\def\bea#1\eea{\begin{align}#1\end{align}} 

\newcommand{\bef}{\begin{figure}[htb]\centering}
\newcommand{\eef}{\end{figure}}

\newcommand{\nn}{\nonumber}

\newcommand{\GeV}{\rm GeV}

\newcommand{\as}{\alpha_s}

\renewcommand\cleardoublepage{%
 \clearpage
 \ifodd\value{page}\else\stepcounter{page}\fi
}

\def\L{\Lambda}

\def\z{\zeta}

\def\<{\langle}
\def\>{\rangle}

\def\b{\beta}
\def\g{\gamma}  \def\G{\Gamma}
\def\d{\delta}  
   \def\L{\Lambda}

\def\m{\mu}

\def\z{\zeta}

\def\({\left(}
\def\[{\left[}
\def\){\right)}
\def\]{\right]}

\def\sin{\hbox{sin}}
\def\ln{\hbox{ln}}

\def \le { \left    }
\def \ri { \right }

\newcommand{\bt}{b}
\newcommand{\qt}{q_{\perp}}
\newcommand{\kt}{k_{\perp}}
\newcommand{\pt}{p_{\perp}}

\newcommand{\eref}[1]{Eq.~(\ref{e.#1})}

\newcommand{\fref}[1]{Fig.~\ref{#1}}

\begin{document}
	
\title{Global analysis of the Sivers functions at NLO+NNLL in QCD}
	
\author[a]{Miguel G. Echevarria,}
\author[b,c,d]{Zhong-Bo Kang}
\author[b,c]{and John Terry}
\affiliation[a]{Dpto. de F\'isica y Matem\'aticas, Universidad de Alcal\'a, 28805 Alcal\'a de Henares (Madrid), Spain}
\affiliation[b]{Department of Physics and Astronomy, University of California, Los Angeles, California 90095, USA}
\affiliation[c]{Mani L. Bhaumik Institute for Theoretical Physics, University of California, Los Angeles, California 90095, USA}
\affiliation[d]{Center for Frontiers in Nuclear Science, Stony Brook University, Stony Brook, NY 11794, USA}
\emailAdd{m.garciae@uah.es}
\emailAdd{zkang@physics.ucla.edu}
\emailAdd{johndterry@physics.ucla.edu}

\abstract{We perform global fit to the quark Sivers function within the transverse momentum dependent~(TMD) factorization formalism in QCD. We simultaneously fit Sivers asymmetry data from Semi-Inclusive Deep Inelastic Scattering (SIDIS) at COMPASS, HERMES, and JLab, from Drell-Yan lepton pair production at COMPASS, and from $W/Z$ boson at RHIC. This extraction is performed at next-to-leading order (NLO) and next-to-next-to leading logarithmic (NNLL) accuracy. We find excellent agreement between our extracted asymmetry and the experimental data for SIDIS and Drell-Yan lepton pair production, while tension arises when trying to describe the spin asymmetries of $W/Z$ bosons at RHIC. We carefully assess the situation, and we study in details the impact of the RHIC data and their implications through different ways of performing the fit. In addition, we find that the quality of the description of $W/Z$ vector boson asymmetry data could be strongly sensitive to the DGLAP evolution of Qiu-Sterman function, besides the usual TMD evolution. We present discussion on this and the implications for measurements of the transverse-spin asymmetries at the future Electron Ion Collider.}
\maketitle
\section{Introduction}\label{Introduction}
%%%%%%%%%%%%%%%%%%%%%%%%%%%%%%%%%%%%%%%%%%%%%%%%%%%%%%%%%%%%%

One of the most important discoveries in hadronic physics over the past decades has been the measurements of large spin asymmetries in hadronic interactions \cite{Bunce:1976yb,Kane:1978nd}. 
These experimental measurements eventually lead to the conclusions that not only are QCD dynamics important for describing experimental data; but that these experimental measurements can be used to probe the internal structure of hadrons.
For the past forty years, a major focus of the hadronic physics community has been precision extractions of the distribution functions which describe this internal structure~\cite{Boer:2011fh,Accardi:2012qut,Lin:2017snn,Proceedings:2020eah}.
In particular, the Sivers function~\cite{Sivers:1989cc,Sivers:1990fh}, which provides the transverse momentum distribution of unpolarized quarks in a transversely polarized proton via a correlation between the transverse momentum of the quark and the transverse spin of the proton, has received considerable attention in recent years.
By studying the Sivers function, major advancements have been made in the understanding of the spin-transverse momentum correlation and factorization theorems. 
For instance, theoretical investigation of the Sivers function led to the discovery that this function observes modified universality between semi-inclusive deep inelastic scattering (SIDIS) and Drell-Yan process~\cite{Brodsky:2002cx,Collins:2002kn,Boer:2003cm,Kang:2011hk,Kang:2009bp}.
Roughly speaking, this effect occurs because the phase which is produced from the re-scattering of the unpolarized quark and the color remnant field of the initial-state hadron is opposite between these two processes.
A fundamental goal of the future Electron Ion Collider (EIC)~\cite{Accardi:2012qut} will be high precision determination of these so-called transverse momentum dependent distribution functions (TMDs) over a wide range of energy scales, i.e. the so-called quantum three-dimensional (3D) imaging of the hadrons. 

While the extraction of TMDs is an essential ingredient in describing transverse momentum dependent observables, high precision determination of these distributions functions has remained a challenge. 
The Sivers function and all other TMDs are non-perturbative objects. 
These TMDs must then be either computed on a lattice~\cite{Lin:2017snn,Lin:2020rut}, or fitted from spin asymmetry data with the use of TMD factorization theorems~\cite{Collins:2011zzd,Ji:2004wu,Collins:1981uk,GarciaEchevarria:2011rb}. 
The TMD factorization theorems are valid in the region where $\qt/Q\ll 1$ where $\qt$ is the transverse momentum resolution scale and $Q$ is the relevant hard scale of the collision.
In this region, the cross section can be factorized in terms of transverse momentum dependent parton distribution functions (TMDPDFs) and/or transverse momentum dependent fragmentation functions (TMDFFs), and perturbatively calculable short distance hard coefficients.
In this paper, we rely on the TMD factorization theorems for SIDIS and Drell-Yan processes. 

Despite the challenges involved with fitting TMDs, tremendous progress has been made in the field over the past few years.
In particular, the focus of the field has been to increase the perturbative accuracy of the extractions of the TMDs. 
In \cite{Bacchetta:2017gcc,Pisano:2018skt} global extractions of the unpolarized TMDPDFs and TMDFFs were performed from SIDIS and Drell-Yan data at leading order (LO) and next-to-leading logarithmic (NLL) accuracy. 
In \cite{Scimemi:2017etj} the unpolarized TMDPDFs were extracted at next-to-next-to leading order (NNLO) and next-to-next-to leading logarithm (NNLL) accuracy. 
Recently in~\cite{Bacchetta:2019sam} the TMDPDFs were extracted at NNLO+N$^3$LL accuracy from Drell-Yan data; while in \cite{Scimemi:2019cmh} the TMDPDFs and TMDFFs were extracted simultaneously from SIDIS and Drell-Yan data at NNLO+N$^3$LL in which the authors further include target mass corrections as well as $\qt/Q$ power corrections. Progress has also been made in understanding the predictive power of the TMD factorization formalism in different kinematic regions~\cite{Boglione:2019nwk,Grewal:2020hoc}, and in matching with the collinear factorization~\cite{Collins:2016hqq,Gamberg:2017jha,Cammarota:2020qcw}.

In this paper, we perform the first fit at NLO+NNLL to the Sivers function, one of the most known spin-dependent TMDs. Previously, the highest precision extraction of the Sivers asymmetry has been at LO+NLL in~\cite{Echevarria:2014xaa,Bacchetta:2020gko}. While the focus of phenomenology for unpolarized TMDs is the effects of the TMD evolution, the DGLAP evolution of twist-three function, the collinear counterpart that enters the TMD evolution for spin-dependent TMDs, introduces additional complications for fits to transverse spin-asymmetry data. For example, in the study of TMD Sivers functions with TMD evolution, the collinear twist-three Qiu-Sterman functions arise. The evolution of Qiu-Sterman function has been studied extensively in the literature~\cite{Kang:2008ey,Zhou:2008mz,Vogelsang:2009pj,Braun:2009mi,Kang:2012em,Kang:2012ns,Schafer:2012ra,Ma:2012ye,Dai:2014ala}, however a method of performing the full evolution of this function has not been well established. 
Nevertheless in the extractions of the Sivers functions in the literature, two approximate schemes for performing this evolution have been used in the literature. For example, in~\cite{Bacchetta:2020gko}, the DGLAP evolution of the Qiu-Sterman function is treated to be the same as the  unpolarized PDF. On the other hand, in~\cite{Sun:2013hua}, the authors use a large-$x$ approximation for the splitting kernel~\cite{Braun:2009mi,Kang:2012em} in the evolution equation of the Qiu-Stermn function. In this paper, we carefully compare the impact of these two schemes on the extraction of the Sivers function.

We perform the first global extraction of the Sivers function from all different processes, including SIDIS at HERMES, COMPASS, and JLab, Drell-Yan lepton pair at COMPASS, and $W/Z$ production at RHIC. To perform the fit, we note that a large number of experimental data are available. At HERMES, the Sivers function has been probed by measuring both pion and kaon production in SIDIS on a proton target~\cite{Airapetian:2009ae}. At COMPASS, the Sivers asymmetries have been measured in \cite{Adolph:2012sp} for unidentified charged hadron production from the proton target, with a re-analysis of this data in \cite{Adolph:2016dvl}. The measurements with a deuteron target are presented in~\cite{Alekseev:2008aa}. The Sivers function has also been probed for a neutron target at JLab for pion production in \cite{Qian:2011py}. To test the modified universality prediction, Drell-Yan Sivers asymmetries have been measured at COMPASS \cite{Aghasyan:2017jop} for virtual photon (or lepton pair) production at relatively small energy scales of $Q\sim$ a few GeV, as well as RHIC \cite{Adamczyk:2015gyk} for $W$ and $Z$ production at much large energy scales, $Q\sim M_{W/Z}$. 

The rest of the paper is organized as follows. In Sec.~\ref{sec:theory}, we summarize the relevant TMD factorization formalism for SIDIS and Drell-Yan processes. In Sec.~\ref{Non-pert}, we first discuss our non-perturbative parameterizations for the unpolarized TMDPDFs and TMDFFs, and benchmark them with the SIDIS hadron multiplicity and Drell-Yan cross section data. We then present our non-perturbative parametrization for the Sivers function, and discuss how we perform the DGLAP evolution of the Qiu-Sterman function. In Sec.~\ref{Fit Results}, we present our fit results, where we explore several different ways for performing the fit. In Sec.~\ref{sid_dy} we present the results of a simultaneous fit to the low energy data from SIDIS and the COMPASS Drell-Yan data. In Sec.~\ref{RHIC} we study the impact of the high energy data from RHIC. In Sec.~\ref{DGLAP} we study the impact of the DGLAP evolution scheme for the Qiu-Sterm function on the fit. In Sec.~\ref{global} we present the global fit where we include Sivers asymmetry data from all processes. In Sec.~\ref{EIC} we give predictions for Sivers asymmetry at the EIC. We conclude our paper in Sec.~\ref{Conclusions}.

%%%%%%%%%%%%%%%%%%%%%%%%%%%%%%%%%%%%%%%%%%%%%%%%%%%%%%%%%%%%%
\section{Formalism}
\label{sec:theory}
In this section, we provide the TMD factorization formalism for the Sivers asymmetry. We begin in Sec.~\ref{SIDIS Formalism} with the SIDIS formalism, while in Sec.~\ref{DY Formalism} and \ref{DY-Vector} we present the formalism for Drell-Yan lepton pair and $W/Z$ boson production, respectively. 
\subsection{Sivers Formalism in SIDIS}\label{SIDIS Formalism}
The differential cross section for SIDIS, $e(\ell)+p\left(P,\textbf{S}_{\perp}\right) \rightarrow e\left(\ell '\right)+h\left(P_{h}\right)+X$, where $\textbf{S}_{\perp}$ is the transverse spin vector of the polarized nucleon, can be written as the following form~\cite{Bacchetta:2006tn,Ji:2004wu}
\begin{align}\label{e.cross}
    \frac{d\sigma}{d\mathcal{PS}} = \sigma_0^{\mathrm{DIS}}\left[ F_{UU} +\sin(\phi_h-\phi_s) F_{UT}^{\sin(\phi_h-\phi_s)} \right]\,,
\end{align}
where the phase space $d\mathcal{PS} = d x_B\, d Q^2\, d z_h\, d^2 P_{h \perp}$, the electron-proton center-of-mass (CM) energy $S = (P+\ell)^2$ and the exchanged virtual photon momentum $q = \ell'-\ell$ with $Q^2= - q^2$, and the usual SIDIS kinematic variables are defined as
\begin{align}
    x_B = \frac{Q^2}{2 P\cdot q}\,,
    \qquad
    y = \frac{Q^2}{x_B S}\,,
    \qquad
    z_h = \frac{P\cdot P_h}{P\cdot q}\,.
\end{align}
\begin{figure}[hbt!]
    \centering
    \includegraphics[width = 0.5\textwidth]{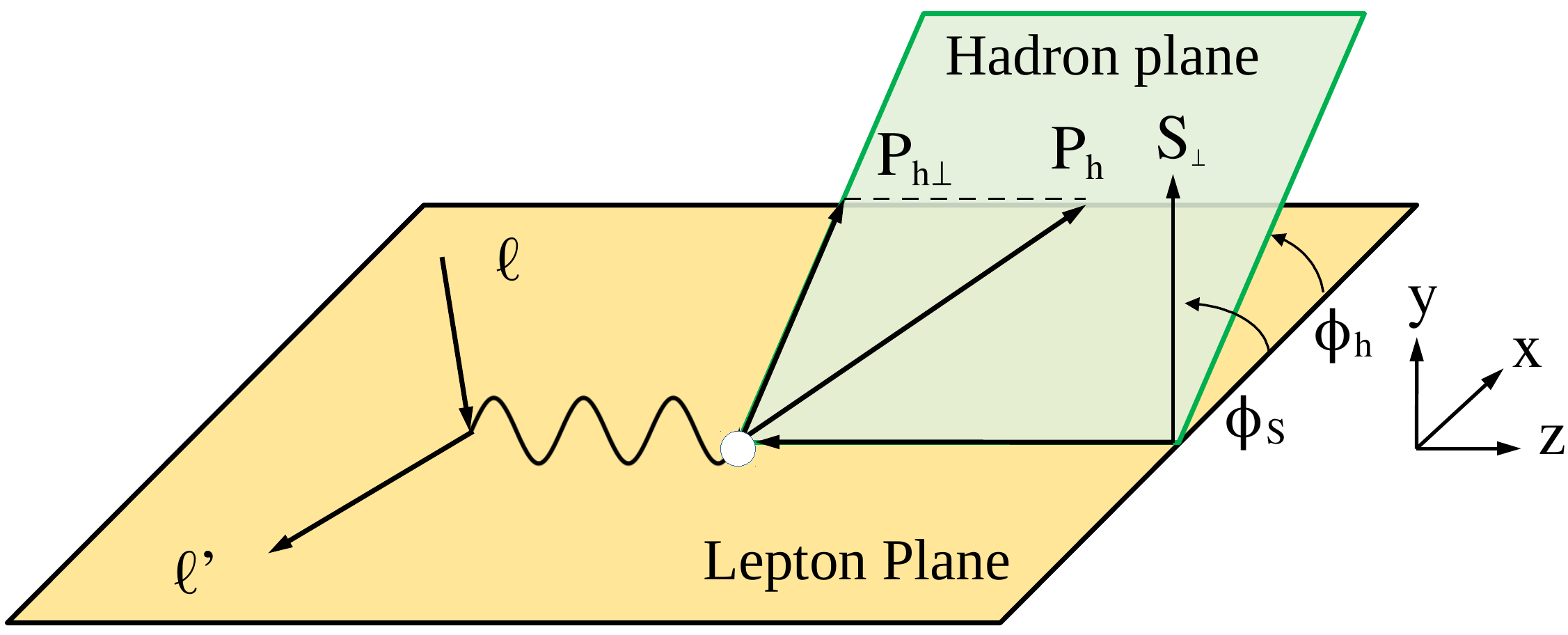}
    \caption{Illustration of azimuthal angles in Semi Inclusive Deep Inelastic Scattering process (SIDIS)}
    \label{f.SIDIS}
\end{figure}
As shown in \fref{f.SIDIS}, the plane which contains the initial and final lepton momentum vectors is the \textit{lepton plane}, while the momentum vectors of the photon and final state hadron form the \textit{hadron plane}. The azimuthal angle of the hadron plane with respect to the lepton plane is denoted $\phi_h$, while the azimuthal angle of the transversely polarized proton spin with respect to the lepton plane is denoted $\phi_s$. We follow the \textit{Trento} conventions~\cite{Bacchetta:2004jz} for the definition of the azimuthal angles. In this expression, $\sigma_0^{\mathrm{DIS}}$ is the leading order (LO) electromagnetic scattering cross section given by
\begin{align}
    \sigma_0^{\mathrm{DIS}} = \frac{2\pi\alpha_{\mathrm{EM}}^2}{Q^4}\left[1+\left(1-y\right)^2\right]\,,
\end{align}
where $\alpha_{\mathrm{EM}}$ is the electromagnetic fine structure constant. 

$F_{UU}$ and $F_{UT}^{\sin(\phi_h-\phi_s)}$ in~\eref{cross} are the unpolarized and transversely polarized structure functions, respectively. The experimentally measured quantity, the Sivers asymmetry, $A_{UT}^{\sin(\phi_h-\phi_s)}$, for this process is given in terms of the structure functions as follows
\begin{align}
    A_{UT}^{\sin(\phi_h-\phi_s)} = \frac{F_{UT}^{\sin(\phi_h-\phi_s)}}{F_{UU}}\,.
\end{align}
The momentum space expression for these structure functions are given by 
\begin{align}
F_{UU}(x_B, z_h, P_{h\perp}, Q) = &\; H^{\mathrm{DIS}}(Q;\mu)
\mathcal{C}^{\rm DIS}\left [f D\right ]\,,
\\
F_{UT}^{\sin(\phi_h -\phi_s)}(x_B, z_h,P_{h\perp}, Q)  = &\; H^{\mathrm{DIS}}(Q;\mu) 
\label{e.sidis_pt} \mathcal{C}^{\rm DIS}\left[-\frac{{\bf {\hat h}}\cdot{\bf k}_\perp}{M}f_{1T}^{\perp} D\right]\,,  
\end{align}
where the hard factor, $H^{\mathrm{DIS}}(Q;\mu)$, is given in \cite{Manohar:2003vb,Idilbi:2005ky} as follows
\begin{align}
    H^{\mathrm{DIS}}(Q; \mu)  = 1+\frac{\alpha_s}{\pi}C_F \left[ \frac{3}{2}\ln\left(\frac{Q^2}{\mu^2}\right) -\frac{1}{2}\ln^2\left( \frac{Q^2}{\mu^2}\right) -4 +\frac{\pi^2}{12}\right]\,. 
\end{align}
In these expressions, we have used the short-hand notation
\begin{align}
\mathcal{C}^{\rm DIS}\left[w A B\right ] = & \sum_q e_q^2 \int d^2 {\bf k}_\perp d^2 {\bf p}_\perp \delta^2\left(z_h {\bf k}_\perp + {\bf p}_\perp - {\bf P}_{h \perp}\right) w({\bf k}_{\perp},{\bf p}_{\perp}) \nn\\
& \times A_{q/p}(x_B, \kt^2;\mu,\zeta_A)\,B_{h/q}(z_h, \pt^2;\mu,\zeta_B)\, ,
\end{align}
for the convolution integrals. In these expressions $e_q$ is the fractional electric charge for the quarks. $\textbf{k}_{\perp}$ represents the transverse momentum of the quark relative to the nucleon, while ${\bf p}_\perp$ is the transverse momentum of the final state hadron relative to the fragmenting quark. $\hat{\textbf{h}} = \textbf{P}_{h\perp}/P_{h\perp}$ is the unit vector which points in the direction of the final-state hadron transverse momentum and $M$ is the mass of the struck nucleon. $f_{q/p}(x_B,k_\perp^2;\mu,\zeta)$ is the unpolarized TMDPDF, while $f_{1T,q/p}^{\perp}(x_B, \kt^2;\mu,\zeta)$ is the SIDIS Sivers function and $D_{h/q}(z_h, \pt^2;\mu,\zeta)$ is the unpolarized TMDFF.
In these expressions $\mu$ and $\zeta$ are the renormalization and rapidity (Collins-Soper) scales~\cite{Collins:2011zzd}, which are used to regulate ultraviolet and rapidity divergences, respectively.
Moreover, the rapidity scales obey the relation $\zeta_A \zeta_B=Q^4$ in the TMD region.

The expressions for the structure functions are simplified by going to the $\bt$-space, the Fourier conjugate space to the transverse momentum space. In the $\bt$-space, these expressions become
\begin{align}
     F_{UU}(x_B, z_h,P_{h\perp}, Q)  = & H^{\mathrm{DIS}}(Q;\mu) \sum_q e_q^2 \int_0^{\infty} \frac{\bt\, d\bt}{2\pi} J_0\left(\frac{\bt P_{h \perp}}{z_h} \right) \nn \\
    & \times
    f_{q/p}(x_B,\bt;\mu,\zeta_A)D_{h/q}(z_h,\bt;\mu,\zeta_B)\,,
\\
     F_{UT}^{\sin(\phi_h-\phi_s)}(x_B, z_h,P_{h\perp}, Q) = & H^{\mathrm{DIS}}(Q;\mu) \sum_q e_q^2 \int_0^{\infty} \frac{\bt^2\, d\bt}{4\pi}J_1\left(\frac{\bt P_{h \perp}}{z_h} \right) \nn\\
    & \times f_{1 T, q/p}(x_B,\bt;\mu,\zeta_A)D_{h/q}(z_h,\bt;\mu,\zeta_B)  \, .
\end{align}
Here the $\bt$-space TMDs are defined as
\begin{align}
    f_{q/p}(x,\bt;\mu,\zeta) =& \int d^2{\bf k}_{\perp}e^{-i{\bf k}_{\perp}\cdot {\bf b}}f_{q/p}(x,\kt^2;\mu,\zeta)\,,
\\
    D_{h/q}(z,\bt;\mu,\zeta) = & \int \frac{d^2{\bf p}_{\perp}}{z^2}e^{-i{\bf p}_{\perp}\cdot {\bf b}/z} D_{h/q}(z,\pt^2;\mu,\zeta)\,,
\\
    f_{1T,q/p}^{\perp\, \alpha\,\rm SIDIS}(x,\bt;\mu,\zeta)     =&
    \frac{1}{M}\int d^2{\bf k}_\perp\, k_{\perp}^{\alpha}\, e^{-i{\bf k}_{\perp}\cdot {\bf b}} f^{\perp\, \rm SIDIS}_{1 T, q/p}(x,\kt^2;\mu,\zeta)
    \nn\\
    \equiv &\left(\frac{ib^\alpha}{2}\right) f_{1T,q/p}^{\perp}(x,\bt;\mu,\zeta)\,.
    \label{e.SIDIS-Sivers}
\end{align}
At small $b$ where $1/b \gg \Lambda_{\rm QCD}$, one can perform an operator product expansion (OPE) of these functions in terms of their collinear counterparts:
\begin{align}
    f_{q/p}&(x,\bt;\mu,\zeta) =  \Big[C_{q\leftarrow i}\otimes f_{i/p}\Big]\left(x,\bt;\mu,\zeta\right) \,,
    \label{e.PDF}
    \\
    D_{h/q} & (z,\bt;\mu,\zeta) = \frac{1}{z^2} \Big[\hat{C}_{i\leftarrow q}\otimes D_{h/i}\Big]\left(z,\bt;\mu,\zeta\right) \,,
    \label{e.FF}
    \\
    f_{1T,q/p}^{\perp} & (x, \bt;\mu,\zeta) = \Big[\bar{C}_{q\leftarrow i} \otimes T_{F\, i/p}\Big](x,\bt;\mu,\zeta) \,,
    \label{e.Sivers}
\end{align}
where $f_{i/p}(x, \mu)$, $D_{h/i}(z, \mu)$ and $T_{F\, i/p}(x_1, x_2, \mu)$ are the collinear PDF, FF and the Qiu-Sterman function, respectively. The operator $\otimes$ denotes the convolution over the parton momentum fractions and are given by
\begin{align}
    \Big[C_{q\leftarrow i}\otimes f_{i/p}\Big]\left(x,\bt;\mu,\zeta\right) = & \int_x^{1} \frac{d\hat{x}}{\hat{x}} C_{q\leftarrow i}\left(\frac{x}{\hat{x}},\bt;\mu,\zeta \right) f_{i/p}\left(\hat{x}; \mu\right)\,,
\end{align}
for $f_{i/p}$ and likewise for $D_{h/i}$. 
In these expressions, the sum over the index $i=q, g$ is implicit. The convolution in the case of the Sivers function is more complicated, since it involves two kinematic variables $\hat{x}_1$ and $\hat{x}_2$:
\begin{align}
&\Big[\bar{C}_{q\leftarrow i} \otimes T_{F\, i/p}\Big](x,\bt;\mu,\zeta) = 
\int_{x}^{1} \frac{d\hat{x}_1}{\hat{x}_1}\frac{d \hat{x}_2}{\hat{x}_2} \bar{C}_{q\leftarrow i}(x/\hat{x}_1,x/\hat{x}_2,\bt;\m,\z) \, 
T_{F\, i/p}(\hat{x}_1,\hat{x}_2;\m)
\,.
\end{align}
The $C$ functions in the above equations are the Wilson coefficient functions, and their expressions at NLO are given in Appendix.~\ref{Coefficient}.

Several comments are in order for the case of the Sivers function. First, although the coefficient function for general scales $\mu$ and $\zeta$ are quite complicated, it becomes much simpler when one chooses the canonical scales $\mu = \sqrt{\zeta} = \mu_b = c_0/b$, with $c_0 = 2e^{-\gamma_E}$ and $\gamma_E$ the Euler constant. such scales are referred to as the natural scale of the TMDs. Second, there are different conventions/normalization for the Qiu-Sterman function. In our case, we first follow the \textit{Trento} convention~\cite{Bacchetta:2004jz} for the quark Sivers function and then the convention for the Qiu-Sterman function is such that the coefficient $\bar C$ function at leading order in Eq.~\eqref{e.Sivers} is a simple delta function. Our convention is related to the so-called first transverse moment of the Sivers function~\cite{Boer:2003cm,Cammarota:2020qcw}
\begin{align}
    f_{1T\, q/p}^{\perp\, (1)}(x;Q) = -\frac{1}{2 M} T_{F\, q/p}(x,x; Q)\,. 
    \label{e.f1T1}
\end{align}
Third, in principle the convolution in Eq.~\eqref{e.Sivers} receives contribution not only from the Qiu-Sterman function which is a quark-gluon-quark twist-3 correlator, but also the so-called twist-3 three-gluon correlator. Since the three-gluon correlator is not well-known at the moment in phenomenology, we neglect all contributions from gluon to quark splitting in the Sivers function~\cite{Dai:2014ala,Scimemi:2019gge}. Finally, \eref{SIDIS-Sivers} is only defined for the Sivers function in SIDIS. Thus if one changes to the Sivers function in Drell-Yan, one should include an additional minus sign in the last line of this expression. 

The large logarithms present in Wilson coefficient functions are resummed in the renormalization group evolution of TMDs from the natural scale $\mu_i^2 = \zeta_i = \mu_{b}^2$ to the hard scale $\mu_f^2 = \zeta_f = Q^2$. Such a TMD evolution is encoded in the exponential factor, $\exp\left[-S\right]$, with the so-called Sudakov form factor $S$. The perturbative part of the Sudakov form factor is given by 
\begin{align}
    S_{\rm pert}(\bt;\mu_i,\zeta_i,\mu_f,\zeta_f) =& \int_{\mu_{i}}^{\mu_f} \frac{d\mu'}{\mu'}\left[ \gamma^{V}+\Gamma_{\rm cusp}\,\ln\left(\frac{\zeta_f}{{\mu'}^2}\right) \right]
    +D(\bt;\mu_i)\ln\left(\frac{\z_f}{\z_i}\right)\,,
\end{align}
where $\Gamma_{\rm cusp}$ and $\gamma^{V}$ are the cusp and non-cusp anomalous dimensions, respectively, and $D$ is the rapidity anomalous dimension (Collins-Soper kernel) \cite{Collins:2011zzd,Echevarria:2016scs}.
In this paper, we perform the resummation of these logarithms up to NNLL. 
All information on the anomalous dimensions up to NNLL are given in Appendices \ref{Anom} and \ref{sec:app}. 

When $b$ becomes large and thus $\mu_b\lesssim \Lambda_{\rm QCD}$, the TMD evolution runs into the non-perturbative region. We follow the usual $b_*$-prescription~\cite{Collins:1984kg} that introduces a cut-off value $b_{\rm max}$ and allows for a smooth transition from perturbative to non-perturbative region,
\bea
b_* = \bt/\sqrt{1+\bt^2/b_{\rm max}^2}\,,
\eea
with $b_{\rm max} = 1.5$ GeV$^{-1}$. With the introduction of $b_*$ in the Sudakov form factor, the total Sudakov form factor can be written as the sum of perturbatively calculable part and non-perturbative contribution. The final expressions for the structure functions are given by
\begin{align}
    & F_{UU} (x_B, z_h,P_{h\perp}, Q) = H^{\mathrm{DIS}}(Q;Q)  \int_0^{\infty} \frac{d\bt\, \bt}{2\pi} J_0\left(\bt \qt \right) \sum_q e_q^2 \\
    &\hspace{1in} \times 
    \Big[C_{q\leftarrow i}\otimes f_{i/p}\Big]\left(x_B,b_*; \mu_{b_*},\mu_{b_*}^2\right) \frac{1}{z_h^2}\Big[\hat{C}_{j\leftarrow q}\otimes D_{h/j}\Big]\left(z_h,b_*;\mu_{b_*},\mu_{b_*}^2\right)
    \nonumber\\
    &\hspace{1in} \times 
    \exp\Big[
    -2S_{\rm pert}(b_*;\mu_{b_*},\mu_{b_*}^2,Q,Q^2)
    -S^f_{\rm NP}(x_B,\bt;Q_0,Q)
    -S^D_{\rm NP}(z_h,\bt;Q_0,Q)\Big]\, \nn , 
\\
    & F_{UT}^{\sin(\phi_h-\phi_s)} (x_B, z_h,P_{h\perp}, Q) = H^{\mathrm{DIS}}(Q;Q)  \int_0^{\infty} \frac{d\bt \, \bt^2 }{4\pi}J_1\left(\bt \qt \right) \sum_q e_q^2 \\
    &\hspace{1in}\times 
    \Big[\bar{C}_{q\leftarrow i} \otimes T_{F\, i/p}\Big](x_B,b_*;\mu_{b_*},\mu_{b_*}^2) \frac{1}{z_h^2} \Big[\hat{C}_{j\leftarrow q}\otimes D_{h/j}\Big]\left(z_h,b_*; \mu_{b_*},\mu_{b_*}^2\right)
    \nonumber \\
    &\hspace{1in} \times 
    \exp\Big[
    -2S_{\rm pert}(b_*;\mu_{b_*},\mu_{b_*}^2,Q,Q^2)
    -S^s_{\rm NP}(x_B,\bt;Q_0,Q)
    -S^D_{\rm NP}(z_h,\bt;Q_0,Q)\Big]\, \nn ,
\end{align}
where we have replaced $\mu_b$ by $\mu_{b_*} = c_0/b_*$, and $Q_0$ is the reference scale of the TMDs. The functions $S_{\rm NP}^f$, $S_{\rm NP}^D$, and $S_{\rm NP}^s$ are the corresponding non-perturbative Sudakov form factors for the unpolarized TMDPDF, TMDFF, and the Sivers function, respectively, and they will be given in the next section. Note that in these expressions we have introduced the vector ${\bf q}_{\perp}=-{\bf P}_{h\perp}/z_h$, while $\qt = |{\bf q}_{\perp}|$ denotes its magnitude. 
\subsection{Sivers Formalism in Drell-Yan}
\label{DY Formalism}
For Drell-Yan scattering, $p(P_A,\textbf{S}_{\perp})+p(P_B)\to [\gamma^*(q)\to]\ell^+\ell^-+X$, the differential cross section with the relevant terms is given in \cite{Arnold:2008kf,Kang:2009sm,Anselmino:2009st,Huang:2015vpy} by the expression
\begin{align}
\frac{d\sigma}{d\mathcal{PS}} = \sigma_0^{\mathrm{DY}} \left[ W_{UU} +\sin(\phi_q-\phi_s)W_{UT}^{\sin(\phi_q-\phi_s)}\right]\,,
\label{e.fact_dy}
\end{align}
where $d\mathcal{PS} = dQ^2\, dy\, d^2\qt$, $y$ is the rapidity of the lepton pair while ${\bf \qt}$ and $Q$ are the transverse momentum and invariant mass of the virtual photon, respectively. Here, $W_{UU}$ and $W_{UT}^{\sin(\phi_q-\phi_s)}$ are the unpolarized and transversely polarized structure functions. Note that we have deviated from the notation in \cite{Kang:2009sm} by writing the Drell-Yan structure functions as $W$ in order to differentiate them from the SIDIS structure function. The leading order electro-magnetic scattering cross section is given by
\begin{align}
\sigma_0^{\mathrm{DY}} = \frac{4\pi  \alpha_{\mathrm{EM}}^2}{3SQ^2N_C}\,,
\end{align}
where $S = (P_A+P_B)^2$ is the center of mass energy squared and $N_C=3$ is the number of color. 
\begin{figure}[hbt!]
    \centering
    \includegraphics[width = 0.5\textwidth]{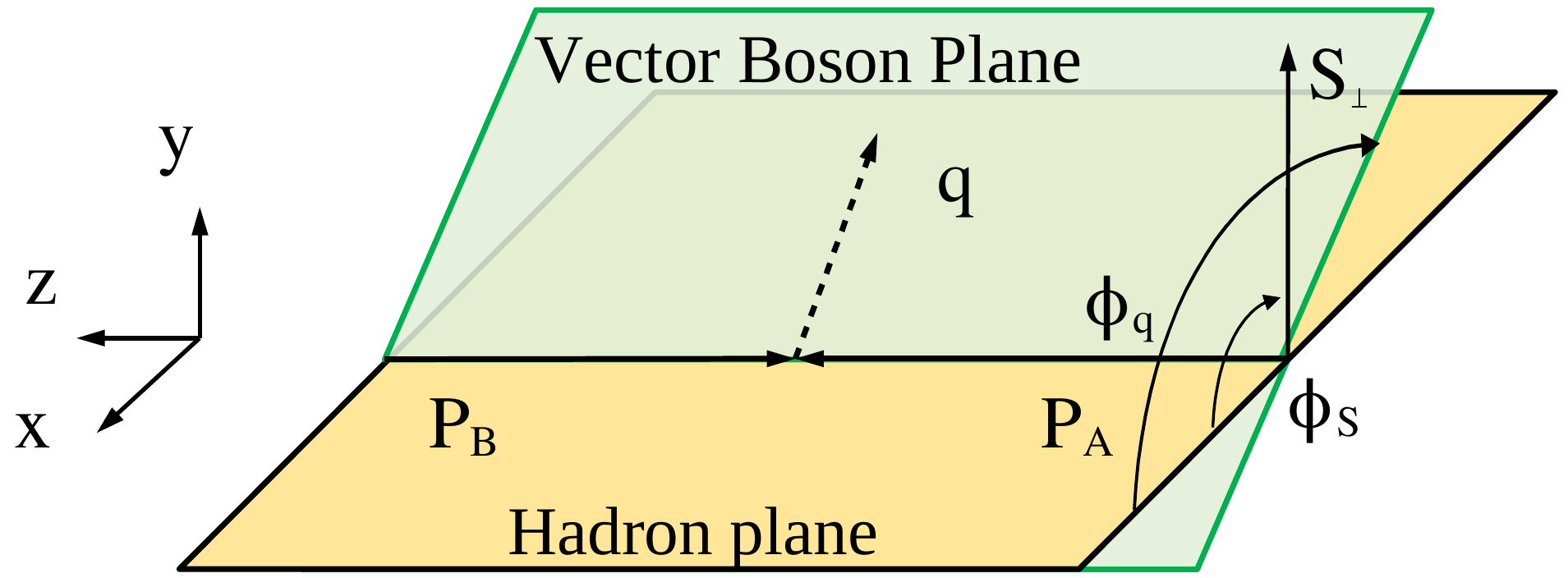}
    \caption{Illustration of Drell-Yan vector boson production in collisions of hadrons $A$ and $B$. The transversely polarized hadron $A$ is moving in $+z$-direction, while the unpolarized hadron $B$ is along $-z$-direction. We denote the vector boson using a dotted line.
    }
    \label{f.Drell-Yan}
\end{figure}

As shown in \fref{f.Drell-Yan}, the plane which is perpendicular to the spin vector $S_{\perp}$ and which also contains the initial hadrons forms the \textit{hadron plane}. The plane which contains the hadron momenta and which contains the vector boson (i.e. $\gamma^*$ here) momentum generates the \textit{vector boson plane}. We use the convention that the polarized hadron moves in the $z$ direction while $S_\perp$ moves in the $y$ direction. We note that the convention for the $x$ and $z$ axes must be reversed in order to compare with the COMPASS Drell-Yan data. For the Drell-Yan production, $\phi_q$, the azimuthal angle of the vector boson, and $\phi_s$, the azimuthal angle of $\bf{S}_{\perp}$ generate the $\sin(\phi_q - \phi_s)$ modulation for this process. 

Analogous to the asymmetry in SIDIS, the Drell-Yan Sivers asymmetry can be written in terms of the structure function as~\footnote{Note that another single spin asymmetry denoted as $A_N$ for Drell-Yan process has also been frequently used in the literature, which is related to the Sivers asymmetry defined here by a minus sign: $A_{N} = -A_{UT}^{\sin(\phi_q-\phi_s)}$. For details, see~\cite{Kang:2009sm}.}
\begin{align}
    A_{UT}^{\sin(\phi_q-\phi_s)} = \frac{W_{UT}^{\sin(\phi_q-\phi_s)}}{W_{UU}}.
    \label{e.AUT_V}
\end{align}
In the TMD formalism, these structure functions are given by the following expressions
\begin{align}
W_{UU}(x_a,x_b,\qt,Q) = &\; H^{\mathrm{DY}}(Q;\mu)
\mathcal{C}^{\rm DY}\left[f\;f\right]\,,
\\
W_{UT}^{\sin(\phi_q-\phi_s)}(x_a,x_b,\qt,Q)  =& \;H^{\mathrm{DY}}(Q;\mu)  \mathcal{C}^{\rm DY}\left[\frac{{\bf {\hat q}}_{\perp}\cdot{\bf k}_{a \perp}}{M}\;f_{1T}^{\perp}\;f\right]\,.
\end{align}
For Drell-Yan process, the above convolution in the structure functions is given by
\bea
    \mathcal{C}^{\rm DY}\left[w A B\right]   = &  \sum_q e_q^2 
    \int d^2 {\bf k}_{a \perp} d^2{\bf k}_{b \perp} \delta^2\left({\bf k}_{a \perp} + {\bf k}_{b \perp} - {\bf q}_{\perp}\right) 
    w({\bf k_{a \perp}},{\bf k_{b \perp}})
\nn \\
    & \times A_{q/A}(x_a, k_{a\perp}^2;\mu,\z_A)
    \,B_{\bar{q}/B}(x_b,k_{b \perp}^2;\mu,\z_B)  \,,
\eea
where $x_a$ and $x_b$ are the momentum fractions of the hadrons carried by the quarks and are given by
\begin{align}
    x_a = \frac{Q}{\sqrt{S}}e^y 
    \,,
    \qquad
    x_b = \frac{Q}{\sqrt{S}}e^{-y}
    \,.
\end{align}
The usual Feynman-$x$ is related to $x_{a,b}$ as follows $x_F = x_a - x_b$, which will be used in the next section. On the other hand, ${\bf k}_{a \perp}$ and ${\bf k}_{b \perp}$ are the transverse momenta of the parton relative to their corresponding nucleon. The hard function is given in~\cite{GarciaEchevarria:2011rb} by 
\begin{align}
H^{\rm DY}(Q; \mu)  =  1+\frac{\alpha_s}{\pi}C_F \left[\frac{3}{2}
\ln\left(\frac{Q^2}{\mu^2}\right) -\frac{1}{2}\ln^2\left( \frac{Q^2}{\mu^2}\right) +\frac{7}{12}\pi^2 - 4 \right]\,. 
\end{align}

The expressions for the structure functions can once again be simplified by going to the $\bt$-space. At this point, it might be important to emphasize again that the Sivers function $f_{1T}^{\perp}$ above for the Drell-Yan process differs by a sign from that in SIDIS in~\eref{sidis_pt}:
\bea
f_{1T}^{\perp\, \rm DY}(x, k_\perp^2; \mu, \zeta) = -f_{1T}^{\perp\, \rm SIDIS}(x, k_\perp^2; \mu, \zeta)\,.
\label{eq:sign}
\eea
This will lead to slightly different definition for the Sivers function in the $\bt$-space:
\bea
    f_{1T,q/p}^{\perp\, \alpha\,\rm DY}(x,\bt;\mu,\zeta) 
    =&
    \frac{1}{M}\int d^2{\bf k}_\perp\, k_{\perp}^{\alpha}\, e^{-i{\bf k}_{\perp}\cdot {\bf b}}  f^{\perp\, \rm DY}_{1 T, q/p}(x,\kt^2;\mu,\zeta)
    \nn\\
    \equiv &\left(- \frac{ib^\alpha}{2}\right) f_{1T,q/p}^{\perp}(x,\bt;\mu,\zeta)\,.
 \eea   
Note the additional minus sign in the second line of the equation, in comparison with the corresponding SIDIS expression in Eq.~\eqref{e.SIDIS-Sivers}. The final expressions for the $\bt$-space structure functions are given by
\begin{align}
    & W_{UU} (x_a,x_b,\qt,Q) =   H^{\mathrm{DY}}(Q;Q) \int  \frac{d\bt \, \bt}{2\pi} J_{0}(\bt\qt) \sum_q e_q^2 
    \\
    &\hspace{1in}\times 
    \Big[C_{q\leftarrow i}\otimes f_{i/A}\Big] \left(x_a,b_*; \mu_{b_*},\mu_{b_*}^2\right) \Big[C_{\bar{q}\leftarrow j}\otimes f_{j/B}\Big] \left(x_b,b_*; \mu_{b_*},\mu_{b_*}^2\right)
    \nn \\
    &\hspace{1in}\times 
    \exp\Big[
    -2S_{\rm pert}(b_*;\mu_{b_*},\mu_{b_*}^2,Q,Q^2)
    -S^f_{\rm NP}(x_a,\bt;Q_0,Q)
    -S^f_{\rm NP}(x_b,\bt;Q_0,Q)\Big] \nn \,, 
\\
    & W_{UT}^{\sin(\phi_q-\phi_s)} (x_a,x_b,\qt,Q) =  H^{\mathrm{DY}}(Q;Q) \int  \frac{d\bt \, \bt^2}{4\pi}J_{1}(\bt\qt)\sum_q e_q^2
    \\
    &\hspace{1in}\times 
    \Big[\bar{C}_{q\leftarrow i} \otimes T_{F\, i/p}\Big](x_a,b_*;\mu_{b_*},\mu_{b_*}^2) \Big[C_{\bar{q}\leftarrow j}\otimes f_{j/B}\Big] \left(x_b,b_*;\mu_{b_*},\mu_{b_*}^2\right) \nn \\
    &\hspace{1in} \times 
    \exp\Big[
    -2S_{\rm pert}(b_*;\mu_{b_*},\mu_{b_*}^2,Q,Q^2)
    -S^s_{\rm NP}(x_a,\bt;Q_0,Q)
    -S^f_{\rm NP}(x_b,\bt;Q_0,Q)\Big] \nn \,.
\end{align}
Note that in the second expression, we have already taken into account the sign change in the Sivers functions between DY and SIDIS processes in Eq.~\eqref{eq:sign}. 

\subsection{Sivers formalism for $W/Z$ Production}\label{DY-Vector}
The case for $W/Z$ boson production in the proton-proton collisions is similar to the case for virtual photon production. In this case, the hard scale $Q$ is set equal to the mass of the produced vector boson, $Q=M_{W,\, Z}$. The expression for the differential cross section is given by
\begin{align}
\frac{d\sigma_{V}}{d\mathcal{PS}} = \sigma_0^{\mathrm{V}} \left[ W_{UU,V} +\sin(\phi_q-\phi_s)W_{UT,V}^{\sin(\phi_q-\phi_s)}\right]\,,
\end{align}
where the phase space $d\mathcal{PS} = dy\, d^2\qt$ and $V=W,\, Z$. The leading-order scattering cross sections are given by
\begin{align}
    \sigma_0^W =& \frac{\sqrt{2}\pi G_F M_W^2}{S N_C}\,,
    \\
    \sigma_0^Z =& \frac{\sqrt{2}\pi G_F M_Z^2}{S N_C}\,,
\end{align}
where $G_F$ is the Fermi weak coupling constant. On the other hand, the structure functions are given by
\begin{align}
    & W_{UU,V}(x_a,x_b,\qt,Q) = H^{\mathrm{DY}}(Q;Q) \int  \frac{d\bt \, \bt}{2\pi} J_{0}(\bt\qt) \sum_{q,q'} e_{qq',V}^2 \\
    &\hspace{1in}\times
    \Big[C_{q\leftarrow i}\otimes f_{i/A}\Big] \left(x_a,b_*; \mu_{b_*},\mu_{b_*}^2\right) \Big[C_{q'\leftarrow j}\otimes f_{j/B}\Big] \left(x_b,b_*; \mu_{b_*},\mu_{b_*}^2\right)\nonumber \\
    &\hspace{1in}
    \times \exp\Big[
    -2S_{\rm pert}(b_*;\mu_{b_*},\mu_{b_*}^2,Q,Q^2)- S^f_{\rm NP}(x_a,\bt;Q_0,Q)
    - S^f_{\rm NP}(x_b,\bt;Q_0,Q)\Big] \nn \,,
\\
    & W_{UT,V}^{\sin(\phi_q-\phi_s)} (x_a,x_b,\qt,Q) = H^{\mathrm{DY}}(Q;Q) \int  \frac{d\bt \, \bt^2}{4\pi}J_{1}(\bt\qt) \sum_{q,q'} e_{qq',V}^2 \\
    &\hspace{1in}\times
    \Big[\bar{C}_{q\leftarrow i} \otimes T_{F\, i/p}\Big](x_a,b_*;\mu_{b_*},\mu_{b_*}^2) \Big[C_{q'\leftarrow j}\otimes f_{j/B}\Big] \left(x_b,b_*;\mu_{b_*},\mu_{b_*}^2\right) \nonumber \\
    &\hspace{1in}
    \times \exp\Big[
    -2S_{\rm pert}(b_*;\mu_{b_*},\mu_{b_*}^2,Q,Q^2)
    -S^s_{\rm NP}(x_a,\bt;Q_0,Q)
    -S^f_{\rm NP}(x_b,\bt;Q_0,Q)\Big] \nn \,, 
    \label{e.UTV}
\end{align}
where we have 
\begin{align}
    e_{qq',W}^2 = |V_{qq'}|^2\,,
    \qquad
    e_{qq',Z}^2 = \left(V_q^2+A_q^2\right) \delta_{qq'}\,.
\end{align}
Here $|V_{qq'}|^2$ is the CKM matrix, while $V_q$ and $A_q$ are the vector and axial couplings of the $Z$ boson to a quark of flavor $q$. Just like Eq.~\eqref{e.AUT_V} in the last section, the asymmetry can be written as a ratio of these structure functions in the exactly same form.

%%%%%%%%%%%%%%%%%
\section{Non-Perturbative Parameterization}\label{Non-pert}
Now that we have included all of the perturbative elements of the Sivers asymmetry, we begin discussing the non-perturbative contributions to the Sivers function.
As we have seen in the previous section, the Sivers asymmetry depends not only on the Sivers functions but the unpolarized TMDs as well.
Therefore, in order to isolate the fit to affect only the Sivers function from these experimental data, it is first necessary to fix the non-perturbative evolution of the unpolarized TMDs.
In Sec.~\ref{Unpolarized}, we choose a parameterization for the unpolarized TMDPDF and TMDFF from a previous extraction and use this formalism to describe unpolarized SIDIS and Drell-Yan data. 
In Sec.~\ref{Polarized} we provide the details of our numerical scheme for the Sivers function.
\subsection{Numerical Scheme for Unpolarized TMDs}\label{Unpolarized}
\begin{figure*}[hbt!]
    \centering
    \includegraphics[width = 0.75\textwidth]{\FigPath/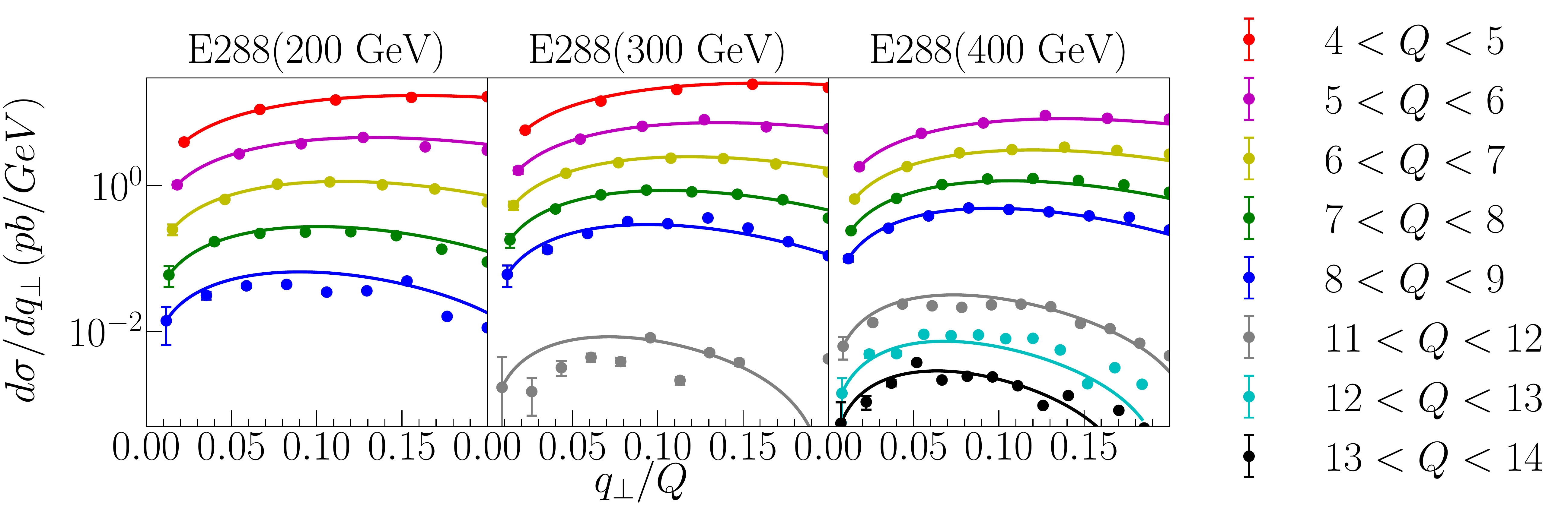}
    \caption{The experimental data for Drell-Yan lepton pair production measured by the E288 collaboration~\cite{Ito:1980ev} plotted as a function of $q_\perp/Q$ are compared with the normalized theoretical curve. Different colors represent different invariant mass of the lepton pair from $4<Q<5$, $5<Q<6$, $6<Q<7$, $7<Q<8$, $8<Q<9$, $11<Q<12$, $12<Q<13$, $13<Q<14$ GeV, respectively. Three panels correspond to different energies for incident proton beams: 200 GeV (left), 300 GeV (middle), and 400 GeV (right).}
    \label{f.E288}
\end{figure*}
\begin{figure*}[hbt!]
    \centering
    \includegraphics[width = 0.42\textwidth]{\FigPath/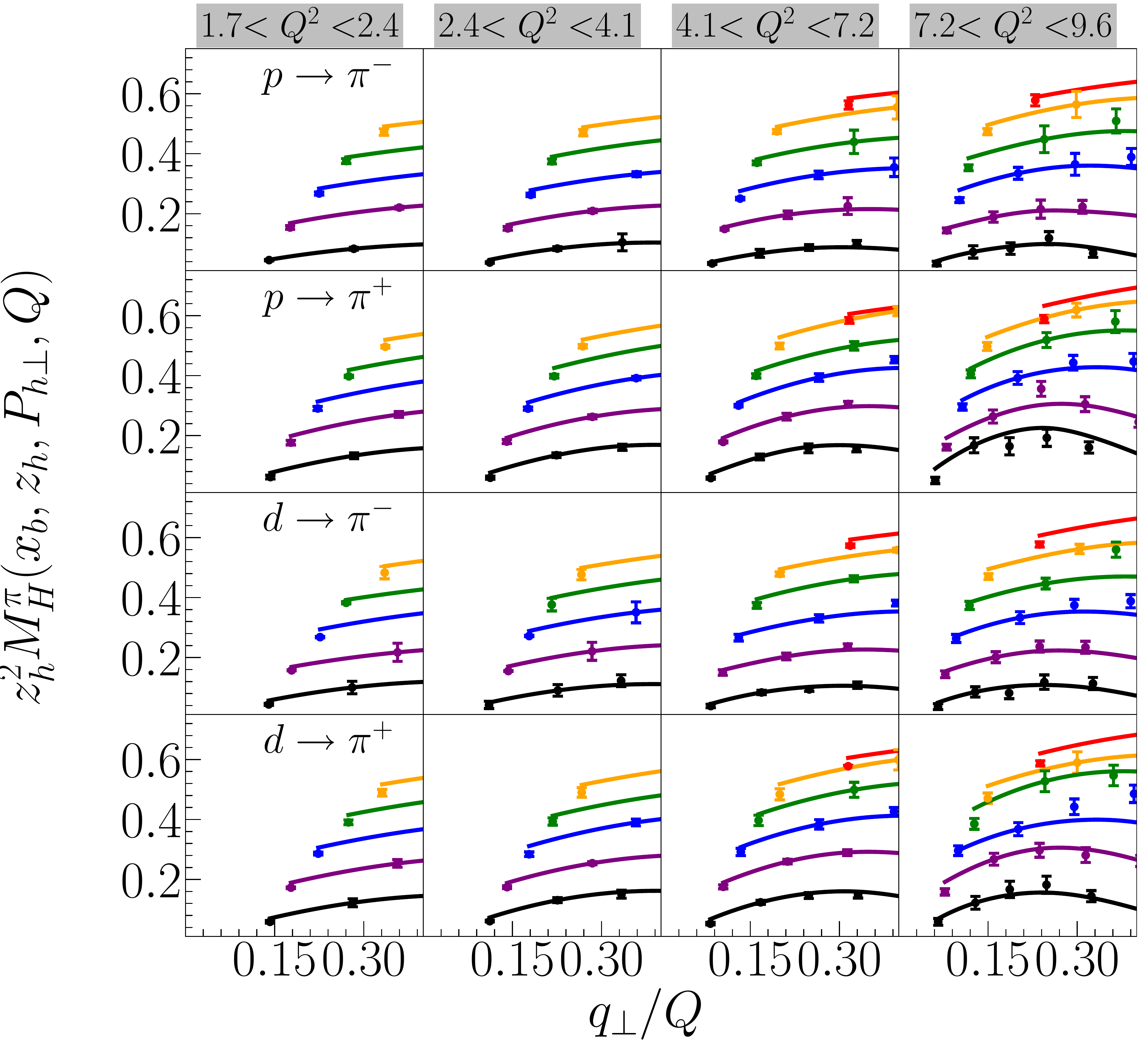}
    \includegraphics[width = 0.57\textwidth]{\FigPath/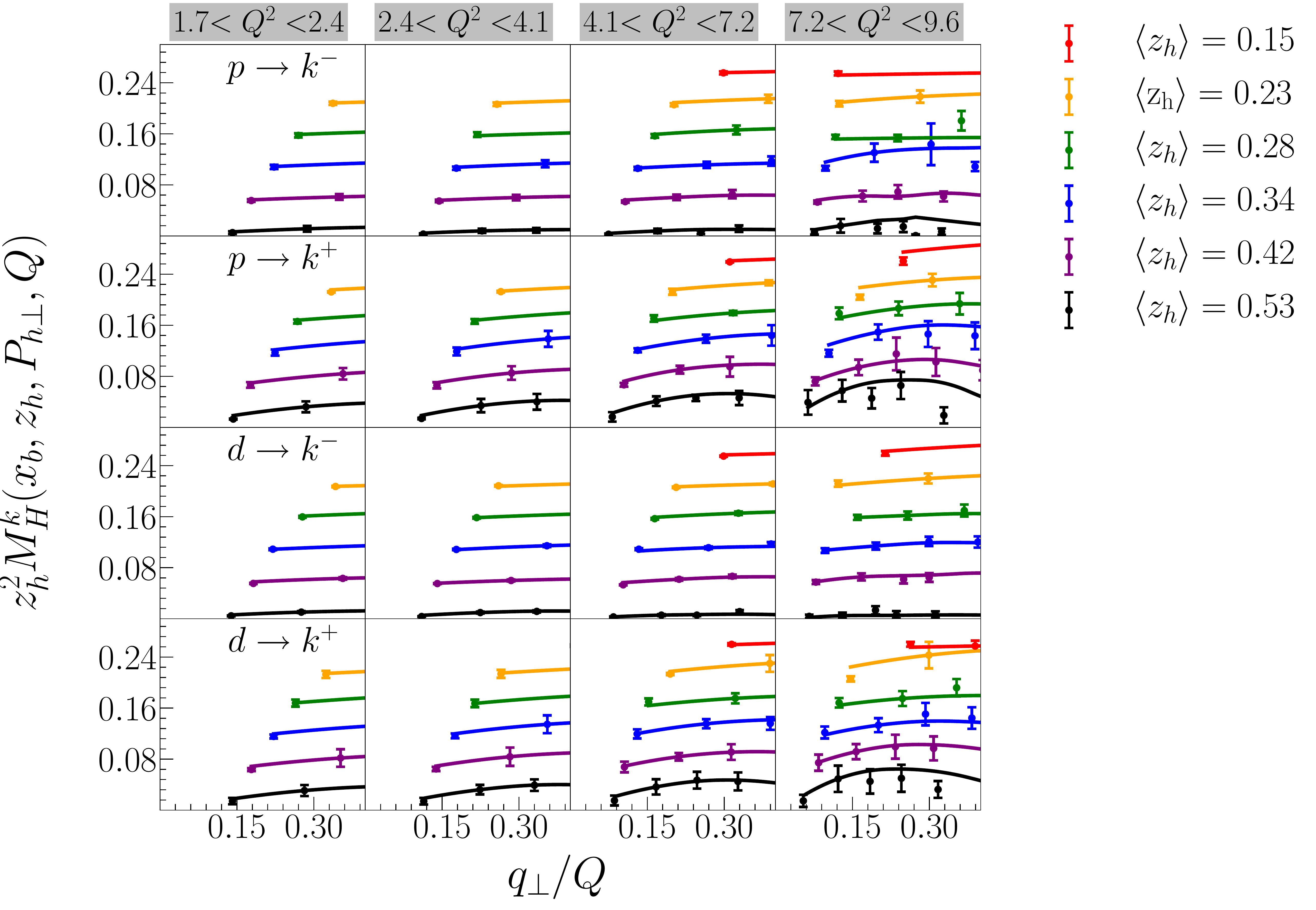}
    \caption{Left panel: The HERMES multiplicity data in \cite{Airapetian:2012ki} for pion production from either a proton (denoted as $p\to \pi$) or deuteron (denoted as $d\to \pi$)  target. For better presentation, the data is offset by 
     0.0 for $\langle z_h \rangle = 0.53$, 
     0.1 for $\langle z_h \rangle = 0.42$, 
     0.2 for $\langle z_h \rangle = 0.34$, 
     0.3 for $\langle z_h \rangle = 0.28$, 
     0.4 for $\langle z_h \rangle = 0.23$, 
 and 0.5 for $\langle z_h \rangle = 0.15$. Right panel: The HERMES multiplicity data for kaon production. The offsets are half of the offsets from the pions.}
    \label{f.H_Mult}
\end{figure*}
\begin{figure*}[hbt!]
    \centering
    \includegraphics[width = \textwidth]{\FigPath/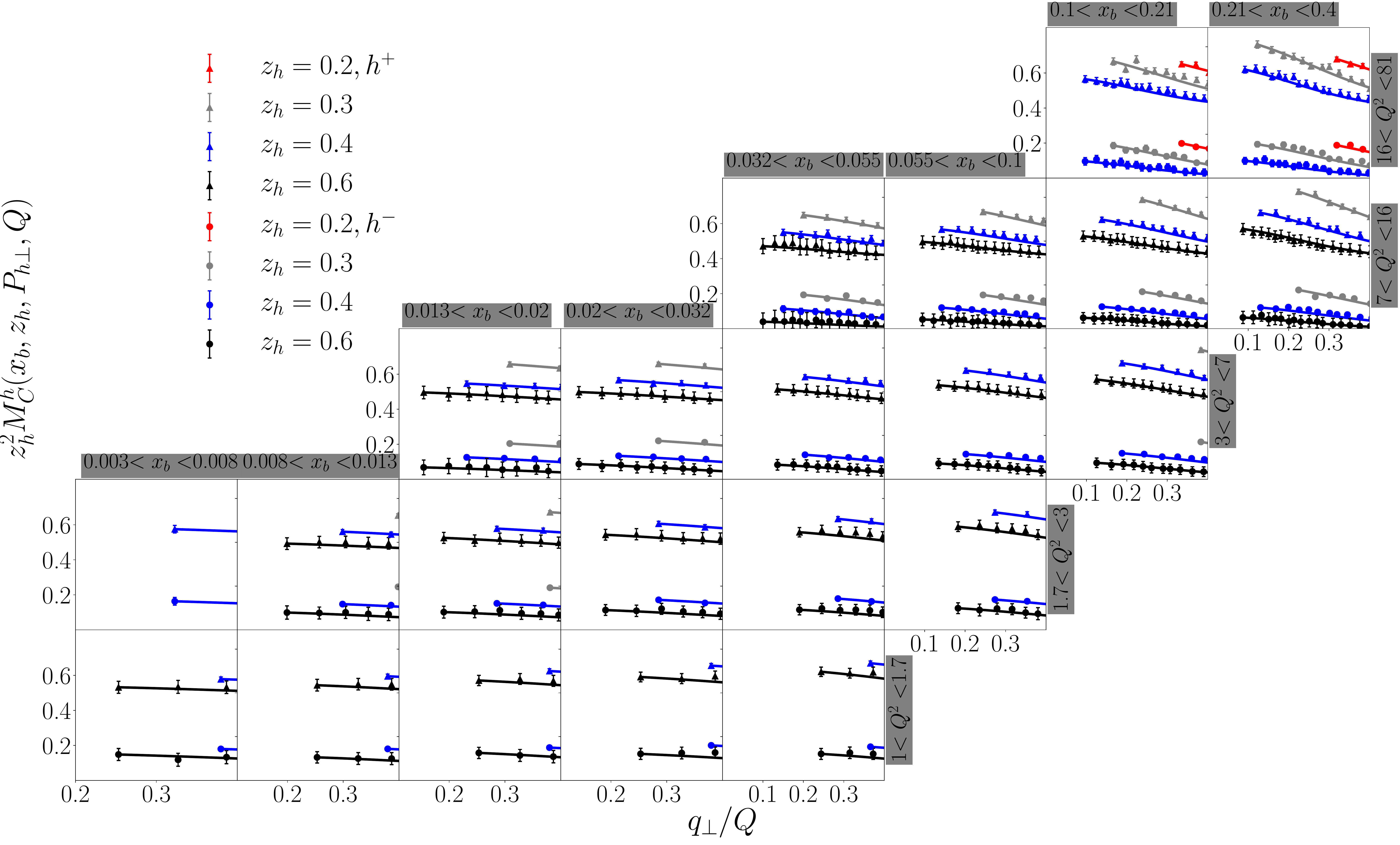}
    \caption{The COMPASS multiplicity in~\cite{Aghasyan:2017ctw} for charged hadron production from a deuteron target is compared with the normalized theory curve. The triangular points represent the $h^+$ data points while the circular data points represent the $h^-$ data points. For better presentation, the $h^+$ data is offset by a factor of $0.4$. }
    \label{f.C_Mult}
\end{figure*}
The non-perturbative evolution functions for the unpolarized TMDs have been extracted widely in the literature. 
Because we perform a simultaneous fit between SIDIS and Drell-Yan data in this paper, the appropriate parameterizations for the unpolarized TMDs are those that have also been obtained in simultaneous fits. 
Furthermore since we perform our fit at NLO+NNLL, the optimal parameterization is one that has been obtained at the same perturbative order.

Simultaneous extractions from SIDIS and Drell-Yan data have been performed in \cite{Su:2014wpa,Bacchetta:2017gcc,Pisano:2018skt,Scimemi:2019cmh} \footnote{We note that the fits \cite{Su:2014wpa,Bacchetta:2017gcc,Pisano:2018skt} all introduced normalization corrections in the fitting procedure so that the shape of the asymmetry is described but the size is not. 
Currently a systematic way of addressing these normalization issues has not been addressed.
While this issue has remained a challenge for unpolarized fit, this issue is not present in asymmetry data. 
For example in \cite{Echevarria:2014xaa,Bacchetta:2020gko} LO+NLL fits were performed to the Sivers asymmetry without issues while in \cite{Cammarota:2020qcw} the Sivers function have been extracted using a Gaussian.}. 
In \cite{Su:2014wpa} the extraction was performed at NLO+NLL.
Similarly in \cite{Bacchetta:2017gcc,Pisano:2018skt} the extraction was performed at LO+NLL. 
In \cite{Scimemi:2019cmh}, the authors performed the fit of the unpolarized data at NNLO+N$^3$LL level, where they further included both $m/Q$ and $\qt/Q$ power corrections. This could introduce additional complications when performing the fit to the Sivers asymmetry, since those power corrections are likely to be different for spin-dependent cross sections. 

In view of the current status, we choose the non-perturbative parametrization in~\cite{Su:2014wpa} for the unpolarized TMDs in our study at NLO+NNLL accuracy. We will first verify that such a parametrization describes the unpolarized experimental data well. 
From \cite{Su:2014wpa}, the non-perturbative factors in Eqs.~\eqref{e.PDF} and \eqref{e.FF} have the following form
\begin{align}
    S_{\rm NP}^f(\bt;Q_0,Q) =& \frac{g_2}{2}\ln{\frac{Q}{Q_0}}\ln{\frac{\bt}{b_*}}+g_1^f \bt^2\,,
    \label{e.NP_upol_f}
    \\
    S_{\rm NP}^D(z,\bt;Q_0,Q) =& \frac{g_2}{2}\ln{\frac{Q}{Q_0}}\ln{\frac{\bt}{b_*}} + g_1^D \frac{\bt^2}{z^2} \,.
    \label{e.NP_upol_D}
\end{align}
The factors which contain $g_1^f$ and $g_1^D$ contain information on the Gaussian width of the TMDs in momentum space at the initial scale $Q_0$, while the factor which involves $g_2$ controls how the TMDs evolve from $Q_0$ to $Q$. The latter is universal to all TMDs~\cite{Collins:2011zzd} and will enter into our discussion in the Sivers non-perturbative parameterization. The values of the parameters that were obtained in this reference are given by
\begin{align}
    g_1^f = 0.106\,,
    \qquad
    g_1^D = 0.042\,,
    \qquad
    g_2 = 0.84\, .
    \label{e.g12}
\end{align}
Note that in the expression of \eref{NP_upol_f}, the non-perturbative parameterization is independent of $x$. Thus we have dropped explicit dependence on the variable $x$. At this point, it is important to note that for the COMPASS Drell-Yan data in \cite{Aghasyan:2017ctw}, the asymmetry was measured for $\pi+p$ scattering. In \cite{Wang:2017zym} the pion TMDPDF was extracted from the experimental data in~\cite{Conway:1989fs} and it was found that $g_1^f=0.082$ for pions.

To perform numerical calculations, we choose to use  \texttt{HERA\_NLO\_as\_118} parametrization in \cite{Abramowicz:2015mha} for the collinear parton distribution functions. For the collinear pion fragmentation function, $D_{\pi/q}(z_h, \mu_{b_*})$, we use the DSS14 parameterization \cite{deFlorian:2014xna}. While for the collinear kaon fragmentation function $D_{K/q}(z_h,\mu_{b_*})$, we use the DSS17 parameterization in \cite{deFlorian:2017lwf}. For unidentified charged hadrons, we follow the work in \cite{Scimemi:2019cmh} to use the approximation $D_{h/q}(z,\mu_{b_*}) = D_{\pi/q}(z,\mu_{b_*})+D_{K/q}(z,\mu_{b_*})$. 

To demonstrate that this parameterization describes the unpolarized TMDs, we now compare this numerical scheme with the unpolarized TMD data. We start this comparison by examining a sample of Drell-Yan data in order to check the validity of the scheme for the TMDPDF. We note that the Drell-Yan Sivers asymmetry data which enters into our fit from COMPASS and RHIC do not contain so-called fiducial cuts. In order to avoid complications associated with these cuts on Drell-Yan data, we choose to benchmark our expression for the unpolarized cross section against the E288 data~\cite{Ito:1980ev}, which also does not contain fiducial cuts, see Tab.~2 of \cite{Bacchetta:2019sam}. For E288, the target nucleus is Copper. In order to describe the Copper TMDPDF, we use nuclear modification prescription in \cite{Eskola:1998df}. In \fref{f.E288}, we plot the theoretical curve against the experimental data~\cite{Ito:1980ev}, as a function of $q_\perp/Q$. For each bin, we have normalized the theory such that the theory and data are equal at the first point. Different colors represent different invariant mass of the lepton pair from $4<Q<5$, $5<Q<6$, $6<Q<7$, $7<Q<8$, $8<Q<9$, $11<Q<12$, $12<Q<13$, $13<Q<14$ GeV, respectively. Three panels correspond to different energies for incident proton beams: 200 GeV (left), 300 GeV (middle), and 400 GeV (right). We find that the parameterization of \cite{Su:2014wpa} is well-suited at describing the shape of the Drell-Yan data.

To check the validity of our scheme for the unpolarized TMDFFs, we now examine the HERMES multiplicity defined as
\begin{align}
    M^h_{\rm H}(x_B,z_h,P_{h \perp},Q^2) = \frac{\left(d\sigma/dx_B dz_h dQ^2dP_{h\perp}\right)}{\left(d\sigma_{\rm DIS}/dx_B dQ^2\right)}\,,
    \label{e.multiplicity}
\end{align}
where the superscript $h$ denotes the species of the final state observed hadron, and the subscript ``H'' represents the HERMES data. We also study the COMPASS multiplicity data, which has a slightly different convention and is given by 
\bea
M^h_{\rm C} = 2P_{h\perp} M^h_{\rm H}\,,
\eea
where the subscript ``C'' denotes the COMPASS data and $M^h_{\rm H}$ is defined in \eref{multiplicity}. On the other hand, the denominator in Eq.~\eqref{e.multiplicity} is the inclusive DIS cross section and is given by
\begin{align}
    \frac{d\sigma_{\rm DIS}}{dx_B dQ^2} = \frac{\sigma_0^{\rm DIS}}{x_B}\left[F_2(x_B,Q^2)-\frac{y^2}{1+(1-y)^2} F_L(x_B, Q^2)\right]\,,
\end{align}
where $F_2$ is the usual DIS structure function while $F_L$ is the longitudinal structure function. For their precise definitions see \cite{Aaron:2012qi}. We compute the denominator at the NLO by using the APFEL library \cite{Bertone:2013vaa}.

In the left panel of \fref{f.H_Mult} we plot the HERMES pion multiplicity data~\cite{Airapetian:2012ki} as a function of $\qt/Q$ along with the numerical results for the theory. In the right panel of this figure we plot kaon multiplicity data and theory. As shown in the figure, different colors represent different average $z_h$ values from $\langle z_h\rangle = 0.15, \, 0.23, \, 0.28, \, 0.34, \, 0.34, \, 0.42, \, 0.53$, respectively. In these plots, we have normalized the theory so that data is equal to the theory at the second point of each data set~\footnote{Without normalizing to the second point of the data, we find that the overall normalization factor is around $2$ for each data set, which is consistent with the results of \cite{Su:2014wpa}.}. In \fref{f.C_Mult}, we plot the COMPASS multiplicity data~\cite{Aghasyan:2017ctw} for charged hadron production from a deuteron target along with the numerical results of our scheme. The triangular points represent the $h^+$ data points while the circular data points represent the $h^-$ data points. Here again, different colors represent different $z_h = 0.2,\, 0.3, \, 0.4, \, 0.6$, respectively. From these plots, we find that the presented parameterization work very well at describing the shape of the multiplicity data for both HERMES and COMPASS data, indicating that the scheme for the TMDFFs are valid. 

\subsection{Numerical Scheme for Sivers Function}\label{Polarized}
Now that the non-perturbative evolution for the unpolarized TMDs have been fixed, we present the numerical scheme for the Sivers function in our fit. Analogous to the unpolarized TMDPDF, we take the polarized non-perturbative parameterization
\begin{align}
    S_{\rm NP}^s(\bt;Q_0,Q) = \frac{g_2}{2}\ln{\frac{Q}{Q_0}}\ln{\frac{\bt}{b_*}}+g_1^T \bt^2\,.
\end{align}
As we have emphasized in the previous section, the parameter $g_2$ is spin-independent and thus we take the same value as in the unpolarized TMDs in Eq.~\eqref{e.g12}. On the other hand, we introduce the parameter $g_1^T$, which describes the Gaussian width of the momentum space distribution for the Sivers function and will be a fit parameter. We once again note that since this parameterization is independent of $x$, we will drop its explicit dependence in future notation. 

For the Qiu-Sterman function $T_{F\, q/p}$, we find that the parameterization in \cite{Echevarria:2014xaa} is still the most economical choice, which sets $T_{F\, q/p}(x, x, \mu_0)$ to be proportional to the unpolarized PDF $f_{q/p}(x,\mu_0)$ at some initial scale $\mu_0$:
\begin{align}
    T_{F\, q/p}(x,x,\mu_0) = \mathcal{N}_q(x)f_{q/p}(x,\mu_0)\,,
\end{align}
with $\mathcal{N}_q(x)$ given by 
\begin{align}
    \mathcal{N}_q(x) = N_q\frac{\left( \alpha_q+\beta_q \right)^{\left( \alpha_q+\beta_q \right)}}{\alpha_q^{\alpha_q} \beta_q^{\beta_q}}x^{\alpha_q}(1-x)^{\beta_q}\,.
\end{align}
Note that $\mathcal{N}_q(x)$ characterizes the non-perturbative collinear physics of the Qiu-Sterman function and is to be fit from the experimental data. In this expression, the parameters $\alpha_u$ and $N_u$ are used to fit the up quarks. $\alpha_d$ and $N_d$ are the fit parameters for the down quarks and $N_{\bar{u}}$, $N_{\bar{d}}$, $N_{s}$, $N_{\bar{s}}$, $\alpha_{sea}$ are for sea quarks and $\beta_q = \beta$ is the same for all flavors. This parameterization enforces that the form of the sea quarks is the same while the normalization of each sea quark can vary. Overall we use 11 parameters in total to perform the fit, including $g_1^T$. 

In order to obtain a numerical result for the Sivers function in \eref{Sivers}, DGLAP evolution of the Qiu-Sterman function must be performed from $\mu_0$ to the natural scale, $\mu_{b_*}$. As we have emphasized, the DGLAP evolution of the Sivers function has been studied extensively in the literature, see for instance \cite{Kang:2008ey,Zhou:2008mz,Vogelsang:2009pj,Braun:2009mi,Kang:2012em,Kang:2012ns,Schafer:2012ra,Ma:2012ye,Dai:2014ala}. However, to perform the full evolution of the Qiu-Sterman function is highly nontrivial due to its dependence on two momentum fractions $x_1,\, x_2$ in general~\cite{Kang:2008ey,Kang:2010hg}. Thus in the TMD global analysis, the evolution of the Qiu-Sterman function has been implemented under certain approximations. There are two schemes that are used to perform this evolution in the literature. For both schemes that we discuss in this paper, the relevant DGLAP evolution equation for the Qiu-Sterman function is given by the expression
\begin{align}
    \frac{\partial T_{F\, q/p}(x,x;\mu)}{\partial \ln \mu^2}
    = \frac{\alpha_s(\mu^2)}{2\pi} \left[P^{\rm T}_{q\leftarrow q}\otimes 
    T_{F\, q/p}\right]\left(x;\mu\right)\,.
\end{align}
In the first scheme that we consider, from \cite{Kang:2012em}, the authors show that at large $x$, the transverse spin dynamics leads to a modification to the quark to quark splitting kernel, $P^{\rm T}_{q\leftarrow q}$, with
\bea
P^{\rm T}_{q\leftarrow q}\left(x\right) = P_{q\leftarrow q}\left(x\right) - N_C\,\delta(1-x)\,,
\eea
where $P_{q\leftarrow q}(x)$ is the standard quark to quark splitting kernel for unpolarized PDFs,
\bea
P_{q\leftarrow q}(x) &=  C_F \le[ \frac{1+x^2}{(1-x)_+}+\frac{3}{2}\delta(1-x) \ri]\,.
\eea
This scheme has been used for instance in \cite{Sun:2013hua}.
In the second scheme, for phenomenological purposes, the evolution of the Qiu-Sterman function has often been treated to be the same as the unpolarized collinear PDF, with $P^{\rm T}_{q\leftarrow q}(x) = P_{q\leftarrow q}(x)$. See e.g. Ref.~\cite{Bacchetta:2020gko}.

Apparently, for both cases, we can write the relevant spitting kernel as
\begin{align}
    P^{\rm T}_{q\leftarrow q}\left(x\right) = P_{q\leftarrow q}\left(x\right) - \eta \,\delta(1-x)\,,
    \label{e.TF-evo}
\end{align}
where $\eta$ is a parameter that controls the numerical scheme used to perform the DGLAP evolution. When $\eta = N_C$, the evolution matches the result of \cite{Kang:2012em}. On the other hand, for the second scheme that we consider, we set $\eta = 0$ so that the evolution models the standard DGLAP evolution of the unpolarized PDF.

To solve this evolution equation, it is useful to take the Mellin transform of this expression; for details on Mellin-space evolution, see Sec.~3 in \cite{Vogt:2004ns}.  After performing the Mellin transform of this expression, the evolution equation becomes
\begin{align}
    \frac{\partial}{\partial \ln \mu^2}T_{F\, q/p}(N,\mu)
    = \frac{\alpha_s\left(\mu^2\right)}{2\pi}\, \gamma(N)\,
    T_{F\, q/p}(N,\mu)\,.
    \label{e.TF-evo-N}
\end{align}
In this expression, $T_{F\, q/p}(N,\mu)$ is the Mellin transforms of the Qiu-Sterman function, i.e.
\begin{align}
    T_{F\, q/p}(N, \mu) = \int_0^1 dx \, x^{N-1} \, T_{F\, q/p}(x,x,\mu)\,.
\end{align}
Similarly $\gamma(N)$ is the Mellin transform of $P^{\rm T}_{q\leftarrow q}\left(x\right)$ which can be written as
\begin{align}
    \gamma(N) = \gamma_u(N)-\eta\,.
    \label{e.TF-evo-N-gam}
\end{align}
Here $\gamma_u(N)$ is the Mellin transform of the unpolarized splitting function $P_{q\leftarrow q}\left(x\right)$ and is given by
\begin{align}
    \gamma_u(N) = C_F \left(\frac{3}{2}+\frac{1}{N(N+1)}-2 S_1(N)\right)\,,
    \label{e.gam_upol}
\end{align}
with $S_1(N)$ the harmonic sum function. 

In the region where $\mu_{b_*} < m_b$, the mass of the $b$ quark, the solution of the evolution equation is given by
\begin{align}
    T_{F\, q/p}\left(N,\mu_{b_*}\right) = T_{F\, q/p}\left(N,\mu_0\right) \left (\frac{\alpha_s\left(\mu_{b_*}^2\right)}{\alpha_s\left(\mu_0^2\right)}\right )^{-\gamma(N)/\beta_0(\mu_0)}\,.
    \label{e.Tltmassb}
\end{align}
Here $\beta_0(\mu_0) = 11-2/3 \, n_f(\mu_0)$, where $n_f(\mu_0)$ is the number of active flavors at the scale $\mu_{0}$. In the region where $\mu_{b_*}>m_b$, the solution of the evolution equation is given by
\begin{align}
    T_{F\, q/p}\left(N,\mu_{b_*}\right) = T_{F\, q/p}\left(N,m_b\right) \left (\frac{\alpha_s\left(\mu_{b_*}^2\right)}{\alpha_s\left(m_b^2\right)}\right )^{-\gamma(N)/\beta_0(\mu_{b_*})}\,,
    \label{e.Tgtmassb}
\end{align}
where $T_{F\, q/p}\left(N,m_b\right)$ is given by
\begin{align}
    T_{F\, q/p}\left(N,m_b\right) = T_{F\, q/p}\left(N,\mu_0\right) \left (\frac{\alpha_s\left(m_b^2\right)}{\alpha_s\left(\mu_0^2\right)}\right )^{-\gamma(N)/\beta_0(\mu_0)}\,,
\end{align}
and $n_f(\mu_{b_*})$ is the number of active flavors at the scale $\mu_{b_*}$.

In order to construct the Sivers function in \eref{Sivers} at NLO, there is an additional convolution of the coefficient $C$ function and the Qiu-Sterman function. We find that it is useful to first take its Mellin transform and thus the convolution over the momentum fraction becomes a simple product in Mellin space:
\begin{align}
    f_{1T,q/p}^{\perp q}&\left(N, \bt;\mu,\z\right) = \bar{C}_{q\leftarrow i}(N;\mu,\z) T_{F\, i/p}\left(N,\mu\right)\,,
\end{align}
where $\bar{C}_{q\leftarrow q'}(N,\bt;\mu,\z)$ is the Mellin transform of the Sivers Wilson coefficient function. The NLO Sivers function can then be obtained by numerically taking the inverse Mellin transform of this function,
\begin{align}
    f_{1T,q/p}^{\perp q}(x, \bt;\mu,\z) = &
    \frac{1}{\pi}\int_0^{\infty} dz \mathrm{Im} \left[ e^{i\phi} x^{-c-ze^{i\phi}} f_{1T,q/p}^{\perp q}\left(c+ze^{i\phi}, \bt;\mu,\z\right)\right]\,,
\end{align}
where the parameter $c$ must be taken such that all of the singularities in the function $f_{1T,q/p}^{\perp q}\left(c+ze^{i\phi}, \bt;\mu,\zeta\right)$ lie to the left of the line $x = c$ in the imaginary plane. In our code, we use $c = 2$ which satisfies this criteria. We also take $\phi = \pi/4$ to optimize the numerical integration.

%%%%%%%%%%%%%%%%%%%%%%%%%%%%%%%%%%%%%%%%%%%%%%%%%%%%%%%%%%%%%
\section{Fit Results}\label{Fit Results}
\begin{figure}[hbt!]
    \centering
    \includegraphics[width = 0.5\textwidth]{\FigPath/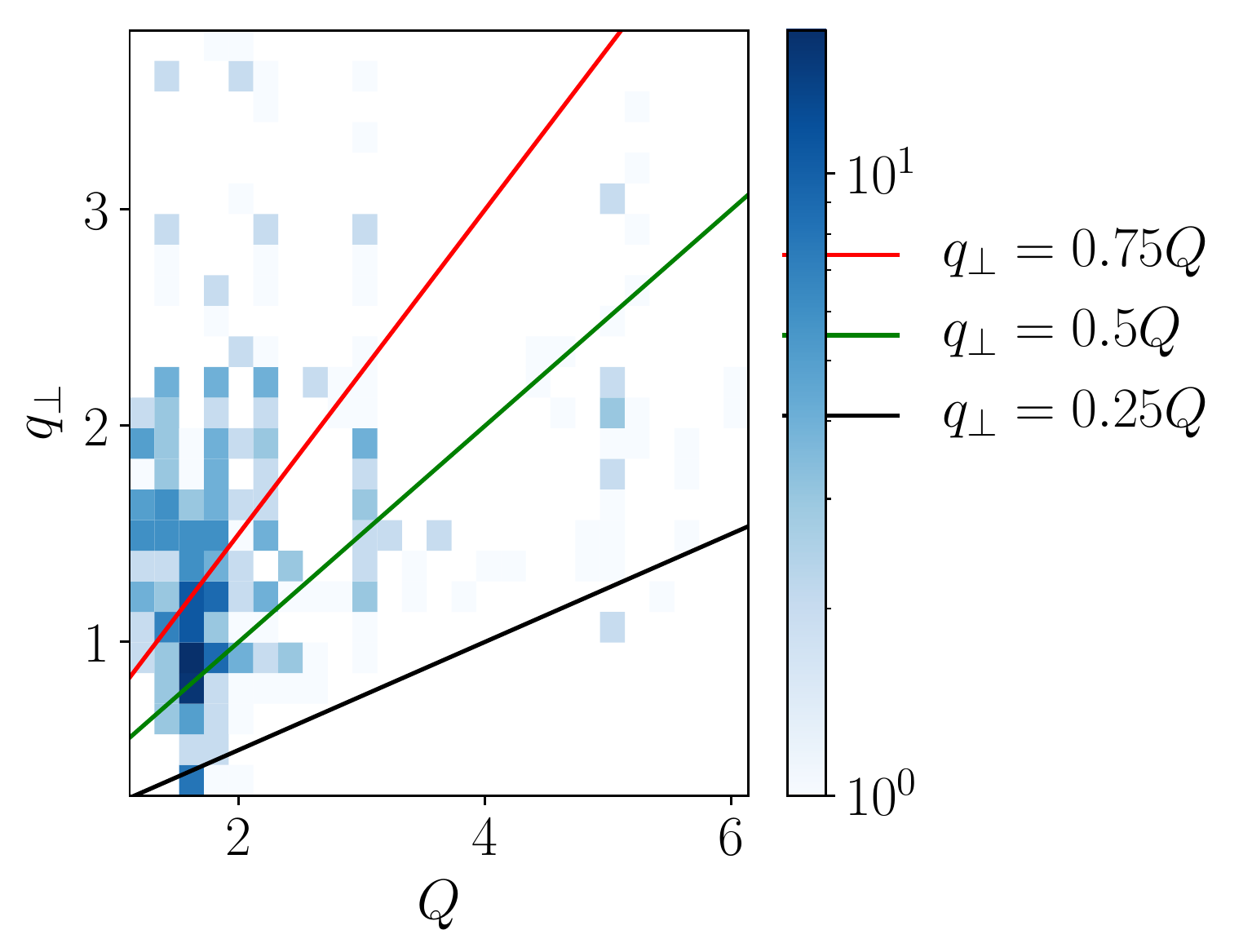}
    \caption{Histogram of the SIDIS data in $\qt$ and $Q$. To obtain this plot, we bin the SIDIS data sets in $q_\perp$ and $Q$. The dark spots indicate a large number of experimental data while the white spots indicate that there are no experimental data. We also plot the line $q_\perp =$ $0.75\,Q$ in red, $0.5\,Q$ in green, and $0.25\,Q$ in black.}
    \label{f.filter}
\end{figure}
In this section, we present the results of three separate extractions of the Sivers function. 
In Sec.~\ref{sid_dy}, we present the result of fit~1, where we consider only the low energy data from SIDIS as well as the COMPASS Drell-Yan data using $\eta = N_C$. 
In Sec.~\ref{DGLAP}, we present the results of fit~2a, where a global extraction is performed using $\eta = N_C$. Furthermore, we perform an extensive study of the impact of the RHIC data.
Finally in Sec.~\ref{global}, we present the results of fit~2b, where we perform a global extraction of the Sivers function with $\eta = 0$.
The extracted parameter values, as well as comparisons with experimental data, are presented for fit~1 and fit~2b in Sec.~\ref{sid_dy} and Sec.~\ref{global}, respectively.

\begin{table}[hbt!]
\begin{center}
 \begin{tabular}{||c | c | c | c | c | c ||} 
 \hline
 fit scheme & SIDIS & Drell-Yan & $W/Z$ & $N_{\rm data}$ & $\eta$ in evolution \\
 \hline
 fit~1 & $\surd$ & $\surd$ & $\times$ & 226 & $N_C$ \\
 \hline
 fit~2a & $\surd$ & $\surd$ & $\surd$ & 243 & $N_C$\\
 \hline
 fit~2b & $\surd$ & $\surd$ & $\surd$ & 243 & $0$\\
 \hline
\end{tabular}
\caption{Description of each of the fits that we present. Fit~1 is presented in Sec.~\ref{sid_dy}, fit~2a is presented in Sec.~\ref{DGLAP}, and fit~2b is presented in Sec.~\ref{global}.}
\label{schemes}
\end{center}
\end{table}
\subsection{Simultaneous Fit to SIDIS and Drell-Yan}\label{sid_dy}
\begin{figure*}[htbp!]
\begin{center}
    \includegraphics[width = \textwidth]{\FigPath/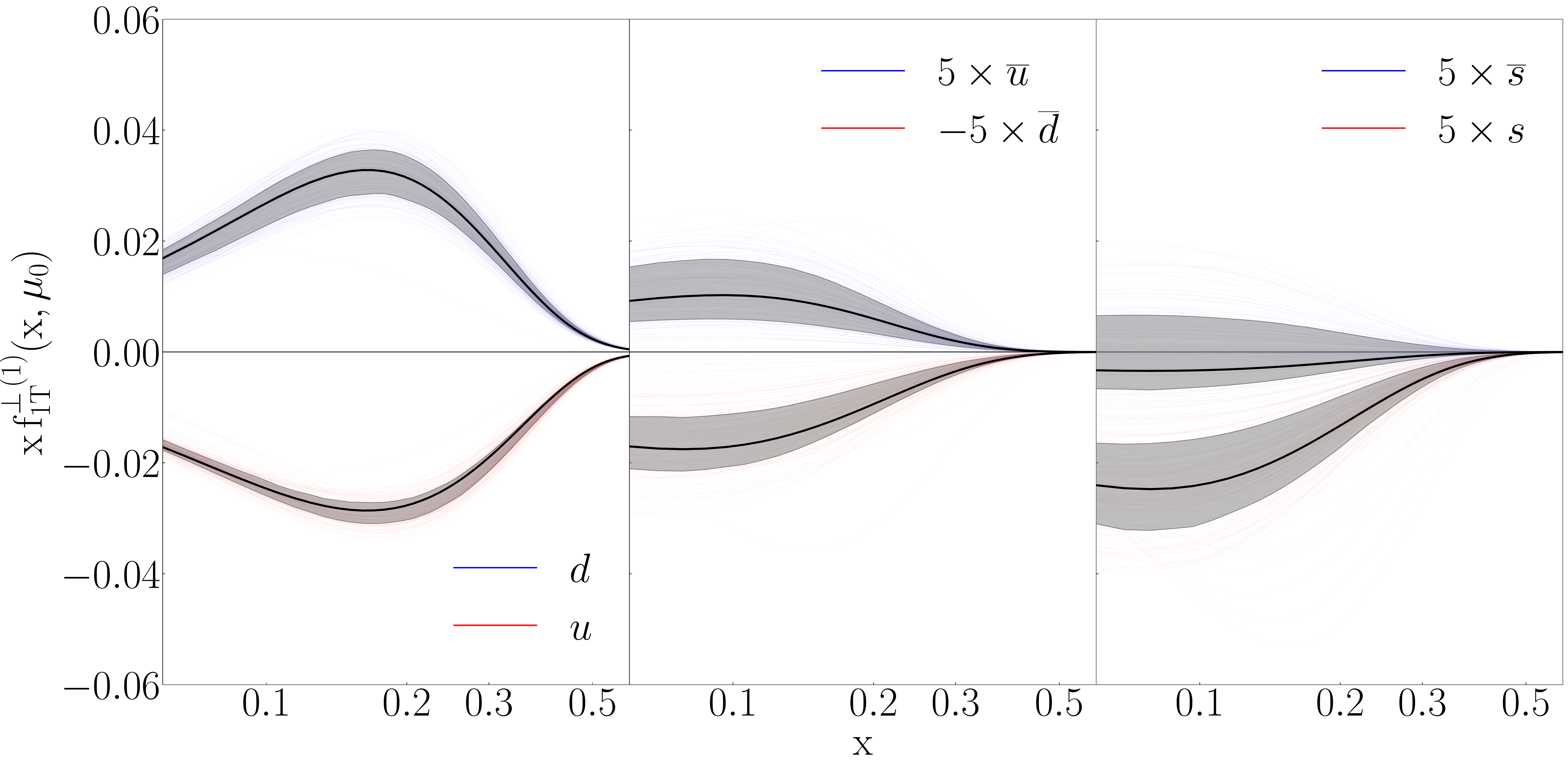}
        \caption{The extracted transverse moment of the Sivers function from fit~1 at $\mu_0 = \sqrt{1.9} $ GeV. The black curve is the fit to the experimental data with no Gaussian noise.}
    \label{f.siversplot}
\end{center}
\end{figure*}
In this section we present a simultaneous fit to measurements of the Sivers asymmetry from SIDIS data sets from JLAB in \cite{Qian:2011py}, HERMES in \cite{Airapetian:2009ae}, COMPASS in \cite{Adolph:2016dvl, Alekseev:2008aa} and the COMPASS Drell-Yan data in \cite{Aghasyan:2017jop}. We note that we do not include the COMPASS data set in \cite{Adolph:2012sp} since the data set in \cite{Adolph:2016dvl} is a re-binning of this set. Furthermore the data set in \cite{Adolph:2016dvl} was projected into two sets of data $z_h>0.1$ and $z_h>0.2$. To avoid fitting correlated data sets, we choose to fit only the $z_h>0.1$ data set. We then compare our prediction for the RHIC asymmetry against the RHIC data. 

While typical kinematic cuts from unpolarized SIDIS fits for instance in \cite{Scimemi:2019cmh} select only data which has $\qt/Q<0.25$, we find that this selection process leaves very few data points for the available Sivers data. In \fref{f.filter} we plot a histogram of the selected data SIDIS data as a function of $\qt$ and $Q$. We find that the cut $\qt/Q<0.25$ leaves only 12 SIDIS data points, while the cut $\qt/Q<0.5$ leaves 97 data points. In fact, we find that the majority of the data has $\qt/Q>0.5$. In order to retain a large enough data set to perform a meaningful fit we perform the cut $\qt/Q<0.75$. Furthermore to restrict the selected data set to the TMD region, we also enforce that the SIDIS data must have $P_{h\perp}<1$ GeV. At the same time in order to avoid the threshold resummation region, we also enforce that $z_h<0.7$. 

In order to perform the fit, we use the \texttt{MINUIT} package~\cite{James:1975dr,James:2004xla} to minimize the $\chi^2$. In this section, we define the $\chi^2$ as
\begin{align}
    \chi^2\left(\left\{a\right\}\right) = \sum_{i=1}^N \frac{\left(T_i\left(\left\{a\right\}\right)-E_i\right)}{\Delta E_i^2}\,,
\end{align}
where $E_i$ are the central values of the experimental measurements, $\Delta E_i$ are the total experimental errors, $T_i\left(\left\{a\right\}\right)$ is the theoretical value at the experimental kinematics, and $\left\{a\right\}$ is a vector containing the fit parameters. 

For this section, we take $\eta = N_C$ to perform the DGLAP evolution of the Qiu-Sterman function, referred to as fit~1 in Tab.~\ref{schemes}. In order to optimize the minimization process, the denominator of our asymmetry is pre-calculated at the beginning of the fit. We also perform pre-calculations for the unpolarized TMDs and use grid interpolation in the numerator of the asymmetry. For the NLO Sivers function, we find that the Mellin space prescription leads to a massive speeds compared to performing the convolution integrals. Furthermore we use the numerical method in \cite{Kang:2019ctl} to perform all Bessel integrals. 

In order to generate an uncertainty band, we follow the work in Ref.~\cite{Pisano:2018skt,Callos:2020qtu} to use the replica method. To generate one replica, we shift each of the the data points by a Gaussian noise with standard deviation corresponding to the experimental error. The fit is performed on the noisy data 200 times as well as the no noise data. This result in 201 sets of stored fit parameters. Using each of the 201 sets of stored parameters, we calculate the asymmetry for each of the included data as well as calculate the first transverse moment of the Sivers function in Eq.~\eqref{e.f1T1} for each of the quark flavors. The uncertainty band is generated at each point by retaining all contribution within the $68\%$ region.

In Table.~\ref{paramtable}, we present the results for the parameter values along with the $\chi^2/d.o.f$ and the parameter uncertainties. The central point that we present for each parameter are the parameter values from the fit with no noise. The parameter uncertainties presented in this fit are obtained by considering only the middle $68\%$ of the 201 sets of parameters. In terms of the quality of the fit, we find an excellent agreement between our fitted theoretical result and the experimental data with a global $\chi^2/d.o.f = 1.032$. In Tab.~\ref{chi2-sid-dy}, we give the value of the $\chi^2/d.o.f$ for each of the sets of data. 
\begin{table}[htb!]
\def\arraystretch{1.25}
  \begin{center}
    \begin{tabular}{c c c c c}
        & \multicolumn{3}{c}{$\chi^2/d.o.f.= 1.032$} & \\
        \hline
        $N_{u}=$         & $ 0.077_{-0.005}^{+0.004}$ GeV & & $\alpha_{u}=$     & $ 0.967_{-0.045}^{+0.028}$\\
        $N_{d}=$         & $-0.152_{-0.016}^{+0.017}$ GeV& & $\alpha_{d}=$     & $ 1.188_{-0.023}^{+0.056}$\\
        $N_{s}=$          & $ 0.167_{-0.051}^{+0.053}$ GeV& & $\alpha_{sea}=$ & $ 0.936_{-0.026}^{+0.069}$\\
        $N_{\bar{u}}=$  & $-0.033_{-0.017}^{+0.016}$ GeV& & $\beta=$       & $5.129_{-0.034}^{+0.017}$\\
        $N_{\bar{d}}=$ & $-0.069_{-0.026}^{+0.019}$ GeV& & $g_1^T\,=$ & $0.180_{-0.070}^{+0.035}$ GeV$^2$\\
        $N_{\bar{s}}=$ & $-0.002_{-0.040}^{+0.047}$  GeV& &
    \end{tabular}
    \caption{Fit parameters for fit~1 in Tab.~\ref{schemes}. The presented values is the parameter value of the fit with no Gaussian noise. The uncertainties for the replicas are generated from the parameter values which lie on the boundary of $68\%$ confidence.}
    \label{paramtable}
  \end{center}
\end{table}
\begin{table}[h!]
\begin{center}
 \begin{tabular}{||c | c | c | c | c | c ||} 
 \hline
 Collab & Ref & Process & $Q_{\rm avg}$& $N_{\rm data}$ & $\chi^2$/$N_{\rm data}$ \\ [0.5ex] 
 \hline\hline
 \multirow{8}{*}{COMPASS} & \multirow{5}{*}{\cite{Alekseev:2008aa}} & $ld \rightarrow lK^0X$ & 2.52& 7 & 0.770 \\
 \cline{3-6}
 & & $ld \rightarrow lK^-X$ & 2.80 & 11 & 1.325 \\
 \cline{3-6}
 & & $ld \rightarrow lK^+X$ & 1.73& 13 & 0.749 \\
 \cline{3-6}
 & & $ld \rightarrow l\pi^-X$ & 2.50& 11 & 0.719 \\
 \cline{3-6}
 & & $ld \rightarrow l\pi^+X$ & 1.69& 12 & 0.578 \\
 \cline{2-6}
 & \multirow{2}{*}{\cite{Adolph:2016dvl}} & $lp \rightarrow lh^-X$ & 4.02& 31 & 1.055 \\
 \cline{3-6}
 & & $lp \rightarrow lh^+X$ & 3.93& 34 & 0.898 \\
 \cline{2-6}
 & \cite{Aghasyan:2017jop}& $\pi^-p \rightarrow \gamma^*X$ & 5.34 & 15 & 0.658 \\
 \hline
 \multirow{6}{*}{HERMES} & \multirow{6}{*}{\cite{Airapetian:2009ae}} & $lp \rightarrow lK^-X$ & 1.70 & 14 & 0.376 \\
 \cline{3-6}
 & & $lp \rightarrow lK^+X$ & 1.73 & 14 & 1.339 \\
 \cline{3-6}
 & & $lp \rightarrow l\pi^0X$ & 1.76 & 13 & 0.997 \\
 \cline{3-6}
 & & $l p \rightarrow l (\pi^+-\pi^-)X$ & 1.73 & 15 & 1.252 \\
 \cline{3-6}
 & & $lp \rightarrow l\pi^-X$ & 1.67 & 14 & 1.498 \\
 \cline{3-6}
 & & $lp \rightarrow l\pi^+X$ & 1.69 & 14 & 1.697 \\
 \hline
 \multirow{2}{*}{JLAB} & \multirow{2}{*}{\cite{Qian:2011py}} & $lN \rightarrow l\pi^+X$ & 1.41 & 4 & 0.508 \\
 \cline{3-6}
 & & $lN \rightarrow l\pi^-X$ & 1.69 & 4 & 1.048 \\
 \hline
 \multirow{3}{*}{\textcolor{red}{RHIC}} & \multirow{3}{*}{\cite{Adamczyk:2015gyk}} & $pp \rightarrow W^+X$ & $M_W$& 8 & 2.189 \\
 \cline{3-6}
 & & $pp \rightarrow W^-X$ & $M_W$ & 8 & 1.684 \\
 \cline{3-6}
 & & $pp \rightarrow Z^0X$ &$M_Z$ & 1 & 3.270 \\
 \hline
  Total & & & & 226 & 0.989 \\
 \hline
\end{tabular}
    \caption{The distribution of experimental after taking the kinematic cuts $\qt/Q<0.75$, $P_{h\perp}<1$ GeV, and $z<0.7$. The column $Q_{\rm avg}$ gives the average hard scale for the measured data set. On the right column, we have included the $\chi^2/N_{\rm data}$ for each set of data from the extraction in fit~1. The RHIC data was not included into the fit. Here we give the $\chi^2/N_{\rm data}$ for the prediction.}
    \label{chi2-sid-dy}
\end{center}
\end{table}

\begin{figure*}[htbp!]
\centering
    \includegraphics[width = 0.48 \textwidth]{\FigPath/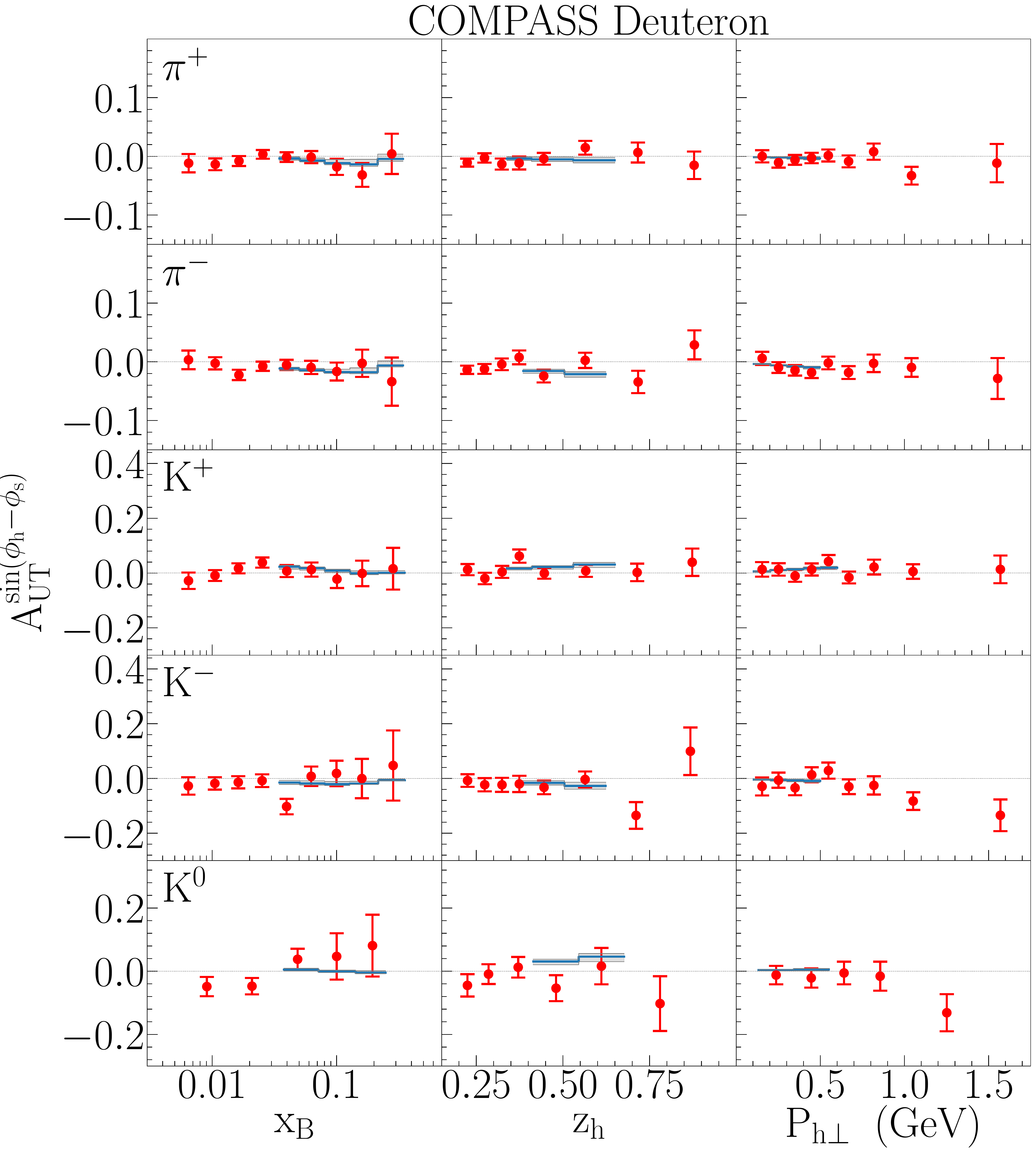}
    \includegraphics[width = 0.48 \textwidth]{\FigPath/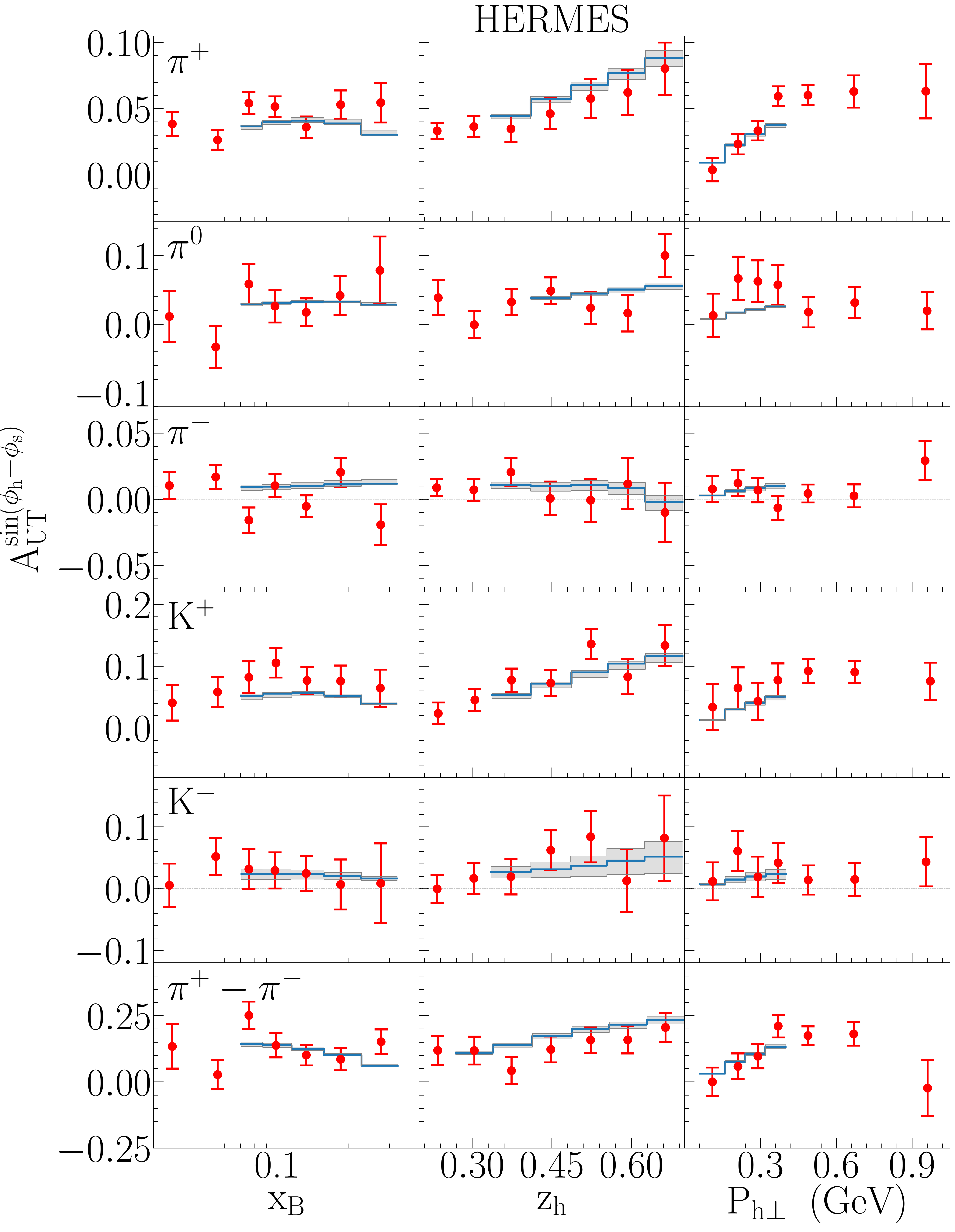}
    \caption{Left: The COMPASS deuteron target measurement \cite{Alekseev:2008aa} for $\pi^+$, $\pi^-$, $K^+$, $K^-$, and $K^0$ from top to bottom, and as a function of $x_B$ (left), $z_h$ (middle), and $P_{h\perp}$ (right). Right: HERMES proton target measurement \cite{Airapetian:2009ae} $\pi^+$, $\pi^0$, $\pi^-$, $K^+$, $K^-$, and $(\pi^+-\pi^-)$ from top to bottom, and as a function of $x_B$ (left), $z_h$ (middle), and $P_{h\perp}$ (right). The data is plotted in red along with the total experimental error. The central curve in blue as well as the uncertainty band in gray are generated using the result from fit~1 in Tab.~\ref{schemes}.}
    \label{C_D+H}
\end{figure*}
\begin{figure*}[htbp!]
    \centering
    \includegraphics[width = 0.48 \textwidth]{\FigPath/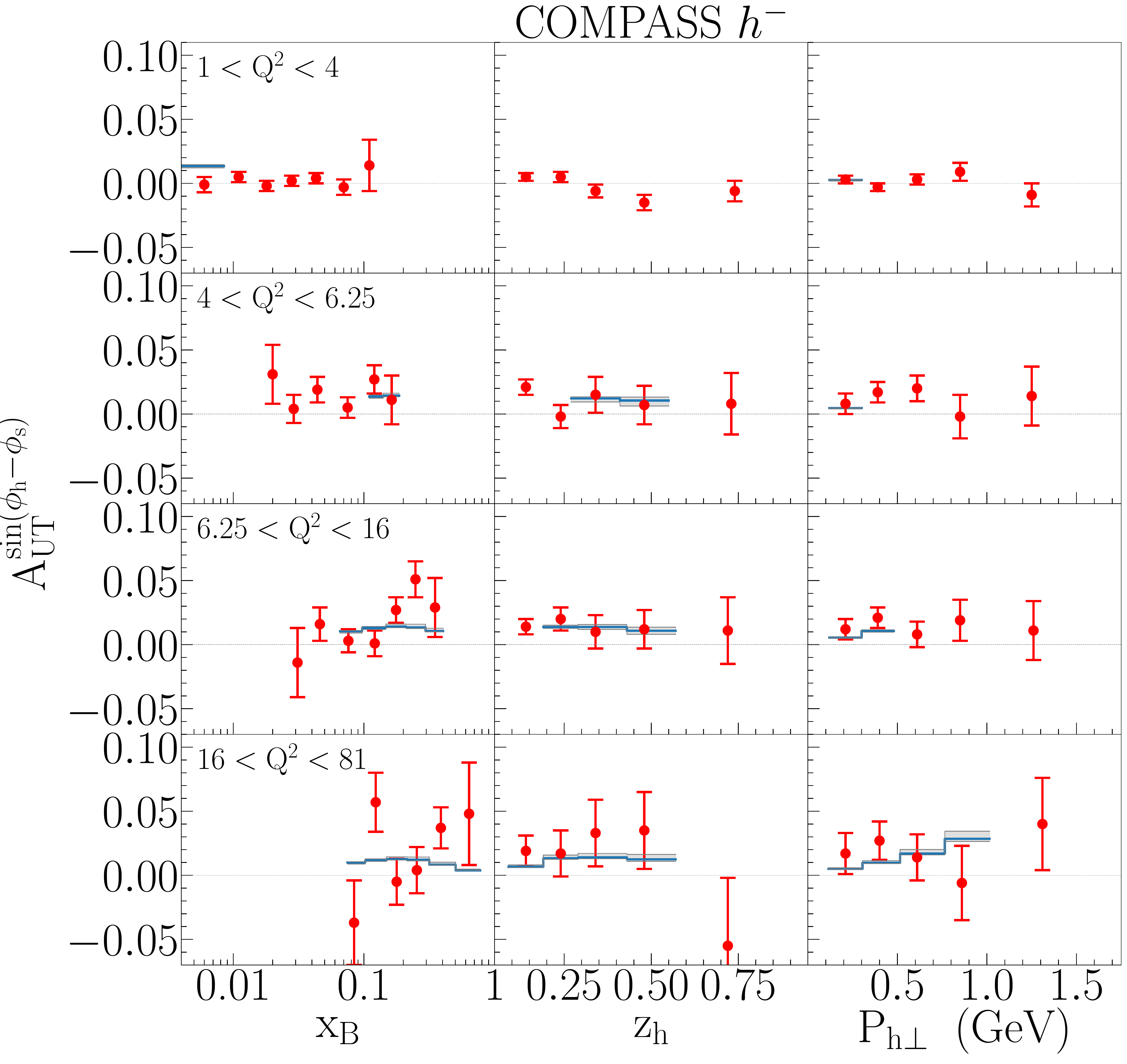}
    \includegraphics[width = 0.48 \textwidth]{\FigPath/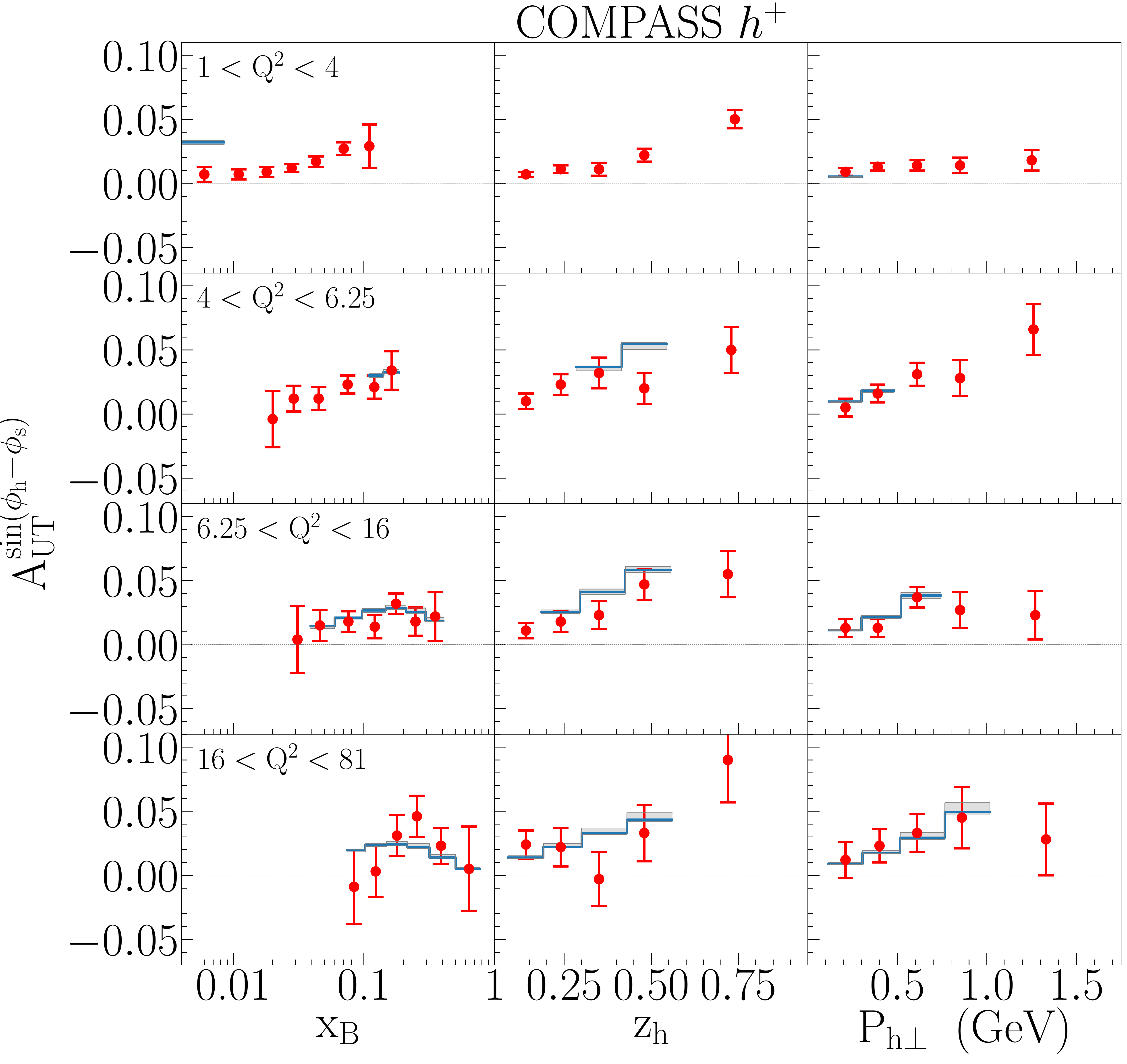}
    \caption{Left: The COMPASS proton target measurement for $h^-$ for $1$ GeV$^2$ $< Q^2< 4$ GeV$^2$, $4$ GeV$^2$ $< Q^2< 6.25$ GeV$^2$, $6.25$ GeV$^2$ $< Q^2< 16$ GeV$^2$, $16$ GeV$^2$ $< Q^2< 81$ GeV$^2$ from top to bottom \cite{Adolph:2016dvl}. Right: Same as the left except for $h^+$ production. The central curve as well as the uncertainty band are generated using the result from fit~1 in Tab.~\ref{schemes}.}
    \label{f.C_h}
\end{figure*}
\begin{figure*}[htbp!]
\centering
\begin{minipage}{.5\textwidth}
    \centering
    \includegraphics[width = 0.5\textwidth]{\FigPath/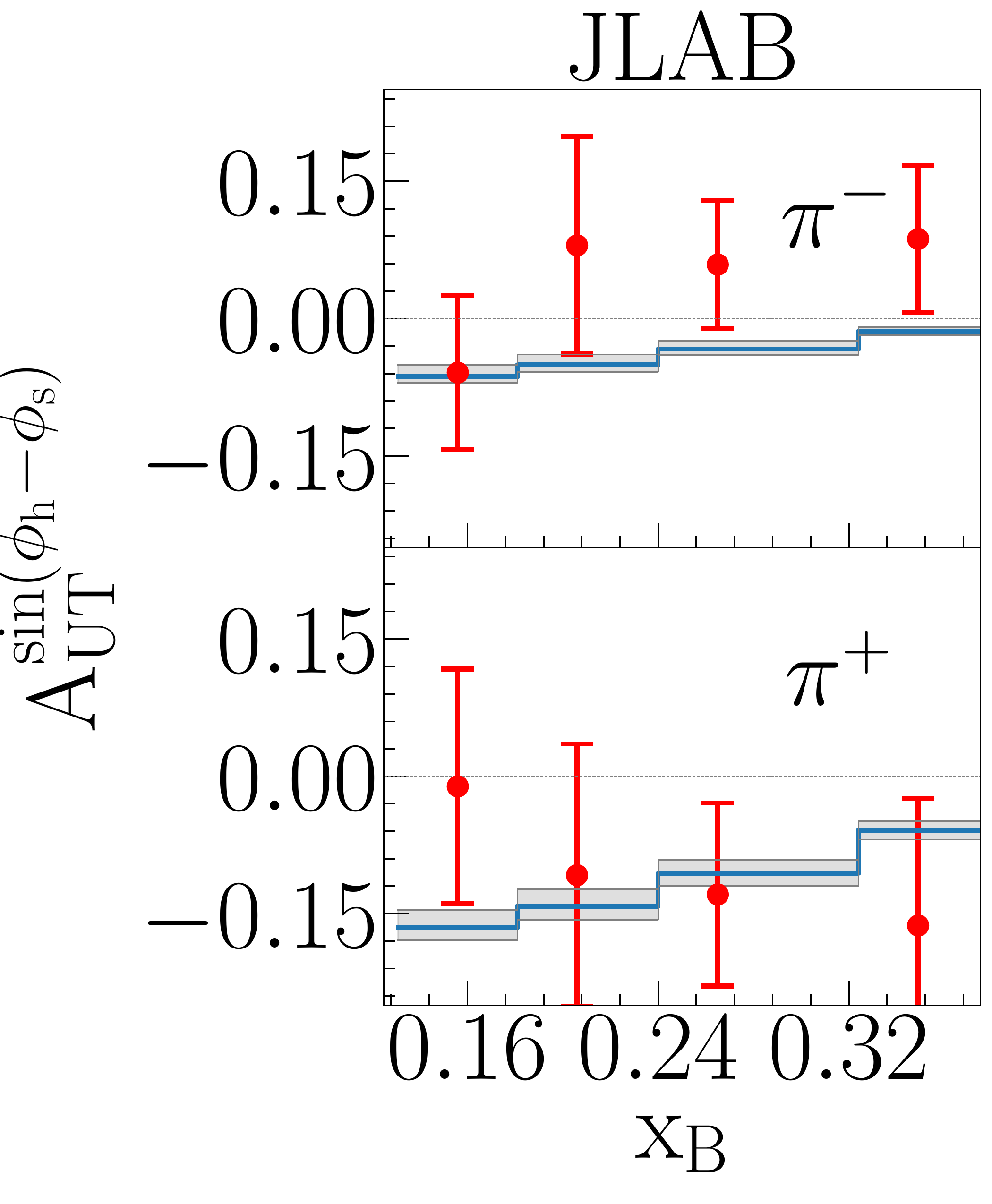}
    \caption{JLab measurement of the Sivers asymmetry for a neutron target \cite{Qian:2011py} as a function of $x_B$. The central curve as well as the uncertainty band are generated using  the result from fit~1.}
    \label{f.J}
\end{minipage}
\end{figure*}
\begin{figure*}
    \centering
    \includegraphics[width = 0.85\textwidth]{\FigPath/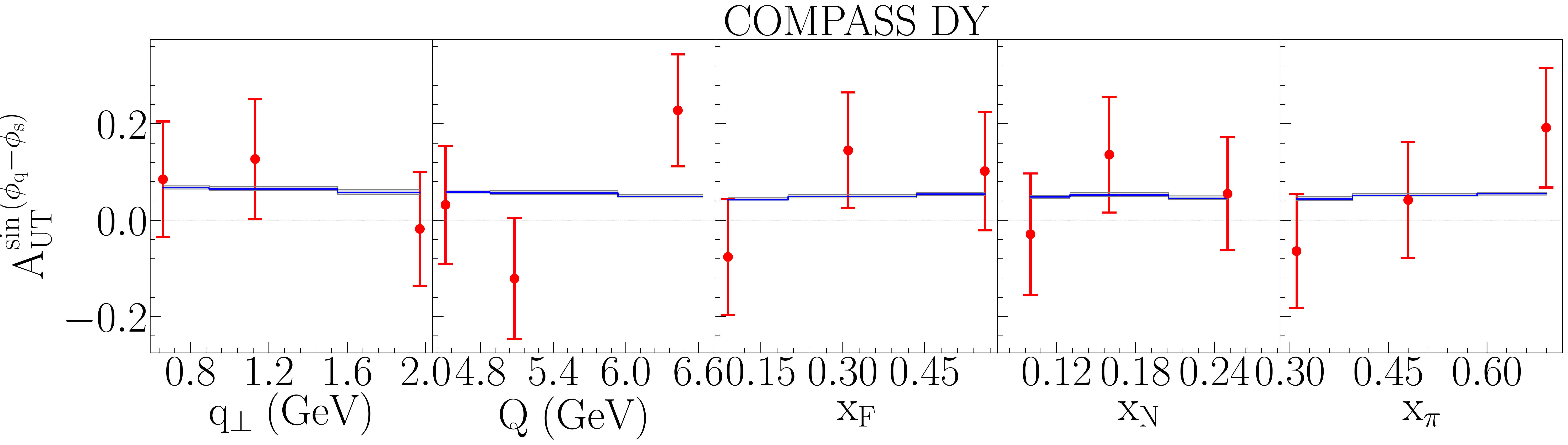}
    \caption{COMPASS Drell-Yan measurement for $\pi^-$-$p$ collision \cite{Aghasyan:2017jop} as a function of $\qt$, $Q$, $x_F$, $x_N$, and $x_{\pi}$ from left to right. The central curve as well as the uncertainty band are generated using the result from fit~1 in Tab.~\ref{schemes}.}
    \label{f.C_DY}
\end{figure*}
\begin{figure*}[htbp!]
\begin{minipage}{.56\textwidth}
    \centering
    \includegraphics[width = \textwidth]{\FigPath/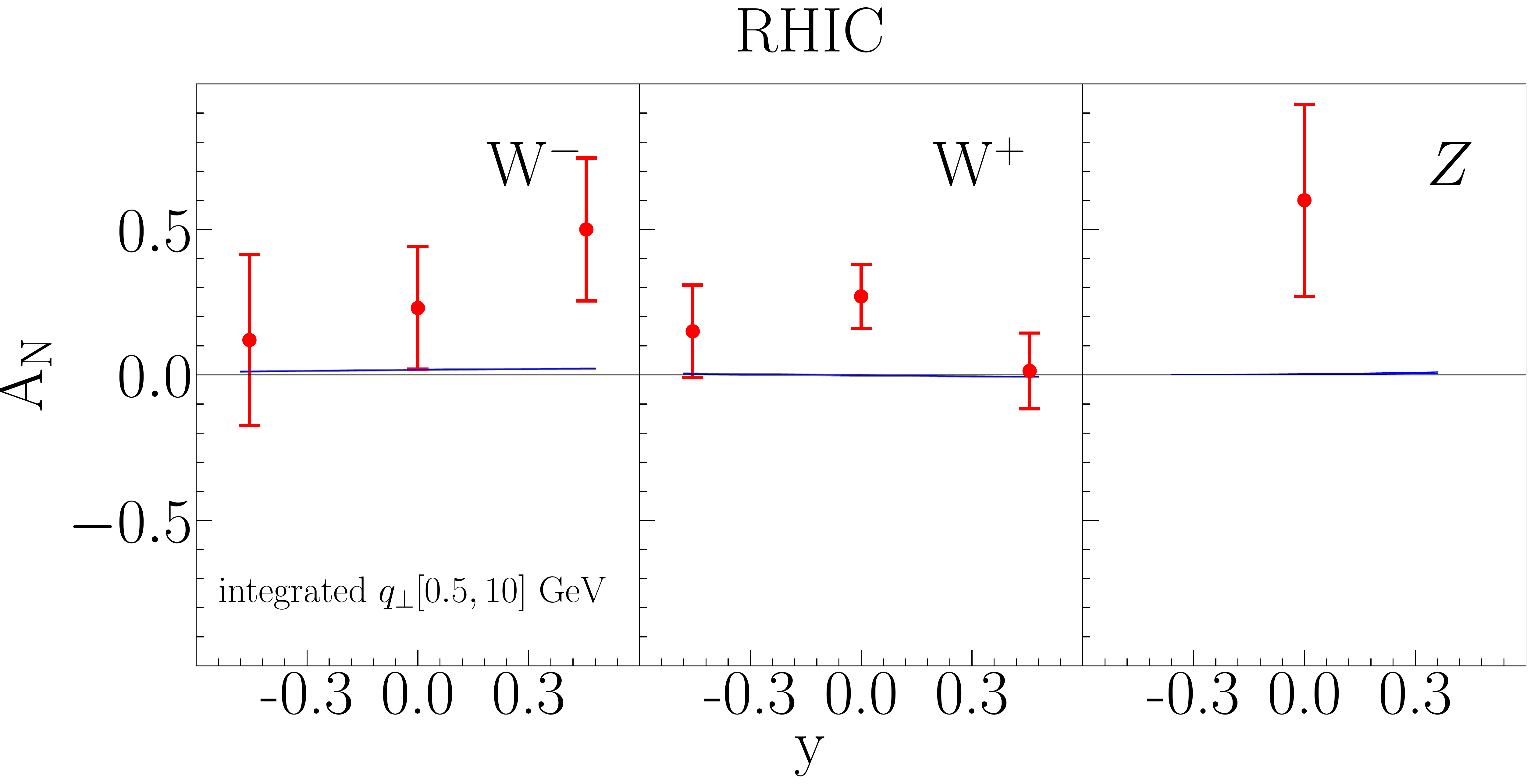}
\end{minipage}%
\begin{minipage}{.42\textwidth}
    \centering
    \includegraphics[width =\textwidth]{\FigPath/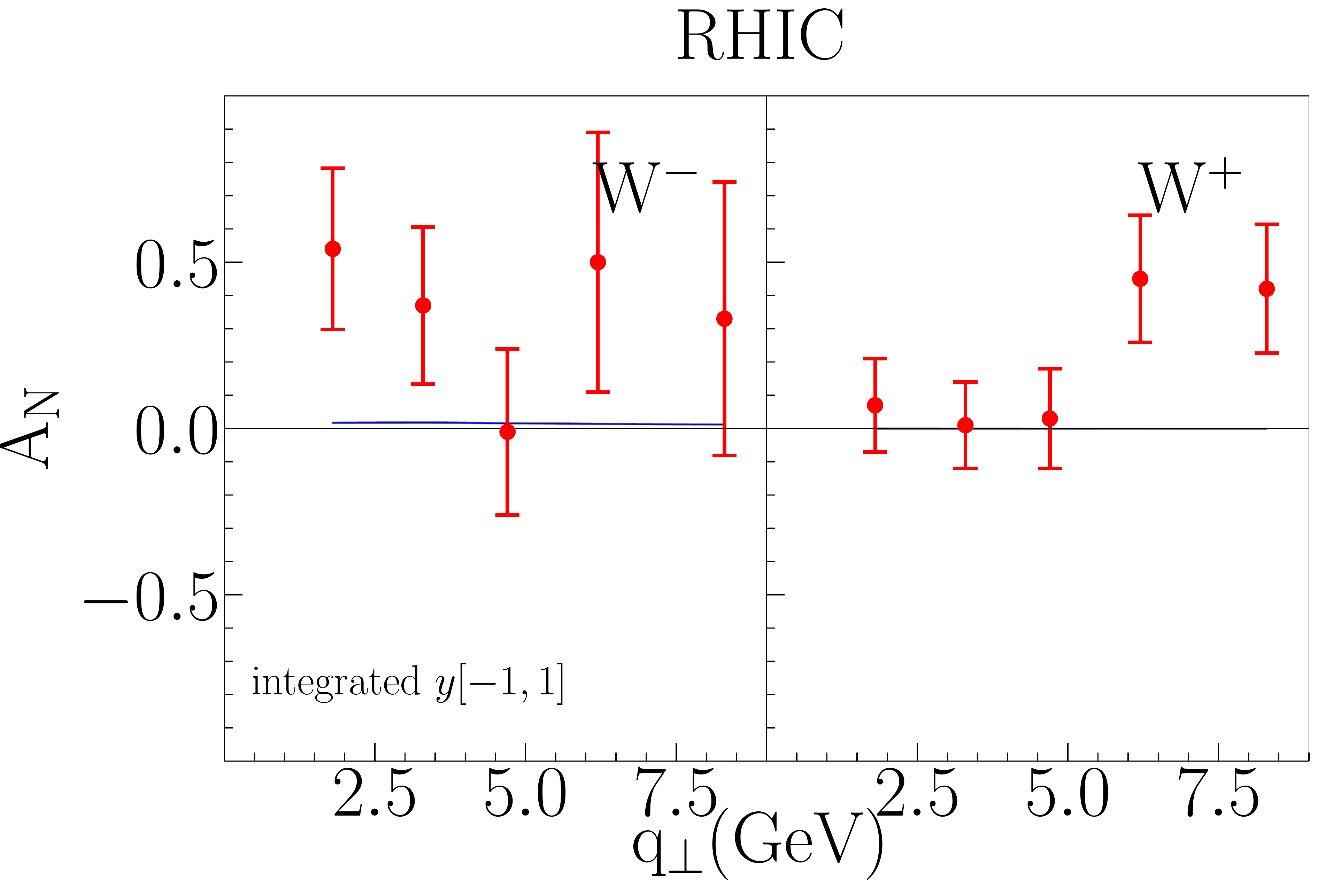}
\end{minipage}
    \caption{Prediction for the Sivers asymmetry for $p+p\rightarrow W/Z$ at $\sqrt{S} = 500$ GeV \cite{Adamczyk:2015gyk} using the result of fit~1 in Tab.~\ref{schemes}. We plot only the central curve from fit~1 here since the size of the uncertainty band is small for this prediction. Left: The $y$ dependent data integrated in $\qt$ from $0.5$ to $10$ GeV. Right: The $\qt$ dependent data integrated in $y$ from $-1$ to $1$.}
    \label{f.R}
\end{figure*}

In \fref{f.siversplot}, we plot the extracted first transverse moment of the proton SIDIS Sivers function at the initial PDF scale, $f_{1T}^{\perp\,(1)}(x, \mu_0)$ with $\mu_0 =\sqrt{1.9}$ GeV as defined in Eq.~\eqref{e.f1T1}. In this figure, we have plotted all $200$ replicas for each of the extracted quark flavors. We again use the middle $68\%$ of the data points in the plot to generate the grey uncertainty band for each of the Sivers moments. For the $\bar{u}$, $s$ and $\bar{s}$-quarks, the Sivers moment have been multiplied by a factor of $5$ while for $\bar{d}$, we have multiplied by a factor of $-5$. We find that the Sivers $d$ function is the largest in magnitude and is positive; while the Sivers $u$ function is nearly as large but is negative. Furthermore we find that the $\bar{u}$ and $\bar{d}$-quark functions are nearly equal to one another in magnitude, both are more than 5 times smaller in magnitude than the valence quarks, and are both positive. For the $s$-quark, we find that the magnitude is approximately 5 times smaller than the valence quarks in magnitude and is negative. Finally for the $\bar{s}$-quark, we find that the magnitude is very small and that the sign is not well determined in this fit.

In Figs.~\ref{C_D+H}, \ref{f.C_h}, and \ref{f.J}, we plot our theoretical curves against the SIDIS data. Fig.~\ref{C_D+H} is for COMPASS deuteron target (left panel) and for HERMES proton target (right panel), and for both pions and kaons. Fig.~\ref{f.C_h} is for charged hadrons from COMPASS proton target. Fig.~\ref{f.J} is for pion production on a neutron target from JLab. Finally in Fig.~\ref{f.C_DY} we plot theoretical curves against the COMPASS Drell-Yan lepton pair data in $\pi^-+p$ collisions. We plot the asymmetry $A_{UT}^{\sin(\phi_q - \phi_s)}$ as a function of transverse momentum $q_\perp$, invariant mass $Q$, Feynman $x_F = x_\pi - x_N$, momentum fraction $x_N$ in the proton target, and momentum fraction $x_\pi$ in the pion target, respectively. The experimental data along with the total experimental uncertainties are plotted in red. The blue curves are the theory curves from the fit with no noise. The uncertainty band in grey is generated from the stored values of the asymmetry for each of the replicas. For each data point, the maximum and minimum value of the asymmetry within the middles $68\%$ are used to generate these error bars. As it is indicated already in Tab.~\ref{chi2-sid-dy} and as it is evident from the figures, the agreement between our theory and SIDIS and Drell-Yan data is very good, although to a less degree with the Drell-Yan data because of the much larger experimental uncertainty. We note that very recently in \cite{Airapetian:2020zzo}, the HERMES collaboration provided additional experimental data for the Sivers asymmetry. Since the HERMES paper has not yet been published, we cannot implement this data into our fit. However, we find that there is very strong agreement between our extracted asymmetry and this new data.

In \fref{f.R}, we plot the prediction for the RHIC data in $p+p$ collisions at $\sqrt{S} =500$ GeV using the extracted Sivers function from this fit. In the left panel, we plot the Sivers asymmetry $A_N$ as a function of rapidity for $W^-$ (left), $W^+$ (middle), and $Z^0$ (right), respectively. We integrate vector boson transverse momentum over $0.5 < q_\perp < 10$ GeV. On the right panel, we plot $A_N$ as a function of $q_\perp$ while we integrate over the rapidity $|y| < 1$. We find that the asymmetry for $W/Z$ for the central fit is at most $2\%$, which is more than an order of magnitude smaller than the central values recorded at RHIC. This leads to a $\chi^2/N_{\rm data}$ of $2.015$ for the prediction for RHIC, as shown in Tab.~\ref{chi2-sid-dy}. Even if one considers the very large error bars in the RHIC data, this comparison seems to indicate some tension between our theory and the RHIC data.

\subsection{Impact of the RHIC data}\label{RHIC}
In this section, we study the impact of the RHIC data to the fit. One possible issue which may be arising in the description of the RHIC data is that while there are a large number of experimental data at small $Q$, there are much less data at RHIC energies. In order to access the impact of the RHIC data, it is therefore convenient to follow the work in \cite{Hou:2016nqm} to introduce a weighting factor to the calculation of the $\chi^2$. Thus in this section, the expression for the $\chi^2$ is given by
\begin{align}
    \chi^2\left(\left\{a\right\}\right) = \sum_{i=1}^N \frac{\left(T_i\left(\left\{a\right\}\right)-E_i\right)}{\Delta E_i^2}+\omega \sum_{i=1}^{N_R} \frac{\left(T_i\left(\left\{a\right\}\right)-E_i\right)}{\Delta E_i^2}\,.
    \label{e.chi2_RHIC}
\end{align}
We also define the $N_{\rm data}$ for this weighted fit as
\begin{align}
    N_{\rm data} = N+\omega N_R\,.
\end{align}
For the first term of \eref{chi2_RHIC}, the sum is performed over all data in the previous section, i.e., all the SIDIS data plus COMPASS Drell-Yan data. In the second term, the sum is performed only over the RHIC data. In this second expression, $\omega$ is the weighting factor. In order to emphasize the contributions of the RHIC data, we choose $\omega = N/N_R = 226/17$ so that the RHIC data and the rest of the experimental data sets are equally weighed in the calculation of the $\chi^2$. Furthermore, in order to perform the DGLAP evolution of the Qiu-Sterman function, we take $\eta = N_C$.

Using this definition of the $\chi^2$, we perform a fit to the selected data. In Tab.~\ref{chi2-all}, we provide the distribution of the $\chi^2$ for this fit. With the addition of the weighting factor, we find that the $\chi^2/N_{\rm data} = 1.888$ for the RHIC is quite large while for the low energy data the $\chi^2/N_{\rm data} = 0.996$. This result indicates that the issue with describing the RHIC data is not that the high energy data has a small number of data points. Rather, it indicates that when using our theoretical assumptions, these sets of data disagree on the properties of the Sivers function.

\begin{table*}[hbt!]
\begin{center}
 \begin{tabular}{||c | c | c | c | c | c | c ||} 
 \hline
 Collab & Ref & Process & $Q_{\rm avg}$ & $N_{\rm data}$ &$\chi^2$/$N_{\rm data}$ &  $\Delta \mathcal{T} (\%)$ \\ [0.5ex]
 \hline\hline
 \multirow{8}{*}{COMPASS} & \multirow{5}{*}{\cite{Alekseev:2008aa}} & $ld \rightarrow lK^0X$ & 2.52 & 7 & 0.755 & 2.09 \\
 \cline{3-7}
 & & $ld \rightarrow lK^-X$ & 2.80 & 11 & 1.687& 1.90 \\
 \cline{3-7}
 & & $ld \rightarrow lK^+X$ & 1.73 & 13 & 0.750& 1.31 \\
 \cline{3-7}
 & & $ld \rightarrow l\pi^-X$ & 2.50 & 11 & 0.863 & 1.77 \\
 \cline{3-7}
 & & $ld \rightarrow l\pi^+X$ & 1.69 & 12 & 0.496 & 1.71 \\
 \cline{2-7}
 & \multirow{2}{*}{\cite{Adolph:2016dvl}} & $lp \rightarrow lh^-X$ & 4.02 & 31 & 0.959 & 6.15 \\
 \cline{3-7}
 & & $lp \rightarrow lh^+X$ & 3.93 & 34 & 0.847 & 4.70 \\
 \cline{2-7}
 & \cite{Aghasyan:2017jop}& $\pi^-p \rightarrow \gamma^*X$ & 5.34 & 15 & 0.659 & 1.03 \\ 
 \hline
 \multirow{6}{*}{HERMES} & \multirow{6}{*}{\cite{Airapetian:2009ae}} & $lp \rightarrow lK^-X$ & 1.70 & 14 & 0.398 & 0.72 \\
 \cline{3-7}
 & & $lp \rightarrow lK^+X$ & 1.73 & 14 & 1.545 & 0.54 \\
 \cline{3-7}
 & & $lp \rightarrow l\pi^0X$ & 1.76 & 13 & 0.962 & 0.59 \\
 \cline{3-7}
 & & $lp \rightarrow l\left(\pi^+-\pi^-\right)X$ & 1.73 & 15 & 1.182 & 0.52\\
 \cline{3-7}
 & & $lp \rightarrow l\pi^-X$ & 1.67 & 14 & 1.571 & 0.63 \\
 \cline{3-7}
 & & $lp \rightarrow l\pi^+X$ & 1.69 & 14 & 1.401 & 0.54\\
 \hline
 \multirow{2}{*}{JLAB} & \multirow{2}{*}{\cite{Qian:2011py}} & $lN \rightarrow l\pi^+X$ & 1.41 & 4 & 0.449 & 0.11\\
 \cline{3-7}
 & & $lN \rightarrow l\pi^-X$ & 1.69 & 4 & 1.725 & 0.11 \\
 \hline
 \multirow{3}{*}{RHIC} & \multirow{3}{*}{\cite{Adamczyk:2015gyk}} & $pp \rightarrow W^+X$ & $M_W$& 8$\omega$  & 2.031 & 48.9 \\
 \cline{3-7}
 & & $pp \rightarrow W^-X$ & $M_W$& 8$\omega$ & 1.583 & 52.0 \\
 \cline{3-7}
 & & $pp \rightarrow Z^0X$ & $M_Z$& $\omega$ & 3.198 & 35.1\\
 \hline
  Total & & & & 452 & 1.444 & \\
 \hline
\end{tabular}
\caption{The distribution of $\chi^2$ for each data set for the fit~2a in Tab.~\ref{schemes}. The column $\Delta \mathcal{T}$ is a measure of the sensitivity of the fit to the DGLAP evolution kernel.}
\label{chi2-all}
\end{center}
\end{table*}

In order to access which one of our theoretical assumptions is responsible for the large $\chi^2$ of the RHIC data, we have performed several tests. Firstly, we have checked whether the quality of the description of the RHIC data was due to the cut on $\qt/Q$. In order to check if quality of the fit is due to the value of this cut, we have performed an additional fit with the cut $\qt/Q<0.5$. We find that this change leads to a $\chi^2/N_{\rm data}$ is 1.885 for the RHIC data. While it would be preferable to perform an fit with $\qt/Q<0.25$, we note that there is not enough data in this region to constrain the parameters of the fit. Because there is no strong improvement in the description of the RHIC data after applying the $\qt/Q<0.5$, we conclude that this cut is not responsible for the disagreement between the data sets.

Another possible assumption that could be causing the large $\chi^2$ of the RHIC data is the assumption that the sea quarks have the same $\alpha$ and $\beta$ parameter. To check this, we have performed a 13 parameter fit with the chosen parameter with the parameters $\alpha_u$, $N_u$, $\beta_{\rm val}$, $\alpha_d$, $N_d$, $N_{\bar{u}}$, $N_{\bar{d}}$, $N_{s}$, $N_{\bar{s}}$, $\alpha_{+}$, $\alpha_{-}$, and $\beta_{\rm sea}$. Here $\alpha_{\bar{d}} = \alpha_{\bar{s}} = \alpha_{+}$ and $\alpha_{s} = \alpha_{\bar{u}} = \alpha_{-}$.  The introduction of the $\alpha_+$ and $\alpha_-$ parameterization decouples the positive and negative sea quarks from one another while the introduction of the parameters $\beta_{\rm val}$ and $\beta_{\rm sea}$ decouples the valance and sea quarks. However, we find that the addition of these parameters lead to a $\chi^2/N_{\rm data}$ is 1.885. This implies that this assumption on the function form is not the issue. 
\begin{figure}[htb!]
    \centering
    \includegraphics[width = 0.66\textwidth]{\FigPath/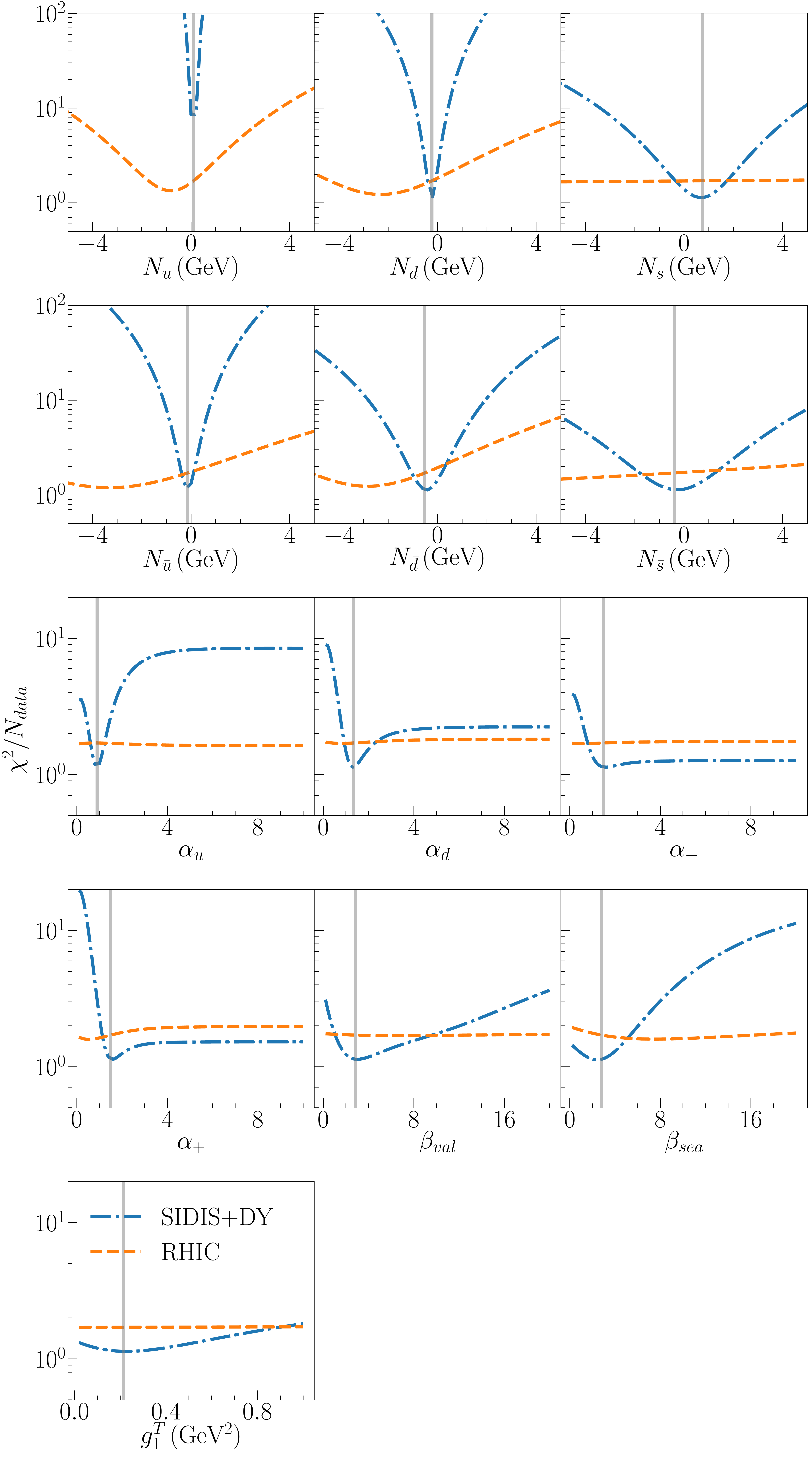}
        \caption{The distribution of $\chi^2/N_{\rm data}$ for each parameter. In each subplot, we vary each parameter about the central value while keeping all other parameters fixed to the optimal values determined by the fit. The gray line is the central value determined from the fit.}
    \label{chi2}
\end{figure}

In order to address the disagreement between the RHIC data and the rest of the data sets, in Fig.~\ref{chi2} we plot the profiles of the $\chi^2/N_{\rm data}$ using the 13 parameter fit. In each plot, we set all but one of the parameters equal to the values which are determined by the fit and we vary the remaining parameter about its best value. The best value determined by the fit is given by a vertical gray line. In this plot, we see that the curves for the RHIC $\chi^2$ do not change much as the $\alpha$, $\beta$, and $g_1^T$ parameters are varied. This indicates that the RHIC data is insensitive to these parameters. On the other hand, we see that when $N_q$ parameters are varied that there are large modifications to the RHIC $\chi^2$. Thus, the RHIC data is sensitive to these parameters. We see from the $N_q$ plots that the RHIC data and the rest of the data sets agree on the sign of the quark-Sivers functions for $N_d$,$N_{\bar{u}}$,$N_{\bar{d}}$, and $N_{\bar{s}}$ while the data sets disagree strongly about the magnitude of the parameters. For $N_{s}$, we see that the RHIC data appears to be insensitive to the sign of this parameter so that the disagreement is not striking. However, we find that the SIDIS and COMPASS Drell-Yan data sets indicate that the sign of the $u$-quark is positive while the RHIC data is indicating that the sign of the $u$-quark is negative. This disagreement is occurring because the fit program is attempting to describe the large positive $A_N$ asymmetry for the $W^+$ RHIC data. Thus in order to describe this data, either the $N_{\bar{d}}$ or $N_{\bar{s}}$ parameters must be large or the sign of the $N_u$ is incorrect. Since the value of the parameter $N_u$ is extremely well constrained by the SIDIS and COMPASS Drell-Yan data while the value of $N_{\bar{d}}$ and $N_{\bar{s}}$ parameters are weakly constrained, we conclude that this sign disagreement will be resolved once the magnitude of $N_{\bar{d}}$ and $N_{\bar{s}}$ parameters are addressed. 

Overall in Fig.~\ref{chi2}, we see the trend that the RHIC data $N_q$ requires much larger values for the $N_q$ parameter than the SIDIS and COMPASS Drell-Yan data. Since the SIDIS and COMPASS Drell-Yan data were gathered at much lower energy scales that the RHIC data, this tension between the sets indicates that the size of the Sivers asymmetry grows as a function of the hard energy scale. This result indicates that the issue in describing the RHIC data appears because of a possible evolution effect. 
Since the perturbative TMD evolution of the Sivers asymmetry is known, this issue is either occurring due to the chosen non-perturbative parameterization of the Sivers function or from the choice of the DGLAP evolution of the Qiu-Sterman function.
RHIC is expected to release the new measurement for $W/Z$ Sivers asymmetry~\cite{Aschenauer:2016our} in the near future in which they have much more statistics and thus smaller experimental uncertainty. The new data will be very valuable in constraining the non-perturbative component of the TMD evolution for the Sivers function. In the next section, we will study the effects of the DGLAP evolution of the Qiu-Sterman function and how they will affect the size of the asymmetry. 
\subsection{Effects of the DGLAP evolution}\label{DGLAP}
In order to examine how the DGLAP evolution of the Qiu-Sterman function affects the size of the asymmetry, we begin by examining \eref{UTV}. The largest contributions to this expression should appear in the region where $\mu_{b_*} \sim Q = M_V$ ~\cite{Collins:1984kg,Qiu:2000hf}. In this region, the size of the asymmetry is roughly proportional to $T_{F\, q/p}\left(N,M_V\right)$ in \eref{Tgtmassb}. To examine how the magnitude of the Qiu-Sterman function evolves in energy, we start from the evolution equation in the moment space in Eqs.~\eqref{e.TF-evo-N} and \eqref{e.TF-evo-N-gam}, and examine the ratio of this function at the two relevant scales $\mu_0$ and $M_V$. One can easily show that this ratio is given by
\begin{align}
    \frac{T_{F\, q/p}\left(N,M_V\right)}{T_{F\, q/p}\left(N,\mu_0\right)} = & 
    \mathcal{N}(\mu_0,M_V)  \left (\frac{\alpha_s\left(m_b^2\right)}{\alpha_s\left(\mu_0^2\right)}\right )^{-\gamma_u(N)/\beta_0(\mu_0)} \left (\frac{\alpha_s\left(M_V^2\right)}{\alpha_s\left(m_b^2\right)}\right)^{-\gamma_u(N)/\beta_0(M_V)}\,,
\end{align}
where $\mathcal{N}(\mu_0,M_V)$ is given by
\begin{align}
    \mathcal{N}(\mu_0,M_V) = & \left (\frac{\alpha_s\left(m_b^2\right)}{\alpha_s\left(\mu_0^2\right)}\right )^{\eta/\beta_0(\mu_0)}\left (\frac{\alpha_s\left(M_V^2\right)}{\alpha_s\left(m_b^2\right)}\right)^{\eta/\beta_0(M_V)}\,.
    \label{e.argument}
\end{align}
From this expression, it becomes clear that when $\eta>0$, the factor $\mathcal{N}$ in \eref{argument} becomes small at large scales. Thus this factor leads to a suppression of the quark-Sivers function at large scales. Thus with this suppression factor, the values of the $N_q$ parameters must be very large in order to describe the RHIC data. On the other hand, when $\eta<0$, the factor $\mathcal{N}$ leads to an enhancement of the asymmetry at large scales. 

\begin{figure}[htb!]
    \centering
    \includegraphics[width = 0.48\textwidth]{\FigPath/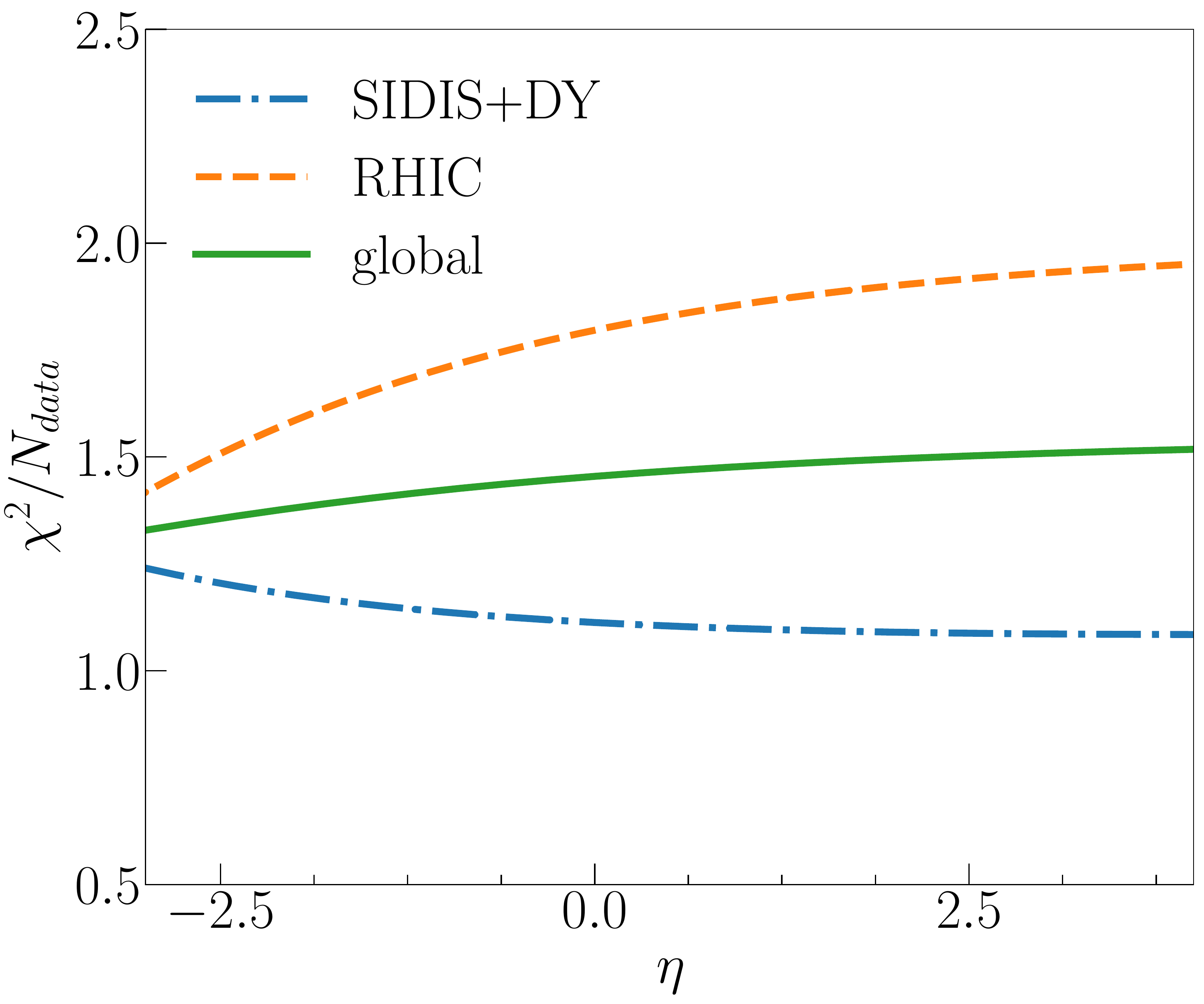}
        \caption{$\chi^2/N_{\rm data}$ profile for the $\eta$ parameter. To generate this plot, we use the parameter values from Sec.~\ref{RHIC} and vary the parameter $\eta$.}
    \label{eta_prof}
\end{figure}

In order to test the sensitivity of each data set to changes in the evolution kernel due to the change in the $\eta$ parameter in Eq.~\eqref{e.TF-evo}, we define the quantity 
\begin{align}
    \Delta \mathcal{T} & = \frac{1}{N_{\rm set}} \sum_{i = 1}^{N_{\rm set}}\left|\frac{T\left(\left\{a\right\},\eta = N_C\right)-T\left(\left\{a\right\},\eta = 0\right)}{T\left(\left\{a\right\},\eta = N_C\right)}\right|\times 100\,, \nn 
\end{align}
which gives the average percent difference between the two theory calculated with $\eta = N_C$ and $\eta = 0$ for a given set. In this expression, $\left\{a\right\}$ are the parameters obtained from the $\eta = N_C$ fit. In Tab.~\ref{chi2-all}, we provide the value for $\Delta \mathcal{T}$ for each data set. We find that the result of the low energy data can vary only within a few percent on the choice of the DGLAP evolution kernel. On the other hand, the high energy RHIC data sets varies by a factor of $~50\%$ when using these different kernels. 

In order to explicitly demonstrate the dependence on the DGLAP evolution scheme, in Fig.~\ref{eta_prof} we plot a profile of the $\chi^2/N_{\rm data}$ as a function of the parameter $\eta$, while the rest of the parameters are fixed as those from scheme fit~2a. As we can see from this plot, $\chi^2/N_{\rm data}$ for RHIC data decreases as $\eta$ decreases. This indicates the RHIC seems to prefer smaller $\eta\sim 0$ or even negative values. This trend is opposite to what is seen in the SIDIS+DY data. Because of the driving from the RHIC data, the global $\chi^2$ seems to favor the evolution scheme with $\eta = 0$ or even negative.
\subsection{Global fit of the Sivers function}\label{global}
\begin{table}[hbt!]
\begin{center}
 \begin{tabular}{||c | c | c | c | c | c ||} 
 \hline
 Collab & Ref & Process & $Q_{\rm avg}$ & $N_{\rm data}$ & $ \chi^2/N_{\rm data}$\\ [0.5ex]
 \hline\hline
 \multirow{8}{*}{COMPASS} & \multirow{5}{*}{\cite{Alekseev:2008aa}} & $ld \rightarrow lK^0X$ & 2.52 & 7 & 0.823 \\
 \cline{3-6}
 & & $ld \rightarrow lK^-X$ & 2.80 & 11 & 0.886 \\
 \cline{3-6}
 & & $ld \rightarrow lK^+X$ & 1.73 & 13 & 0.831 \\
 \cline{3-6}
 & & $ld \rightarrow l\pi^-X$ & 2.50 & 11 & 1.071 \\
 \cline{3-6}
 & & $ld \rightarrow l\pi^+X$ & 1.69 & 12 & 0.596 \\
 \cline{2-6}
 & \multirow{2}{*}{\cite{Adolph:2016dvl}} & $lp \rightarrow lh^-X$ & 4.02 & 31 & 0.975 \\
 \cline{3-6}
 & & $lp \rightarrow lh^+X$ & 3.93 & 34 & 0.988 \\
 \cline{2-6}
 & \cite{Aghasyan:2017jop}& $\pi^-p \rightarrow \gamma^*X$ & 5.34 & 15 & 0.675 \\ 
 \hline
 \multirow{6}{*}{HERMES} & \multirow{6}{*}{\cite{Airapetian:2009ae}} & $lp \rightarrow lK^-X$ & 1.70 & 14 & 0.368 \\
 \cline{3-6}
 & & $lp \rightarrow lK^+X$ & 1.73 & 14 & 2.042 \\
 \cline{3-6}
 & & $lp \rightarrow l\pi^0X$ & 1.76 & 13 & 1.039 \\
 \cline{3-6}
 & & $lp \rightarrow l(\pi^+-\pi^-)X$ & 1.73 & 15 & 1.238\\
 \cline{3-6}
 & & $lp \rightarrow l\pi^-X$ & 1.67 & 14 & 1.318 \\
 \cline{3-6}
 & & $lp \rightarrow l\pi^+X$ & 1.69 & 14 & 1.677\\
 \hline
 \multirow{2}{*}{JLAB} & \multirow{2}{*}{\cite{Qian:2011py}} & $lN \rightarrow l\pi^+X$ & 1.41 & 4 & 0.651 \\
 \cline{3-6}
 & & $lN \rightarrow l\pi^-X$ & 1.69 & 4 & 2.409 \\
 \hline
 \multirow{3}{*}{RHIC} & \multirow{3}{*}{\cite{Adamczyk:2015gyk}} & $pp \rightarrow W^+X$ & $M_W$& 8$\omega$  & 1.929 \\
 \cline{3-6}
 & & $pp \rightarrow W^-X$ & $M_W$& 8$\omega$ & 1.461 \\
 \cline{3-6}
 & & $pp \rightarrow Z^0X$ & $M_Z$& $\omega$ & 3.113\\
 \hline
  Total & & & & 452 & 1.446 \\
 \hline
\end{tabular}
\caption{The distribution of $\chi^2$ for each data set for the fit~2b.}
\label{chi2-all-eta0}
\end{center}
\end{table}
In Sec.~\ref{sid_dy}, we have presented fit~1, which was performed to Sivers asymmetry for SIDIS+DY data at the low energy. The strengths of this extraction are that the theoretical uncertainties were small so that this extraction should describe very well future low energy experiments. However, as we showed in the prediction for the RHIC data, this extraction failed to describe the high energy data. In this section, we present a fit which emphasizes the contributions of the RHIC data in order to allow future predictions for high energy measurements of the Sivers asymmetry.

To emphasize the contributions of the high energy data, we retain the weighted definition of the $\chi^2$ in \eref{chi2_RHIC}. On the other hand, as we have seen in our model, the description of the high energy data from RHIC depends strongly on the choice of the parameter $\eta$. By performing a global fit with $\eta = N_C$, we found that the $\chi^2/N_{\rm data}$ for RHIC was $1.888$. In order to eliminate the suppression from the $-N_C\delta(1-x)$ term in the evolution kernel Eq.~\eqref{e.TF-evo}. In this section, we perform the fit with $\eta = 0$. This fit is referred to as fit~2b in Tab.~\ref{schemes}.

\begin{table}[htb!]
\def\arraystretch{1.25}
  \begin{center}
    \begin{tabular}{c c c c c}
        & \multicolumn{3}{c}{$\chi^2/d.o.f.= 1.482$} & \\
        \hline
        $N_{u}=$         & $ 0.098_{-0.005}^{+0.205}$ GeV& & $\alpha_{u}\;\;\;\;\; =$     & $ 0.821_{-0.205}^{+1.088}$\\
        $N_{d}=$         & $-0.254_{-2.549}^{+0.147}$ GeV&  &$\alpha_{d}\;\;\;\;\;=$     & $ 1.342_{-0.466}^{+4.703}$\\
        $N_{s}=$          & $ 0.754_{-0.148}^{+5.027}$ GeV&  &$\alpha_{sea}\;\;=$ & $ 1.501_{-0.060}^{+1.698}$\\
        $N_{\bar{u}}=$  & $-0.140_{-3.004}^{+0.009}$ GeV&  &$\beta\;\;\;\;\;\;\,=$       & $2.764_{-0.762}^{+2.827}$\\
        $N_{\bar{d}}=$ & $-0.510_{-6.904}^{+0.126}$ GeV & & $g_1^T\;\;\;\;\,=$ & $0.232_{-0.010}^{+0.768}$ GeV$^2$  \\
        $N_{\bar{s}}=$ & $-0.387_{-4.536}^{+0.422}$  GeV&  &
    \end{tabular}
    \caption{Fit parameters for fit~2b in Tab.~\ref{schemes}. The presented values is the parameter value of the fit with no Gaussian noise. The uncertainties for the replicas are generated from the parameter values which lie on the boundary of $68\%$ confidence.}
    \label{paramtableeta0}
  \end{center}
\end{table}
\begin{figure*}[htbp]
    \centering
    \includegraphics[width = \textwidth]{\FigPath/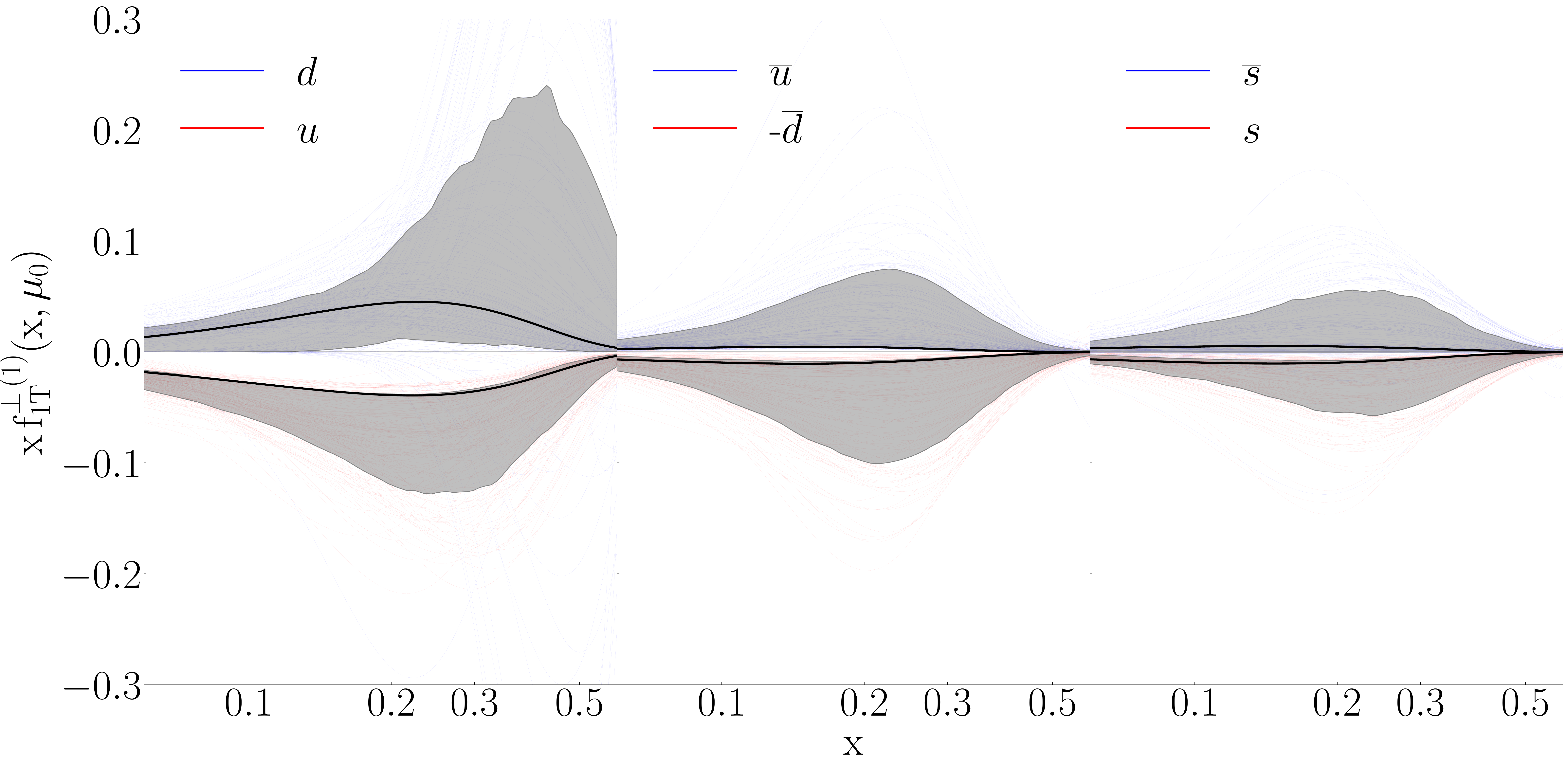}
        \caption{The extracted transverse moment of the Sivers function from fit~2b in Tab.~\ref{schemes} at $\mu_0 = \sqrt{1.9} $ GeV. The black curve is the fit to the experimental data with no Gaussian noise.}
    \label{sivers_eta}
\end{figure*}
\begin{figure*}[hbt]
\centering
    \centering
    \includegraphics[width = 0.48\textwidth]{\FigPath/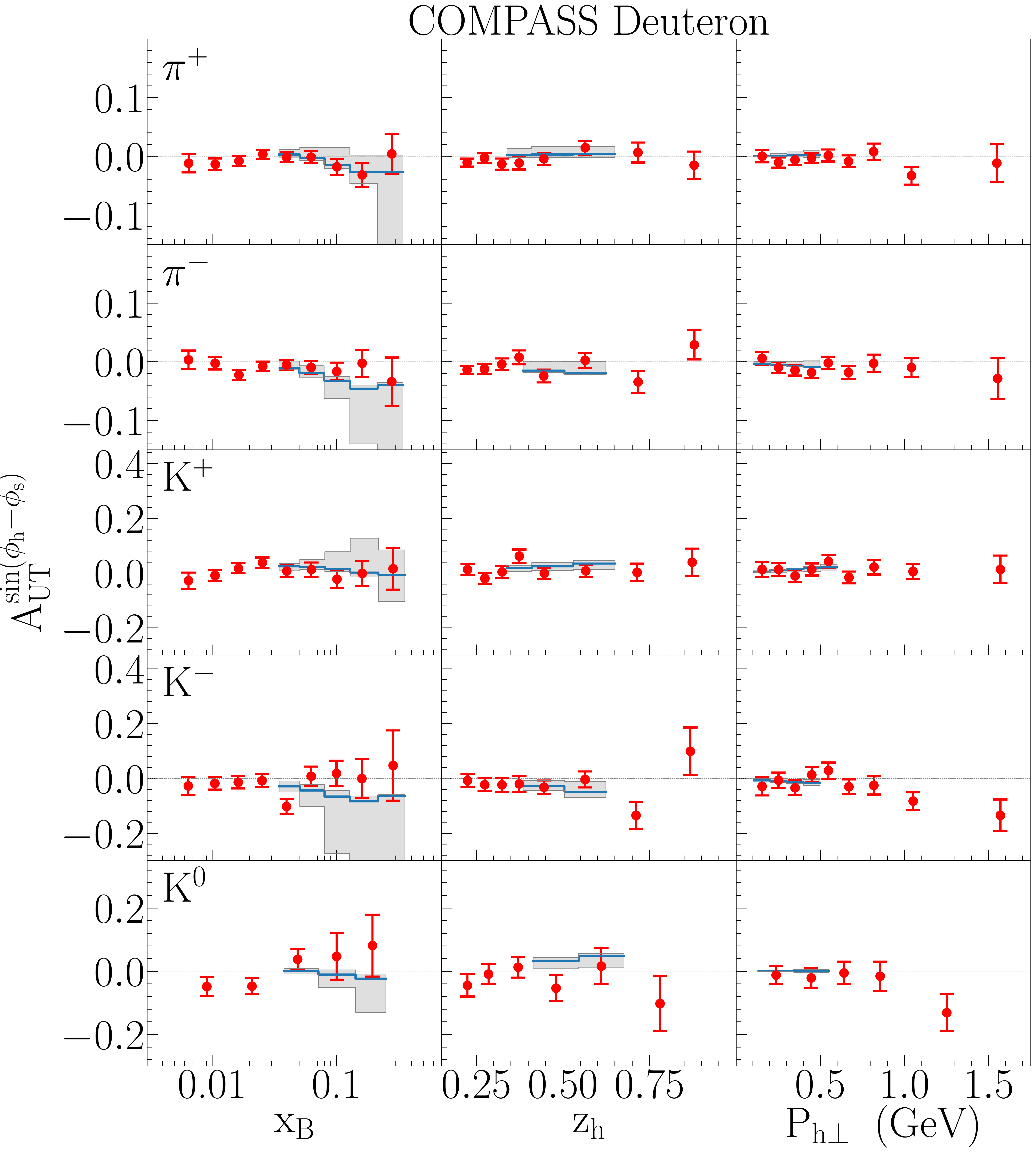}
    \includegraphics[width = 0.48\textwidth]{\FigPath/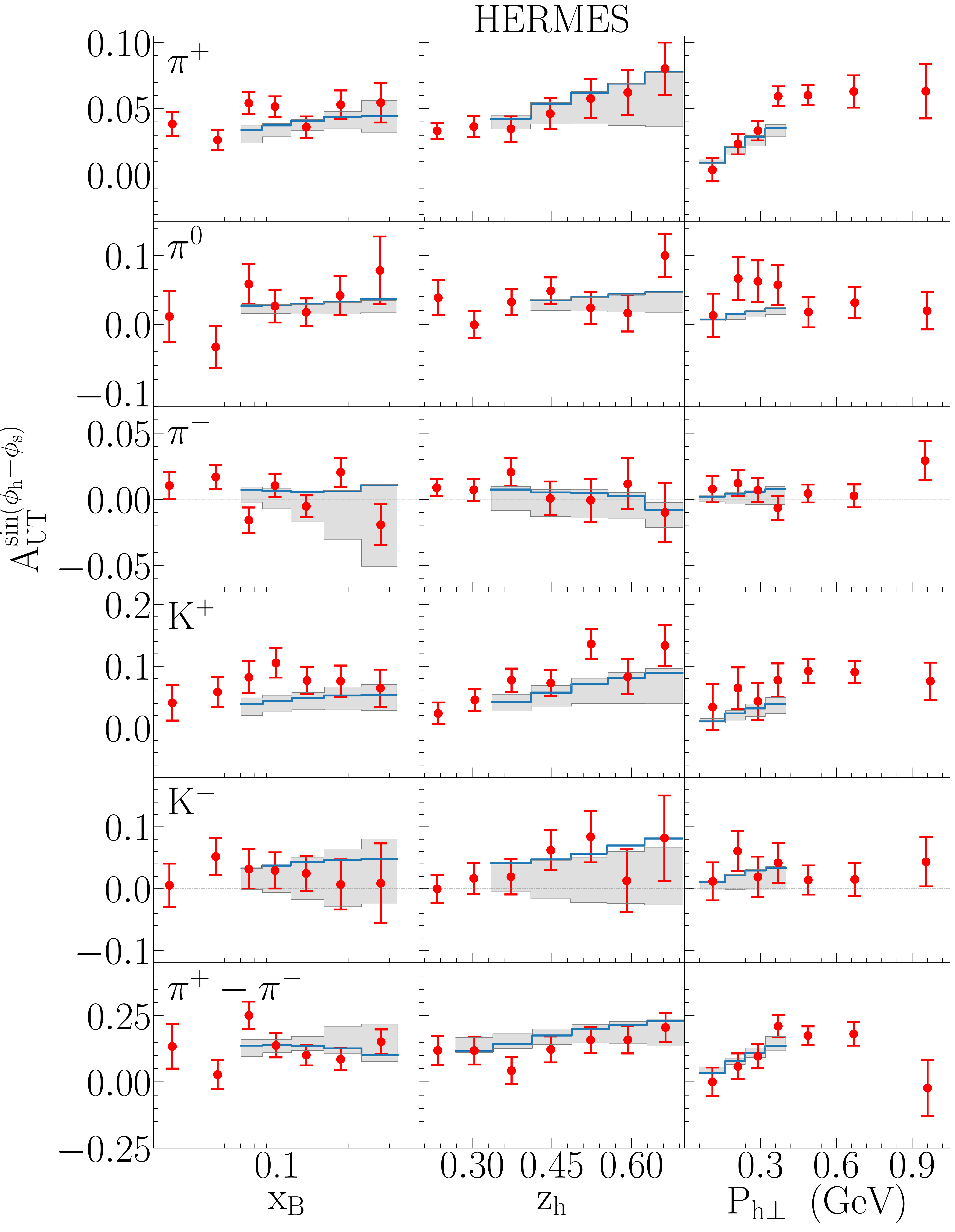}
    \caption{
    Left: The COMPASS deuteron target measurement \cite{Alekseev:2008aa} for $\pi^+$, $\pi^-$, $K^+$, $K^-$, and $K^0$ from top to bottom, and as a function of $x_B$ (left), $z_h$ (middle), and $P_{h\perp}$ (right). Right: HERMES proton target measurement \cite{Airapetian:2009ae} $\pi^+$, $\pi^0$, $\pi^-$, $K^+$, $K^-$, and $(\pi^+-\pi^-)$ from top to bottom, and as a function of $x_B$ (left), $z_h$ (middle), and $P_{h\perp}$ (right). The data is plotted in red along with the total experimental error. The central curve in blue as well as the uncertainty band in gray are generated using the result from fit~2b in Tab.~\ref{schemes}.}
    \label{f.C_D+H_eta}
\end{figure*}

For this fit, we recover a $\chi^2/d.o.f$ of $1.482$ with a $\chi^2/N_{\rm data}$ of $1.778$ for the RHIC data. The parameter values for this fit are given in Tab.~\ref{paramtableeta0} while the distribution of the $\chi^2$ is given in Tab.~\ref{chi2-all-eta0}. We can see from Tab.~\ref{paramtableeta0} that while the extraction of the Sivers function from the low energy data could not resolve the sign of the $s$-quark Sivers function, this fit finds that the $s$-quark should be positive. At the same time, the sign of all other quark functions are consistent with the previous extraction. However, we note that the central values for the $N_q$ parameters are much larger than the previous fit. This is occurring because of the large RHIC asymmetry along with the weighting used in the fit. We see also in this table that the uncertainties in the parameters are very large and tend to skewed in one direction. The magnitude of this uncertainty is due to the large experimental uncertainties in the RHIC data while the skew favors fits which increase the size of the asymmetry. 
\begin{figure*}[htb]
    \centering
    \includegraphics[width = 0.48 \textwidth]{\FigPath/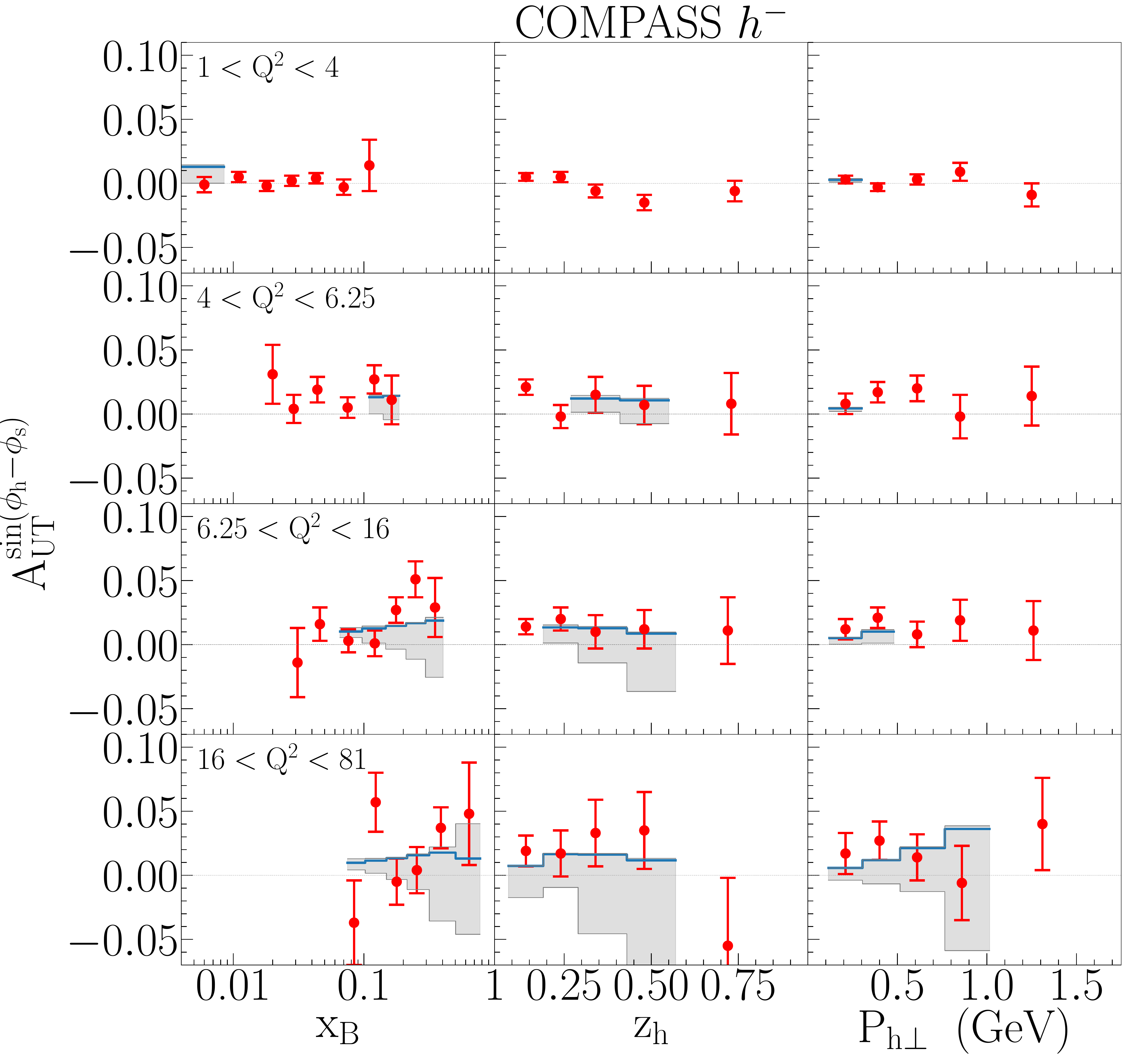}
    \includegraphics[width = 0.48 \textwidth]{\FigPath/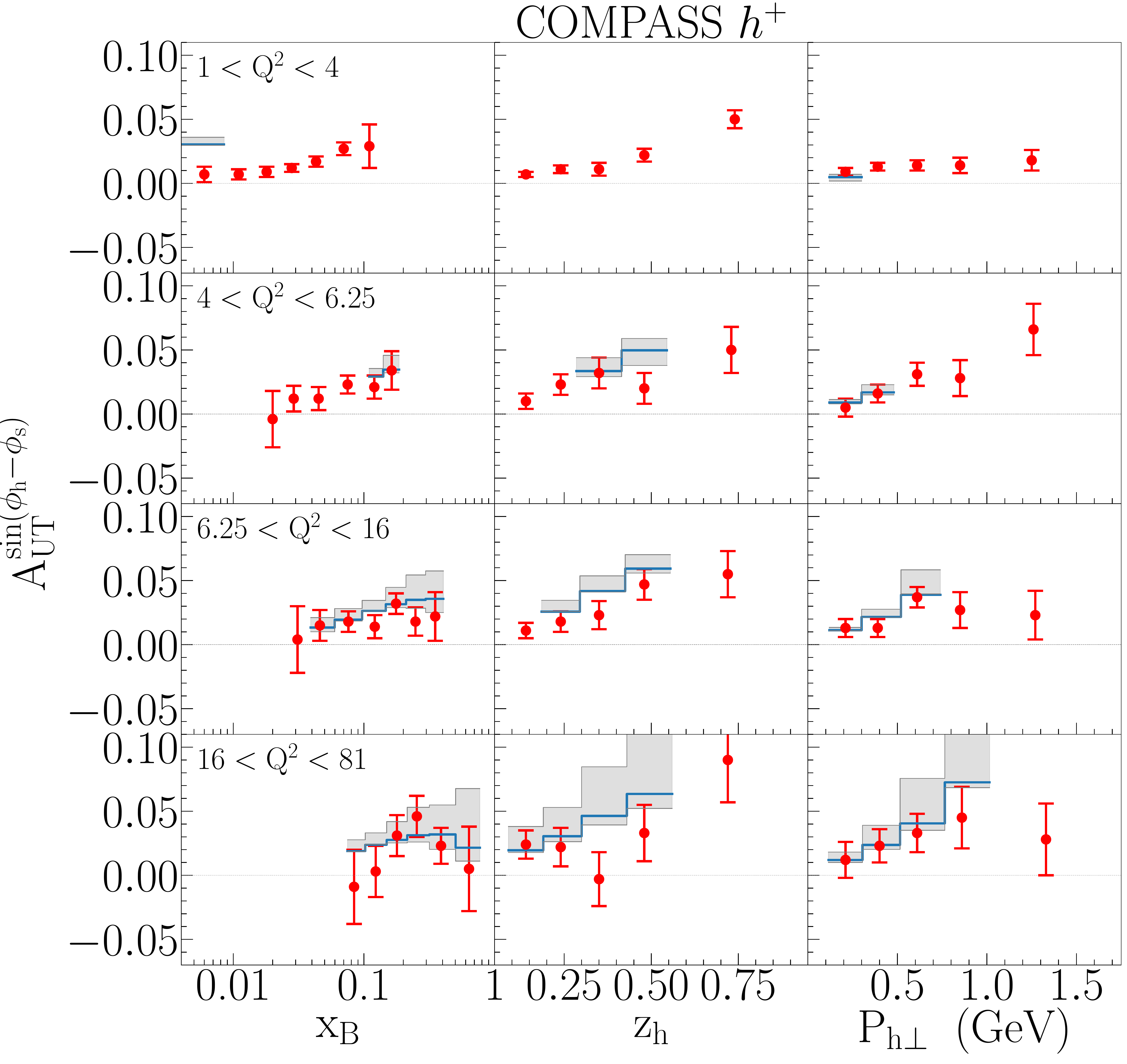}
    \caption{Left: The COMPASS proton target measurement for $h^-$ for $1$ GeV$2$ $< Q^2< 4$ GeV$^2$, $4$ GeV$2$ $< Q^2< 6.25$ GeV$^2$, $6.25$ GeV$2$ $< Q^2< 16$ GeV$^2$, $16$ GeV$2$ $< Q^2< 81$ GeV$^2$ from top to bottom \cite{Adolph:2016dvl}. Right: Same as the left except for $h^+$ production. The central curve and uncertainty band are generated using the result from fit~2b in Tab.~\ref{schemes}.}
    \label{f.C_h_eta}
\end{figure*}
\begin{figure*}[hbt]
\begin{center}
    \begin{minipage}{.5\textwidth}
    \begin{center}
    \includegraphics[width = 0.5\textwidth]{\FigPath/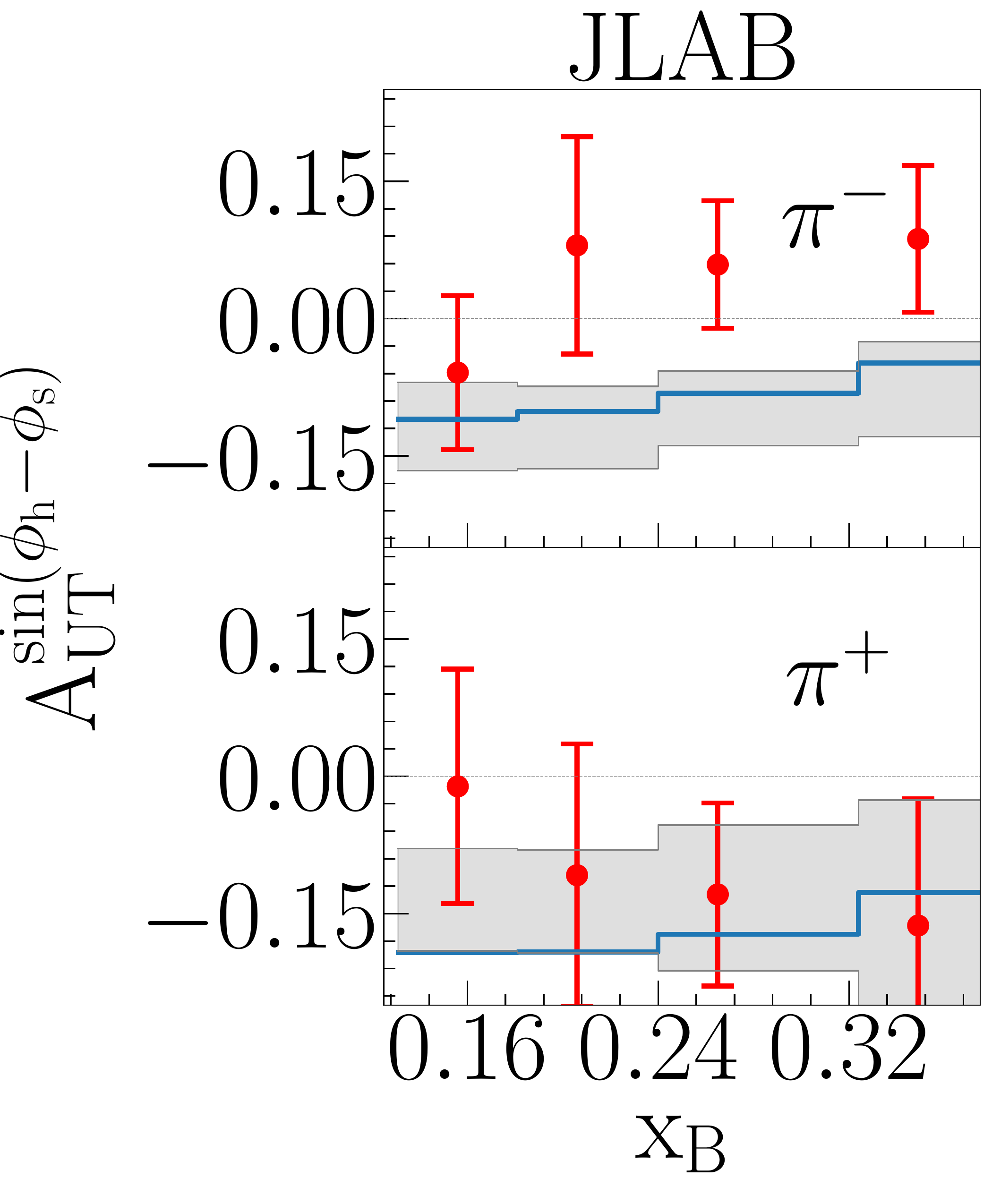}
    \caption{JLab measurement of the Sivers asymmetry for a neutron target \cite{Qian:2011py} as a function of $x_B$. The central curve and uncertainty band are generated using the result from fit~2b in Tab.~\ref{schemes}.}
    \label{f.J_eta}
    \end{center}
    \end{minipage}
\end{center}
\end{figure*}
\begin{figure*}[hbt]
\centering
    \centering
    \includegraphics[width = 0.75 \textwidth]{\FigPath/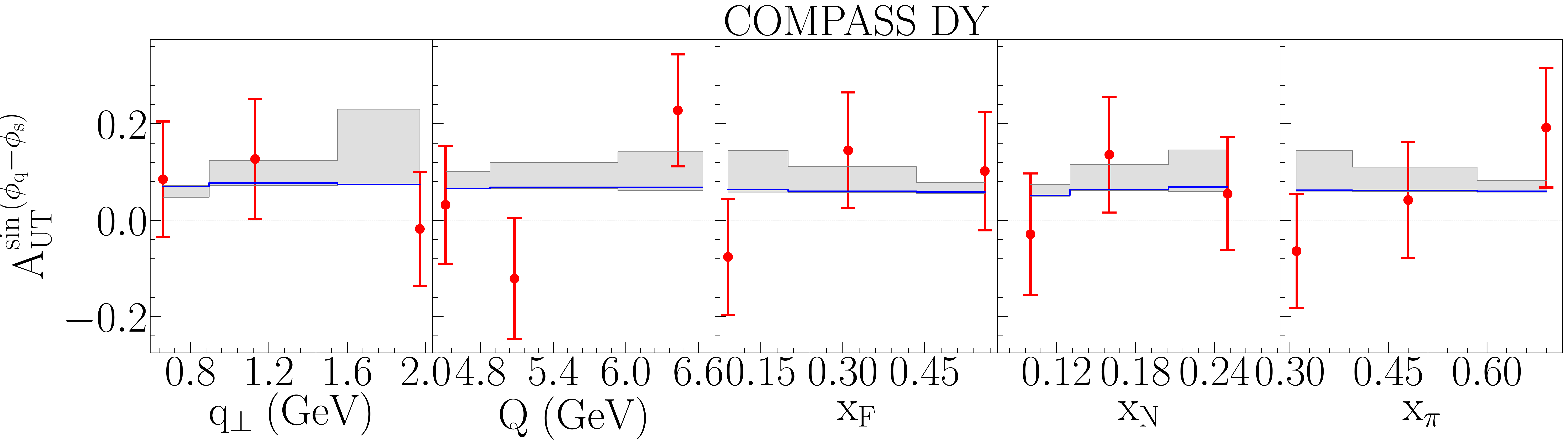}
        \caption{COMPASS Drell-Yan measurement for $\pi^-$-$p$ collision \cite{Aghasyan:2017jop} as a function of $\qt$, $Q$, $x_F$, $x_N$, and $x_{\pi}$ from left to right. The central curve and uncertainty band are generated using the result from fit~2b in Tab.~\ref{schemes}.}
    \label{f.C_DY_eta}
\end{figure*}
\begin{figure*}
\begin{minipage}{.56\textwidth}
    \centering
    \includegraphics[width = \textwidth]{\FigPath/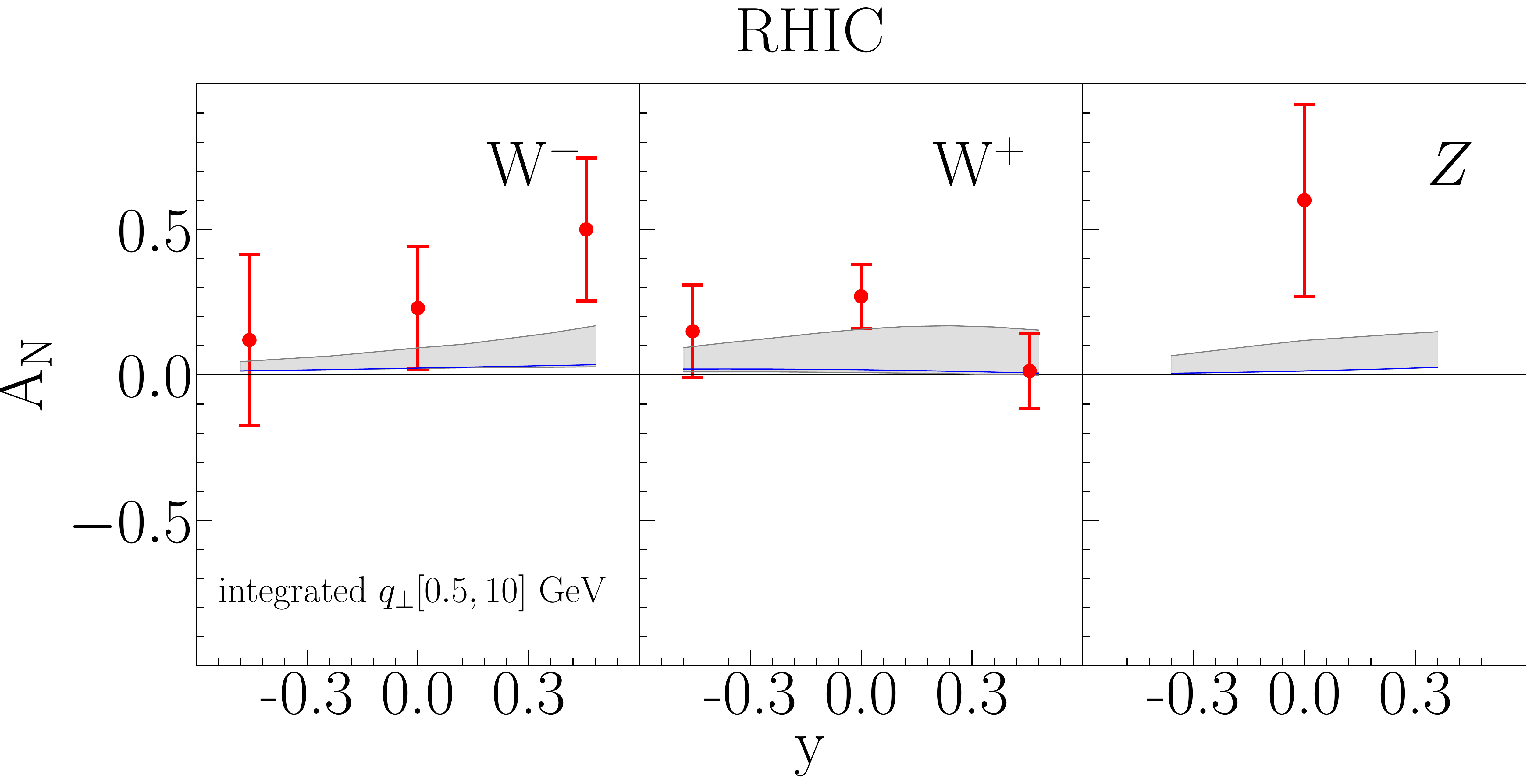}
\end{minipage}%
\begin{minipage}{.42\textwidth}
    \centering
    \includegraphics[width =\textwidth]{\FigPath/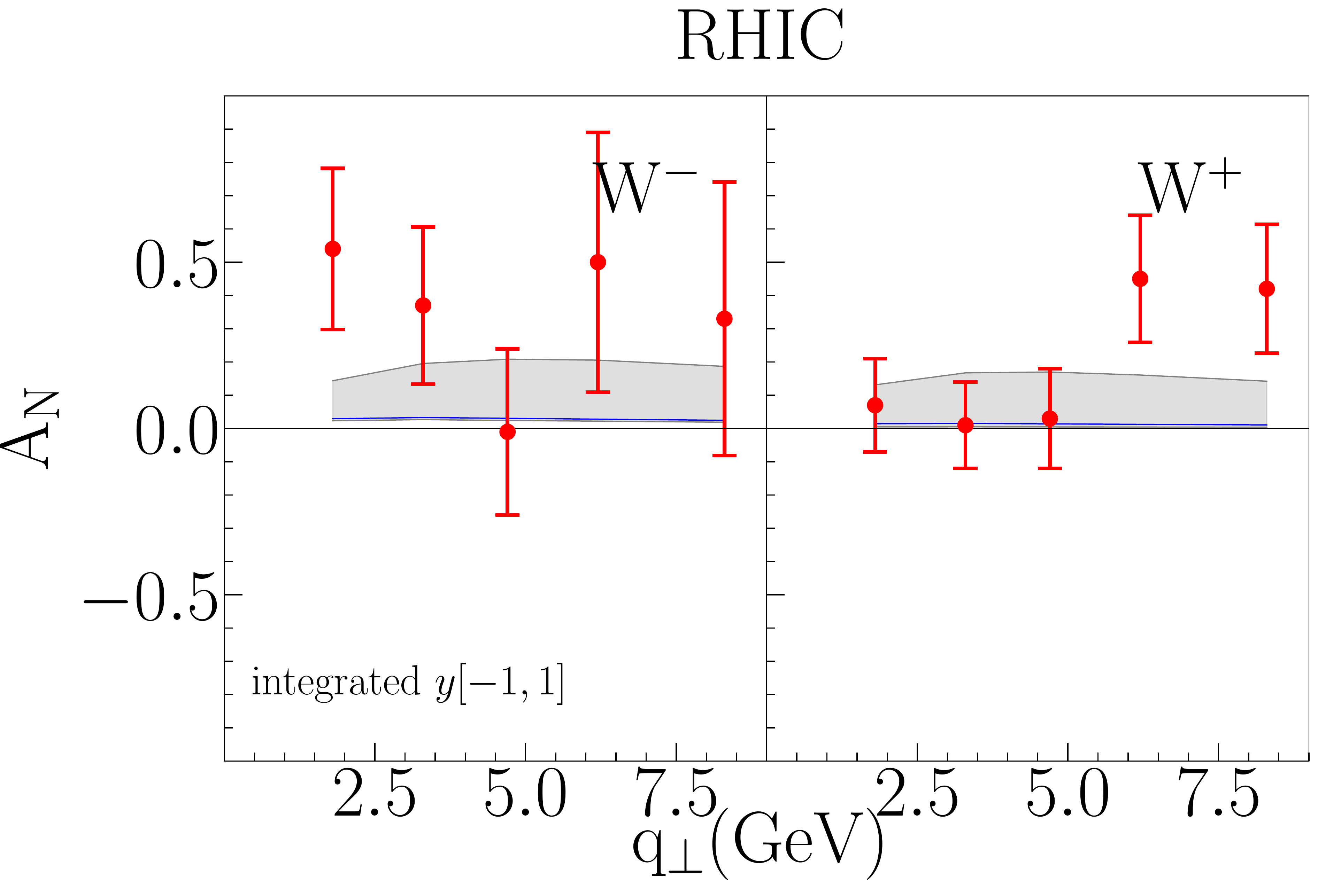}
\end{minipage}
    \caption{The Sivers asymmetry for $p+p\rightarrow W/Z$ at $\sqrt{S} = 500$ GeV \cite{Adamczyk:2015gyk}. The central curve and uncertainty band are generated using the result from fit~2b in Tab.~\ref{schemes}.}
    \label{f.R_eta}
\end{figure*}

In \fref{sivers_eta}, we plot the extracted transverse momentum moment of the Sivers function, $f_{1T}^{\perp (1)}(x, \mu_0)$ as a function of $x$ at the scale $\mu_0 = \sqrt{1.9}$ GeV. The blue curve is the fit to the experimental data with no Gaussian noise, while the grey uncertainty band is generated from the middle 68\% of the curves. In comparison with the extracted Sivers function in Fig.~\ref{f.siversplot} from SIDIS+DY data at low energy, this fit leads to much larger uncertainty band for the Sivers function. The size of the Sivers functions in this fit is also significantly larger. This is of course due to the much larger asymmetries for $W/Z$ bosons measured at RHIC. We note that we have also checked the extracted asymmetry in fit~2b against the new HERMES data in \cite{Airapetian:2020zzo}. We find that there is very strong agreement between this extracted asymmetry and the new data.

In Figs.~\ref{f.C_D+H_eta}, \ref{f.C_h_eta}, and \ref{f.J_eta}, we plot the theoretical curve of this fit against the low energy experimental data for SIDIS Sivers asymmetry. In Fig.~\ref{f.C_DY_eta}, the comparison with the COMPASS Drell-Yan data is presented. While the theoretical uncertainties are much larger than the previous extraction, the fitted asymmetry still describes the this subset of the data very well. Finally, in \fref{f.R_eta}, we plot the fitted asymmetry to the RHIC data. We find that in this scheme, the size of the asymmetry for the central fit can now be up to $5\%$. Overall, this scheme describes the RHIC data much better than the previous extraction. The future RHIC data with much smaller experimental uncertainty will for sure help to reduce the theoretical uncertainties in the extracted Sivers functions, as well as the Sivers asymmetries computed based on these Sivers functions. 

%%%%%%%%%%%%%%%%%%
\section{Predictions for the EIC}\label{EIC}
As we have seen in the previous sections, the choice of DGLAP evolution scheme used for the evolution of the Qiu-Sterman function greatly affects the quality of the fit when considering data at large hard scales. While this issue currently presents difficulties for performing a global extraction of the Sivers function, this effect also presents an opportunity at the future EIC. The EIC will be capable of performing high precision measurements of transverse spin asymmetries at a large range of scales. Experimental data which are collected over these large range of scales can be used to study DGLAP evolution effects of the Qiu-Sterman function.  
\begin{figure}[htb!]
    \centering
    \includegraphics[width =0.48\textwidth]{\FigPath/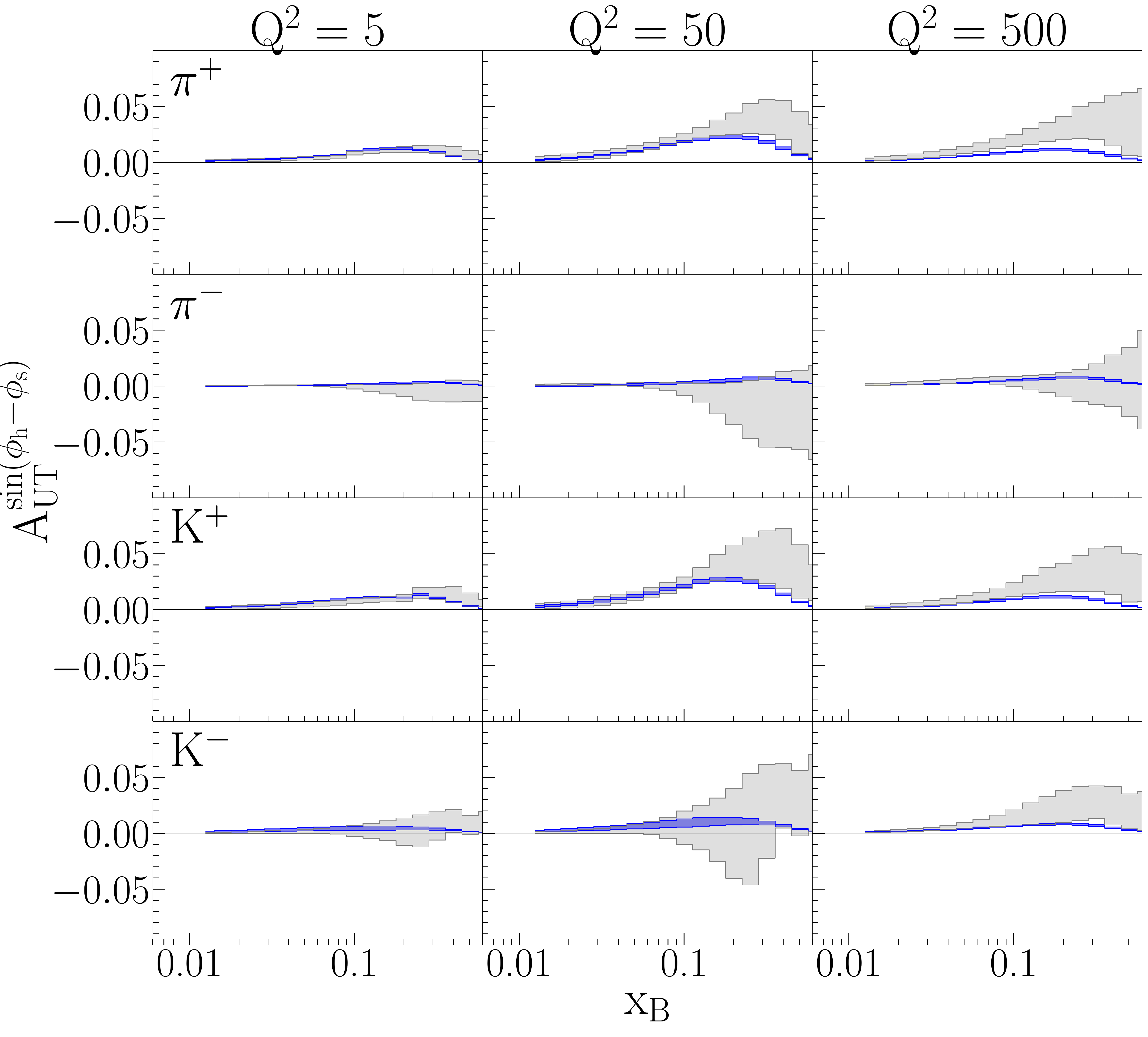}
    \includegraphics[width =0.48\textwidth]{\FigPath/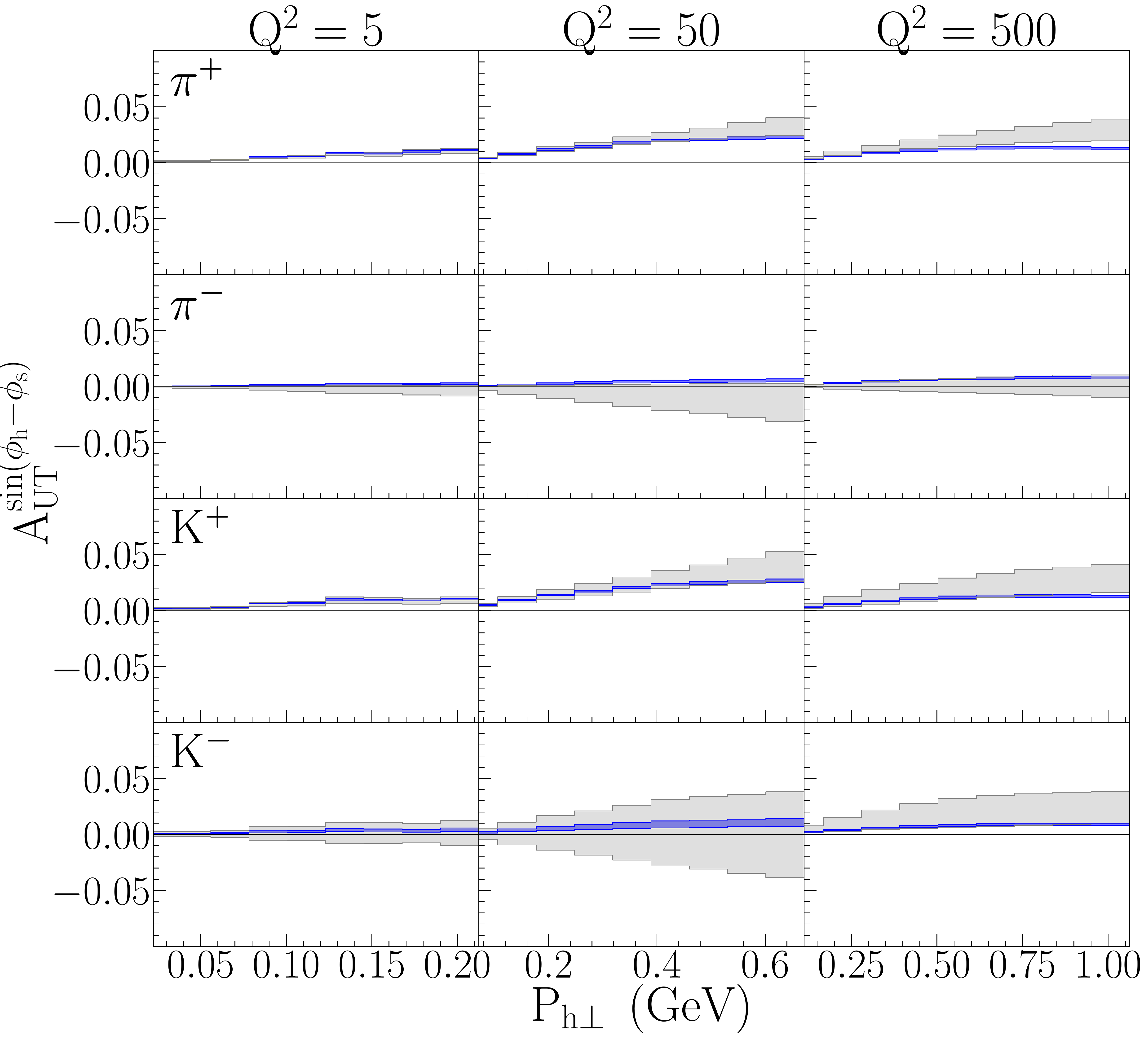}
    \caption{The prediction for the EIC at $\sqrt{S} = 105$~GeV. Left: The $x_B$ dependent prediction at $z_h = 0.5$ and $q_\perp/Q = 0.2$. Right: The $P_{h\perp}$ dependent prediction at $x_B = 0.2$ and $z_h =0.25$. The blue band represents the prediction from the low energy extraction while the blue band represents the prediction from the $\eta = 0$ scheme.}
    \label{f.EIC}
\end{figure}
On the left side of Fig.~\ref{f.EIC}, we plot our prediction for the Sivers asymmetry in SIDIS on a proton target as a function of $x_B$ at $\sqrt{S} = 105$ GeV, $z_h = 0.25$, $\qt/Q = 0.2$ at $Q^2 = 5$, $50$, $500$ GeV$^2$ for $\pi^+$, $\pi^-$, $K^+$, and $K^-$ production. In this figure, we have plotted our prediction for the low energy fit (fit~1 in Tab.~\ref{schemes}) in blue, and the global fit (fit~2b in Tab.~\ref{schemes}) in gray. While this prediction demonstrates the $x$-dependence of our fits, in order to demonstrate the $k_{\perp}$-dependence of our fitted Sivers function, we also make a prediction as a function of $P_{h\perp}$ on the right side of Fig.~\ref{f.EIC}. In this figure, we have used the same kinematics as the left side except that we take $x_B = 0.2$. We see from these curves that the predicted asymmetry for $\pi^-$ and $K^-$ production is small. This behavior is expected because of the suppression by the fractional charge $e_d^2$ for the $d$-quark Sivers function, as well as the cancellation that occurs between the $u$ and $d$-quarks. On the other hand, we predict an asymmetry of a few percent for $\pi^+$ and $K^+$ production in this kinematic region.

We see in these plots that the theoretical curves generated from fit~1 and fit~2b are very similar at $Q^2 = 5$ GeV$^2$. This behaviour occurs because the suppression factor, $\mathcal{N}$ in~\eref{argument}, is close to one at small energies. However, at $Q^2 = 500$ GeV$^2$, the theoretical curves generated from fit~1 and fit~2b can differ by a few percent. This effect presents a great opportunity at the future EIC. Since measurements at large values of $Q^2$ are sensitive to the DGLAP evolution effects of the Qiu-Sterman function, these data may prove useful in phenomenological studies of this evolution. At the same time, these future measurements at the EIC could provide additional statistics for high energy data which will prove useful in reducing the theoretical uncertainties for the extraction of the Sivers asymmetry at large energy scales.

\section{Conclusions}\label{Conclusions}
In this paper, we have performed extractions of the Sivers function for the first time at the NLO+NNLL order. We first perform an extraction from the Sivers asymmetry data measured in SIDIS at HERMES, COMPASS and JLab, and in Drell-Yan lepton pair production at COMPASS. Using this first extraction, we generate a prediction for the Sivers asymmetry of $W/Z$ boson at RHIC kinematics and compare with the experimental data. We find that while the SIDIS and COMPASS Drell-Yan lepton pair production data is very well described by our extraction, that our theoretical curve is much smaller than the RHIC data. We study in great detail the impact of the RHIC data and their implications. For such a purpose, we perform a fit in which we introduce a weighting factor of $\sim 13$ for the RHIC data, so that the RHIC data and the rest of the experimental data sets are equally weighed in the calculation of the $\chi^2$. 
We study how RHIC data are sensitive or insensitive to the non-perturbative parameters in the Sivers function parameterization. In addition, we study in detail the dependence on the choice of the scheme used to perform the DGLAP evolution of the Qiu-Sterman function, the collinear counterpart that enters the TMD evolution formalism for the Sivers function. We investigate the impact of two DGLAP evolution schemes which are commonly used in the extraction of the Sivers function. We find that the scheme which treats the evolution of the Qiu-Sterman function the same as the unpolarized parton distribution function, is better suited for describing the experimental data at RHIC. Using DGLAP evolution scheme, we perform for the first time a global extraction of the Sivers function and find that this scheme improves the description of the RHIC data. While our first fit describes the low energy data extremely well, our second fit describes the RHIC data much better than the first. However, due to the large experimental uncertainties at RHIC, we find that the globally extracted Sivers function has large theoretical uncertainties. We expect the forthcoming RHIC experimental data on $W/Z$ Sivers asymmetry with large statistics and reduced experimental uncertainties would help us better constrain the Sivers function and its evolution. In addition, we make predictions for Sivers asymmetry at the future Electron Ion Collider (EIC). We find that with large range of hard scale $Q$ to be probed at the EIC, the effects due to the DGLAP evolution of the Qiu-Sterman function can be extremely pronounced. Such measurements would present a great opportunity for testing such effects.

Upon publication, the extracted Sivers functions from both fits in this paper will be made available open source at the following link: 
\href{https://github.com/UCLA-TMD/TMD-GRIDS/tree/EKT2020}{https://github.com/UCLA-TMD/TMD-GRIDS/tree/EKT2020}.

%%%%%%%%%%%%%%%%%%%%%%%%%%%%%%%%%%%%%%%%%%%%%%%%%%%%%%%%%%%%%
\section*{Acknowledgements}
We thank A.~Bacchetta, M.~Guzzi, A.~Prokudin, N.~Sato and G.~Schnell for useful discussions. M.G.E. is supported by the Spanish MICINN under Grant No. PID2019-106080GB-C21. Z.K. is supported by the National Science Foundation under Grant No.~PHY-1720486 and CAREER award~PHY-1945471. J.T. is supported by NSF Graduate Research Fellowship Program under Grant No.~DGE-1650604. This work is also supported within the framework of the TMD Topical Collaboration.

%%%%%%%%%%%%%%%%%%%%%%%%%%%%%%%%%%%%%%%%%%%%%%%%%%%%%%%%%%%%%
\appendix
\section{Wilson Coefficient Functions}\label{Coefficient}
The scale dependent TMDPDF quark to quark and gluon to quark Wilson coefficient function is given by \cite{Aybat:2011zv,Kang:2015msa,Collins:2012uy}
\begin{align}
    C_{q\leftarrow q'}(x,\bt;\mu,\zeta)  = & \delta_{qq'} \delta(1-x) +\delta_{qq'}\frac{\alpha_s}{4\pi}
    \Big[2 C_F(1-x)-2 P_{q\leftarrow q}(x) L \nn \\
    &
     -L\le( -3+C_F\le( L+2 L_{\z} \ri) \ri) \delta(1-x) -C_F\frac{\pi^2}{6}\delta(1-x) \Big]\,, 
\\
    C_{q\leftarrow g}(x,\bt;\mu,\zeta) =& \frac{\alpha_s}{\pi}
     \Big[ x(1-x)T_F-\frac{1}{2}P_{q\leftarrow g}(x)L
     \Big ]\,, \nn
\end{align}
where in these expressions, we have used the short-hand
\begin{align}
    L = \ln\left(\frac{\mu^2}{\mu_b^2}\right)\,,
    \qquad
    L_{\zeta} = \ln\left(\frac{\zeta}{\mu^2}\right)\,.
\end{align}
The quark to quark coefficient function for the TMDFF is given by the relation 
\begin{align}
    \hat{C}_{q\leftarrow q'}(z,\bt;\mu,\zeta) = C_{q\leftarrow q'}(z,\bt;\mu,\zeta)|_{L\rightarrow L-\ln(z^2)}
    \,,
\end{align}
while the quark to gluon Wilson coefficient function for the TMD FF is given by 
\begin{align}
    \hat{C}_{g\leftarrow q'}(z,\bt;\mu,\zeta) = \frac{\alpha_s}{2\pi}\le[ C_F z+2P_{g\leftarrow q}(z) \le(\ln(z)-\frac{1}{2}L\ri) \ri]
    \, .
\end{align}
In these expressions, we have introduced the standard collinear splitting kernels
\begin{align}
    P_{q\leftarrow q}(x) &=  C_F \le[ \frac{1+x^2}{(1-x)_+}+\frac{3}{2}\delta(1-x) \ri]
\\
    P_{g\leftarrow q}(x) &=  C_F \frac{1+(1-z)^2}{z}
    \\
    P_{q\leftarrow g}(x) &=  T_F \le[z^2+(1-z)^2\ri]\,.
\end{align}
Finally, the coefficient function for the quark-Sivers function is given by
\begin{align}
\bar{C}_{q\leftarrow q'}(x_1,x_2,\bt;\m,\z)  = &
\d_{qq'}\, \d(1-x_1)\, \d(1-x_2) 
\nn \\
& 
- \frac{\as}{2\pi}\d_{qq'}\, \Bigg\{
-L
\Bigg[\d(1-x_2/x_1)
\Bigg(C_F\bigg(\frac{1+x_1^2}{1-x_1}\bigg)_+
\nn\\
&
- C_A\delta(1-x_1)\Bigg)+\frac{C_A}{2}\Bigg(\d(1-x_2)\frac{1+x_1}{1-x_1}
- \d(1-x_2/x_1)\frac{1+x_1^2}{1-x_1}\Bigg)\Bigg]
\nn\\
&
-\frac{1}{2N_C}\d(1-x_2/x_1)(1-x_1)
\nn \\
& 
+C_F\d(1-x_1)\d(1-x_2)\bigg[
\frac{3}{2}L -L_\zeta L -\frac{1}{2}L^2 
- \frac{\pi^2}{12} 
\bigg]
\Bigg\}
\,,
\end{align}
which for $\mu^2 = \zeta = \mu_{b_*}^2$ reduces to
\begin{align}
\bar{C}_{q\leftarrow q'}(x_1,x_2,\bt;\mu_{b_*},\mu_{b_*}^2)  = & \delta_{qq'} \delta(1-x_1) \delta(1-x_2)-\frac{\alpha_s}{2\pi} \frac{\delta_{qq'}}{2N_C}\d(1-x_2/x_1)(1-x_1)
\,\nn \\
& -\frac{\alpha_s}{2\pi}\delta_{qq'} C_F \frac{\pi^2}{12}\d(1-x_1)\d(1-x_2).
\end{align}

%%%%%%%%%%%%%%%%%%%%%%%%%%%%%%%%
\section{TMD evolution ingredients}\label{Anom}
%%%%%%%%%%%%%%%%%%%%%%%%%%%%%%%%

The following expansions, numbers, etc, can be found in the 2013 PDG \cite{Agashe:2014kda}. First of all, we need the expansion of the strong coupling in terms of $\lqcd$:
\begin{align}
\frac{\as(\m)}{4\pi} &=
\frac{1}{\b_0 x} - \frac{\b_1}{\b_0^3}\frac{\ln x}{x^2} 
%\nonumber \\
%&
+
\frac{\b_1^2}{\b_0^5}\frac{\ln^2x - \ln x - 1}{x^3} +
\frac{\b_2}{\b_0^4}\frac{1}{x^3} + \cdots 
\,,
\end{align}
where $x=\ln\left(\m^2/\lqcd^2\right)$, and the coefficients of the $beta$-function are given as
\begin{align}
\b_0 = &
\frac{11}{3}\,C_A - \frac{4}{3}\,T_F n_f
\,,
\\
\b_1 =&
\frac{34}{3}\,C_A^2 - \frac{20}{3}\,C_A T_F n_f
- 4 C_F T_F n_f
\,,
\\
\b_2 =&
\frac{2857}{54}\,C_A^3 + \left( 2 C_F^2
- \frac{205}{9}\,C_F C_A - \frac{1415}{27}\,C_A^2 \right) T_F n_f
\nn\\
&
+ \left( \frac{44}{9}\,C_F + \frac{158}{27}\,C_A \right) T_F^2 n_f^2\,.
\end{align}
Since we want the resummation up to NNLL, we take the expansion of $\as$ with $\b_0$, $\b_1$ and $\b_2$.
Depending on the number of active flavours, the value of $\lqcd$ changes. For $n_f=4$ we have $\lqcd=0.297~\GeV$, and for $n_f=5$ we have $\lqcd=0.214~\GeV$. The pole-mass for bottom-quark is $m_b=4.7~\GeV$.

The rapidity anomalous dimension, Collins-Soper kernel, is defined perturbatively as
\begin{align}
D(\bt;\mu) &= 
\sum_{n = 1}^{\infty} 
\sum_{k = 0}^{n} 
d^{\left(n,k\right)}\,
\Big(\frac{\alpha_s}{4\pi}\Big)^n
L^k
\,,
\end{align}
where the coefficients up to NNLL are given by
\begin{align}
d^{(1,0)} =& 0\,, 
\qquad 
d^{(1,1)} = \Gamma_0/2\,,
\nn\\
d^{(2,0)} =& C_A C_F\left( \frac{404}{27}-14\zeta_3 \right) -\frac{112}{27} C_F T_F n_f\,,
\nn\\
d^{(2,1)} =& \Gamma_1/2\,,
\qquad
d^{(2,2)} = \Gamma_0\beta_0/4
\,.
\end{align}
On the other hand, in order to describe the perturbative TMD evolution, we want to analytically solve the integral
\begin{align}
\int_{\m_L}^{\m_U}\frac{d\bar\m}{\bar\m}\,
\le(\g^V+\G_{\rm cusp}\,\ln\frac{\m_U^2}{\bar\m^2} \ri)
\,,
\end{align}
where the coefficients of the perturbative expansions of the anomalous dimensions can be found in the below.

%%%%%%%%%%%%%%%%%%%%%%%%%%%%%%%%
\subsection{Integration at NLL accuracy}\label{NLL}
%%%%%%%%%%%%%%%%%%%%%%%%%%%%%%%%
For this order we take $\g_0$, $\G_0$, $\G_1$, $\b_0$ and $\b_1$.
Thus we have:
\begin{align}
C_{\g_0}^{\rm NLL} =&
\int_{\m_L}^{\m_U}\frac{d\bar\m}{\bar\m}\,
\g_0 \frac{\as(\bar\m)}{4\pi} \nonumber \\ 
=&
\frac{\g_0}{2\b_0}\int_{x_L}^{x_U}dx\,
\le(\frac{1}{x} - \frac{\b_1}{\b_0^2}\frac{\ln x}{x^2}\ri) 
\nonumber \\
 = &
\frac{\g_0}{2\b_0}
\le.
\le[
\ln x - \frac{\b_1}{\b_0^2}\le(\frac{-1-\ln x}{x}\ri)
\ri]\ri|_{x_L}^{x_U} 
\\
%%%%%%%%%%%%%%%%
 C_{\G_0}^{\rm NLL} = &
\int_{\m_L}^{\m_U}\frac{d\bar\m}{\bar\m}\,
\G_0\frac{\as(\bar\m)}{4\pi} \ln\frac{\m_U^2}{\bar\m^2} \\
 = &
\frac{\G_0}{2\b_0}\int_{x_L}^{x_U}dx\,
\le(\frac{1}{x} - \frac{\b_1}{\b_0^2}\frac{\ln x}{x^2}\ri) (x_U-x)
\nn\\
= &
\frac{\G_0}{2\b_0}
\le.
\le[
-x+x_U\ln x-\frac{\b_1}{\b_0^2}\le(
-\frac{x_U}{x}-\frac{x_U\ln x}{x}-\frac{\ln^2x}{2}\ri)
\ri]\ri|_{x_L}^{x_U} \nn
\\
%%%%%%%%%%%
 C_{\G_1}^{\rm NLL} = &
\int_{\m_L}^{\m_U}\frac{d\bar\m}{\bar\m}\,
\G_1\le(\frac{\as(\bar\m)}{4\pi}\ri)^2 \ln\frac{\m_U^2}{\bar\m^2} \\
 = &
\frac{\G_1}{2\b_0^2}\int_{x_L}^{x_U}dx\,
\le(\frac{1}{x} - \frac{\b_1}{\b_0^2}\frac{\ln x}{x^2}\ri)^2 (x_U-x)
\nn\\
= &
\frac{\G_1}{2\b_0^2}
\le.
\le[
-\frac{x_U}{x} - \ln x
- 2\frac{\b_1}{\b_0^2}\le(
\frac{1}{x}-\frac{x_U}{4x^2} + \frac{\ln x}{x} - \frac{x_U \ln x}{2x^2} 
\ri)
\ri.\ri.
\nn\\
&\le.\le.
+ \frac{\b_1^2}{\b_0^4} \le(
\frac{1}{4x^2} - \frac{2x_U}{27x^3} + \frac{\ln x}{2x^2} 
- \frac{2x_U\ln x}{9x^3} + \frac{\ln^2x}{2x^2} - \frac{x_U\ln^2x}{3x^3}
\ri)
\ri]\ri|_{x_L}^{x_U}\nn
\end{align}
The final result is then
\begin{align}
\int_{\m_L}^{\m_U}\frac{d\bar\m}{\bar\m}\,
\le(\g^V+\G_{\rm cusp}\,\ln\frac{\m_U^2}{\bar\m^2} \ri)
=
C_{\g_0}^{\rm NLL} + C_{\G_0}^{\rm NLL} + C_{\G_1}^{\rm NLL}
\,.
\end{align}
Be careful with the number of active flavors.
The number of flavors for the $x_U$ that appears inside the integrand is fixed and depends on the value of $\m_U$.
However, depending on the hierarchy between $\m_L$, $\m_U$ and $m_b$ we might have to split the integral in several pieces, and in that case, when we substitute the limits of the integral, $x_L$ and $x_U$, they would have different numbers of active flavors (still the $x_U$ that already appeared in the integrand before the substitutions just depends on the value of $\m_U$).
%%%%%%%%%%%%%%%%%%%%%%%%%%%%%%%%

\subsection{Integration at NNLL accuracy}\label{NNLL}
%%%%%%%%%%%%%%%%%%%%%%%%%%%%%%%%
For this order we take $\g_0$, $\g_1$, $\G_0$, $\G_1$, $\G_2$, $\b_0$, $\b_1$ and $\b_2$.
Thus we have:
\begin{align}
C_{\g_0}^{\rm NNLL}  &= 
\int_{\m_L}^{\m_U}\frac{d\bar\m}{\bar\m}\,
\g_0 \frac{\as(\bar\m)}{4\pi} \\
& = 
\frac{\g_0}{2\b_0}\int_{x_L}^{x_U}dx\,
\le(\frac{1}{x} - \frac{\b_1}{\b_0^2}\frac{\ln x}{x^2} +
\frac{\b_1^2}{\b_0^4}\frac{\ln^2x - \ln x - 1}{x^3} +
\frac{\b_2}{\b_0^3}\frac{1}{x^3}\ri)
\nn\\
&=
\frac{\g_0}{2\b_0}
\le.\le[
\ln x - \frac{\b_1}{\b_0^2}\le(\frac{-1-\ln x}{x}\ri) +\frac{\b_1^2}{\b_0^4}\le(\frac{1}{2x^2}-\frac{\ln^2x}{2x^2}\ri)
+\frac{\b_2}{\b_0^3}\le(\frac{-1}{2x^2}\ri)
\ri]\ri|_{x_L}^{x_U} 
\nonumber \\ 
%%%%%%%%%%%%%%%%%
C_{\g_1}^{\rm NNLL}  &= 
\int_{\m_L}^{\m_U}\frac{d\bar\m}{\bar\m}\,
\g_1 \le(\frac{\as(\bar\m)}{4\pi}\ri)^2 \\
 &= 
\frac{\g_1}{2\b_0^2}\int_{x_L}^{x_U}dx\,
\le(\frac{1}{x} - \frac{\b_1}{\b_0^2}\frac{\ln x}{x^2} +
\frac{\b_1^2}{\b_0^4}\frac{\ln^2x - \ln x - 1}{x^3} +
\frac{\b_2}{\b_0^3}\frac{1}{x^3}\ri)^2
\nn\\
&= 
\frac{\g_1}{2\b_0^2}
\le[
-\frac{1}{x}
+\frac{\b_1^2}{\b_0^4}\le(
-\frac{2}{27 x^3}-\frac{\ln^2(x)}{3 x^3}-\frac{2 \ln(x)}{9 x^3}\ri)
\ri.
\nn\\
&
+\frac{\b_1^4}{\b_0^8}\le(
-\frac{789}{3125 x^5}-\frac{\ln^4(x)}{5 x^5}+\frac{6 \ln^3(x)}{25 x^5}+\frac{43 \ln^2(x)}{125 x^5}-\frac{164\ln(x)}{625 x^5}
\ri) 
\nn\\
&+\frac{\b_2^2}{\b_0^6}\le(
-\frac{1}{5 x^5}
\ri)
-2\frac{\b_1}{\b_0^2}\le(
-\frac{1}{4x^2}-\frac{\ln(x)}{2 x^2}
\ri)
\nn\\
&+2\frac{\b_1^2}{\b_0^4}
\le(\frac{10}{27x^3}-\frac{\ln^2(x)}{3x^3}+\frac{\ln(x)}{9x^3}\ri)
+2\frac{\b_2}{\b_0^3}\le(-\frac{1}{3x^3}\ri)
\nn\\
& -2\frac{\b_1^3}{\b_0^6}\le(
\frac{9}{128x^4}-\frac{\ln^3(x)}{4x^4}+\frac{\ln^2(x)}{16x^4}+\frac{9\ln(x)}{32x^4}\ri)
\nn\\
&
-2\frac{\b_1\b_2}{\b_0^5}\le(-\frac{1}{16x^4}-\frac{\ln(x)}{4x^4}\ri)
\nn\\
&\le.\le.
+2\frac{\b_1^2\b_2}{\b_0^7}\le(
\frac{28}{125x^5}-\frac{\ln^2(x)}{5x^5}+\frac{3\ln(x)}{25x^5}\ri)
\ri]\ri|_{x_L}^{x_U} \nn 
\\
%%%%%%%%%%%%%%%
C_{\G_0}^{\rm NNLL} &=
\int_{\m_L}^{\m_U}\frac{d\bar\m}{\bar\m}\,
\G_0 \frac{\as(\bar\m)}{4\pi} \ln\frac{\m_U^2}{\bar\m^2} \nn \\
 & = 
\frac{\G_0}{2\b_0}\int_{x_L}^{x_U}dx\,
\le(\frac{1}{x} - \frac{\b_1}{\b_0^2}\frac{\ln x}{x^2} +
\frac{\b_1^2}{\b_0^4}\frac{\ln^2x - \ln x - 1}{x^3} +
\frac{\b_2}{\b_0^3}\frac{1}{x^3}\ri) (x_U-x)
\nn\\
&= 
\frac{\G_0}{2\b_0}
\le[
-x+x_U\ln x-\frac{\b_1}{\b_0^2}\le(
-\frac{x_U}{x}-\frac{x_U\ln x}{x}-\frac{\ln^2x}{2}\ri)
\ri.
\nn\\
&
+\frac{\b_1^2}{\b_0^4}\le(
\frac{(\ln(x)+1)((2x-x_U)\ln(x)+x_U)}{2x^2}
\ri)
%\nn\\
%&
\le.\le.
+\frac{\b_2}{\b_0^3}\le(
\frac{1}{x}-\frac{x_U}{2x^2}
\ri)
\ri]\ri|_{x_L}^{x_U}
\\
%%%%%%%%%%%%%%%%%%%
C_{\G_1}^{\rm NNLL} &=
\int_{\m_L}^{\m_U}\frac{d\bar\m}{\bar\m}\,
\G_1\le(\frac{\as(\bar\m)}{4\pi}\ri)^2 \ln\frac{\m_U^2}{\bar\m^2} 
\nn\\
&
=
\frac{\G_1}{2\b_0^2}\int_{x_L}^{x_U}dx\,
\le(\frac{1}{x} - \frac{\b_1}{\b_0^2}\frac{\ln x}{x^2} +
\frac{\b_1^2}{\b_0^4}\frac{\ln^2x - \ln x - 1}{x^3} +
\frac{\b_2}{\b_0^3}\frac{1}{x^3}\ri)^2 (x_U-x)
\nn\\
&=
\frac{\G_1}{2\b_0^2}
\le[
-\frac{x_U}{x} - \ln x
+ \frac{\b_1^2}{\b_0^4} \le(
\frac{1}{4x^2} - \frac{2x_U}{27x^3} + \frac{\ln x}{2x^2} 
- \frac{2x_U\ln x}{9x^3} + \frac{\ln^2x}{2x^2} - \frac{x_U\ln^2x}{3x^3}
\ri)
\ri.
\nn\\
&
+\frac{\b_1^4}{\b_0^8}\le(
\frac{20000 (5 x-4 x_U) \ln^4(x)-4000(25 x-24 x_U) \ln^3(x)-200 (875 x-688 x_U) \ln^2(x)}{400000 x^5}
\ri.
\nn\\
&\le.
+\frac{20(5625 x-5248x_U)\ln(x)+128125 x-100992 x_U}{400000 x^5}
\ri)
\nn\\
&
+\frac{\b_2^2}{\b_0^6}\le(
\frac{1}{4x^4}-\frac{x_U}{5x^5}
\ri)
- 2\frac{\b_1}{\b_0^2}\le(
\frac{1}{x}-\frac{x_U}{4x^2} + \frac{\ln x}{x} - \frac{x_U \ln x}{2x^2} 
\ri)
\nn\\
&
+2\frac{\b_1^2}{\b_0^4}\le(
\frac{9(3x-2x_U)\ln^2(x)+6x_U\ln(x)-27x+20x_U}{54x^3}
\ri)
+2\frac{\b_2}{\b_0^3}\le(
\frac{1}{2x^2}-\frac{x_U}{3x^3}
\ri)
\nn\\
&
-2\frac{\b_1^3}{\b_0^6}\le(
\frac{96(4x-3x_U)\ln^3(x)+72x_U\ln^2(x)+(324x_U-384x)\ln(x)-128x+81x_U}{1152x^4}
\ri)
\nn\\
&
-2\frac{\b_1 \b_2}{\b_0^5}\le(
-\frac{x_U}{16x^4}-\frac{x_U \ln(x)}{4x^4}+\frac{1}{9x^3}+\frac{\ln(x)}{3 x^3}
\ri)
\nn\\
&\le.\le.
+2\frac{\b_1^2 \b_2}{\b_0^7}\le(
\frac{200(5x-4x_U)\ln^2(x)+(480x_U-500x)\ln(x)-1125x+896x_U}{4000x^5}
\ri)
\ri]\ri|_{x_L}^{x_U}
\\
%%%%%%%%%%%%%
C_{\G_2}^{\rm NNLL} &=
\int_{\m_L}^{\m_U}\frac{d\bar\m}{\bar\m}\,
\G_2\le(\frac{\as(\bar\m)}{4\pi}\ri)^3 \ln\frac{\m_U^2}{\bar\m^2} 
\nn\\
&
=
\frac{\G_2}{2\b_0^3}\int_{x_L}^{x_U}dx\,
\le(\frac{1}{x} - \frac{\b_1}{\b_0^2}\frac{\ln x}{x^2} +
\frac{\b_1^2}{\b_0^4}\frac{\ln^2x - \ln x - 1}{x^3} +
\frac{\b_2}{\b_0^3}\frac{1}{x^3}\ri)^3 (x_U-x)
\nn\\
&=
\frac{\G_2}{2\b_0^3}
\le[
-\frac{\b_1^3}{\b_0^6}\le(
-\frac{6x_U}{625x^5}-\frac{x_U \ln^3(x)}{5 x^5}-\frac{3 x_U \ln^2(x)}{25 x^5}-\frac{6 x_U \ln(x)}{125 x^5}+\frac{3}{128 x^4}
\ri.\ri.
\nn\\
&\le.
+\frac{\ln^3(x)}{4 x^4}
+\frac{3 \ln^2(x)}{16 x^4}+\frac{3 \ln(x)}{32 x^4}
\ri)
\nn\\
&
+\frac{\b_1^6}{\b_0^{12}}\le(
\frac{21703 x_U}{131072 x^8}-\frac{x_U \ln^6(x)}{8 x^8}+\frac{9 x_U \ln^5(x)}{32 x^8}+\frac{45 x_U \ln^4(x)}{256 x^8}-\frac{275 x_U \ln^3(x)}{512 x^8}
\ri.
\nn\\
&
-\frac{825 x_U \ln^2(x)}{4096 x^8}+\frac{5319 x_U \ln(x)}{16384 x^8}
-\frac{159580}{823543 x^7}+\frac{\ln^6(x)}{7 x^7}-\frac{15 \ln^5(x)}{49 x^7}
\nn\\
&\le.
-\frac{75 \ln^4(x)}{343x^7}+\frac{1415 \ln^3(x)}{2401 x^7}+\frac{4245 \ln^2(x)}{16807 x^7}-\frac{41931 \ln(x)}{117649 x^7}
\ri)
\nn\\
&
+\frac{\b_2^3}{\b_0^{9}}\le(
\frac{1}{7x^7}-\frac{x_U}{8 x^8}
\ri)
+ \frac{1}{x}-\frac{x_U}{2x^2}
\nn\\
&
-3\frac{\b_1}{\b_0^{2}}\le(
-\frac{x_U}{9x^3}-\frac{x_U \ln(x)}{3 x^3}+\frac{1}{4 x^2}+\frac{\ln(x)}{2 x^2}
\ri)
\nn\\
&
+3\frac{\b_1^2}{\b_0^4}\le(
-\frac{x_U}{32x^4}-\frac{x_U \ln^2(x)}{4 x^4}-\frac{x_U \ln(x)}{8 x^4}+\frac{2}{27 x^3}+\frac{\ln^2(x)}{3x^3}+\frac{2 \ln(x)}{9 x^3}
\ri)
\nn\\
&
+3\frac{\b_1^2}{\b_0^4}\le(
\frac{9x_U}{32 x^4}-\frac{x_U \ln^2(x)}{4 x^4}+\frac{x_U\ln(x)}{8 x^4}
-\frac{10}{27x^3}+\frac{\ln^2(x)}{3x^3}-\frac{\ln(x)}{9 x^3}
\ri)
\nn\\
&
-6\frac{\b_1^3}{\b_0^6}\le(
\frac{29x_U}{625x^5}-\frac{x_U\ln^3(x)}{5 x^5}+\frac{2x_U\ln^2(x)}{25 x^5}
+\frac{29x_U\ln(x)}{125x^5}-\frac{9}{128x^4}+\frac{\ln^3(x)}{4 x^4}
\ri.
\nn\\
&\le.
-\frac{\ln^2(x)}{16x^4}-\frac{9\ln(x)}{32x^4}\ri)
\nn\\
&
+3\frac{\b_1^4}{\b_0^8}\le(
\frac{7 x_U}{648 x^6}-\frac{x_U \ln^4(x)}{6 x^6}
+\frac{x_U \ln^3(x)}{18 x^6}+\frac{7 x_U \ln^2(x)}{36x^6}
+\frac{7 x_U \ln(x)}{108 x^6}-\frac{56}{3125 x^5}
\ri.
\nn\\
&\le.
+\frac{\ln^4(x)}{5 x^5}
-\frac{\ln^3(x)}{25x^5}
-\frac{28 \ln^2(x)}{125 x^5}-\frac{56 \ln(x)}{625 x^5}
\ri)
\nn\\
&
+3\frac{\b_1^4}{\b_0^8}\le(
-\frac{67 x_U}{324 x^6}-\frac{x_U\ln^4(x)}{6 x^6}+\frac{2 x_U \ln^3(x)}{9 x^6}
+\frac{5 x_U \ln^2(x)}{18 x^6}-\frac{13 x_U \ln(x)}{54 x^6}
\ri.
\nn\\
&\le.
+\frac{789}{3125 x^5}
+\frac{\ln^4(x)}{5 x^5}-\frac{6 \ln^3(x)}{25x^5}
-\frac{43 \ln^2(x)}{125 x^5}+\frac{164 \ln(x)}{625 x^5}\ri)
\nn\\
&
-3\frac{\b_1^5}{\b_0^{10}}\le(
-\frac{3263x_U}{117649 x^7}-\frac{x_U \ln^5(x)}{7 x^7}
+\frac{9 x_U \ln^4(x)}{49 x^7}+\frac{85 x_U \ln^3(x)}{343 x^7}
-\frac{431 x_U \ln^2(x)}{2401 x^7}
\ri.
\nn\\
&\le.
-\frac{3263 x_U \ln(x)}{16807x^7}
+\frac{37}{972 x^6}
+\frac{\ln^5(x)}{6 x^6}-\frac{7 \ln^4(x)}{36 x^6}
-\frac{8 \ln^3(x)}{27x^6}+\frac{5\ln^2(x)}{27 x^6}+\frac{37 \ln(x)}{162 x^6}\ri)
\nn\\
&
+3\frac{\b_2}{\b_0^3}\le(
\frac{1}{3 x^3}-\frac{x_U}{4 x^4}
\ri)
-6\frac{\b_1\b_2}{\b_0^5}\le(
-\frac{x_U}{25 x^5}-\frac{x_U\ln(x)}{5x^5}+\frac{1}{16 x^4}+\frac{\ln(x)}{4 x^4}
\ri)
\nn\\
&
+3\frac{\b_1^2\b_2}{\b_0^7}\le(
-\frac{x_U}{108 x^6}-\frac{x_U \ln^2(x)}{6 x^6}-\frac{x_U \ln(x)}{18 x^6}+\frac{2}{125 x^5}+\frac{\ln^2(x)}{5 x^5}+\frac{2 \ln(x)}{25 x^5}
\ri)
\nn\\
&
+6\frac{\b_1^2\b_2}{\b_0^7}\le(
\frac{225(6x-5x_U)\ln^2(x)+(750x_U-810x)\ln(x)-1512x+1250x_U}{6750x^6}
\ri)
\nn\\
&
-6\frac{\b_1^3\b_2}{\b_0^9}\le(
\frac{57x_U}{2401x^7}-\frac{x_U\ln^3(x)}{7x^7}+\frac{4x_U\ln^2(x)}{49x^7}
+\frac{57 x_U \ln(x)}{343 x^7}-\frac{7}{216 x^6}+\frac{\ln^3(x)}{6x^6}
\ri.
\nn\\
&\le.
-\frac{\ln^2(x)}{12x^6}-\frac{7 \ln(x)}{36 x^6}\ri)
\nn\\
&
+3\frac{\b_1^4\b_2}{\b_0^{11}}\le(
-\frac{615x_U}{4096x^8}-\frac{x_U \ln^4(x)}{8 x^8}
+\frac{3 x_U \ln^3(x)}{16 x^8}+\frac{25 x_U \ln^2(x)}{128 x^8}
-\frac{103 x_U \ln(x)}{512 x^8}
\ri.
\nn\\
&\le.
+\frac{2929}{16807 x^7}+\frac{\ln^4(x)}{7x^7}
-\frac{10\ln^3(x)}{49x^7}-\frac{79\ln^2(x)}{343x^7}+\frac{528\ln(x)}{2401 x^7}\ri)
\nn\\
&
+3\frac{\b_2^2}{\b_0^6}\le(
\frac{1}{5x^5}-\frac{x_U}{6x^6}
\ri)
-3\frac{\b_1\b_2^2}{\b_0^8}\le(
-\frac{x_U}{49 x^7}-\frac{x_U \ln(x)}{7x^7}+\frac{1}{36x^6}+\frac{\ln(x)}{6 x^6}
\ri)
\nn\\
&\le.\le.
+3\frac{\b_1^2\b_2^2}{\b_0^{10}}\le(
\frac{35 x_U}{256 x^8}-\frac{x_U \ln^2(x)}{8 x^8}+\frac{3x_U\ln(x)}{32 x^8}
-\frac{54}{343 x^7}+\frac{\ln^2(x)}{7 x^7}-\frac{5 \ln(x)}{49 x^7}\ri)
\ri]\ri|_{x_L}^{x_U}
\end{align}
The final result is then
\begin{align}
\int_{\m_L}^{\m_U}\frac{d\bar\m}{\bar\m}\,
\le(\g^V+\G_{\rm cusp}\,\ln\frac{\m_U^2}{\bar\m^2} \ri)
=
C_{\g_0}^{\rm NNLL} + C_{\g_1}^{\rm NNLL} 
+ C_{\G_0}^{\rm NNLL} + C_{\G_1}^{\rm NNLL} + C_{\G_2}^{\rm NNLL}
\,.
\end{align}

%%%%%%%%%%%%%%%%%%%%%%%%%%%%%%%%%

\section{Evolution of the Hard Matching Coefficient}
\label{sec:app}
%%%%%%%%%%%%%%%%%%%%%%%%%%%%%%%%%

The evolution of the hard matching coefficient $C_V$, which is related to the usual hard function as $H=|C_V|^2$, is given by
\begin{align}\label{gammaV}
\frac{d}{d\ln\mu}\ln C_V(Q^2/\m^2) &=
\g_{C_V}\le(\as(\m),\ln\frac{Q^2}{\m^2} \ri)
\,,
\\
\g_{C_V} &=
\Gamma_{\rm cusp}(\alpha_s)\,\ln\frac{Q^2}{\mu^2}
+ \g^V(\as)
\,,
\end{align}
where the cusp term is related to the evolution of the Sudakov double logarithms and the remaining term with the evolution of single logarithms.
The exact solution of this equation is
\begin{align}\label{CVsol}
C_V(Q^2/\m_f^2) =&
C_V(Q^2/\m_i^2)
 \exp\le[ \int_{\m_i}^{\m_f} \frac{d\bar\m}{\bar\m}\,
\g_{C_V}\le(\as(\bar\m),\ln\frac{Q^2}{\bar\m^2} \ri)\ri]
\nn\\
=&
C_V(Q^2/\m_i^2) \exp\le[ \int_{\as(\m_i)}^{\as(\m_f)} \frac{d\bar\as}{\b(\bar\as)}\,
\g_{C_V}\le(\bar\as\ri)\ri]
\,,
\end{align}
where we have used that $d/d\ln\m=\b(\as)\,d/d\as$, where
$\b(\as)=d\as/d\ln\m$ is the QCD $\b$-function.

Below we give the expressions for the anomalous dimensions and the QCD $\beta$-function, in the $\overline{{\rm MS}}$ renormalization scheme. We use the following expansions:
\begin{align}
\G_{\rm cusp} &=
\sum_{n=1}^{\infty} \G_{n-1} \le( \frac{\as}{4\pi}\ri)^n
\,,
\\
\g^V &= \sum_{n=1}^{\infty} \g^V_{n-1} \le( \frac{\as}{4\pi}\ri)^n
\,,
\\
\b &= -2\as \sum_{n=1}^\infty \b_{n-1} \left( \frac{\as}{4\pi} \right)^n
\,.
\end{align}
The coefficients for the cusp anomalous dimension $\G_{\rm cusp}$ are
\begin{align}
\G_0 = &
4 C_F
\,,
\nn\\
\G_1 = &
4 C_F \le[ \le( \frac{67}{9} - \frac{\pi^2}{3} \ri) C_A - \frac{20}{9}\,T_F n_f \ri]
\,,
\nn\\
\G_2 = &
4 C_F \le[ C_A^2 \left( \frac{245}{6} - \frac{134\pi^2}{27}
+ \frac{11\pi^4}{45} + \frac{22}{3}\,\zeta_3 \right)
+ C_A T_F n_f  \left( - \frac{418}{27} + \frac{40\pi^2}{27}
- \frac{56}{3}\,\zeta_3 \right)\ri.
\nn\\
&\le.
+ C_F T_F n_f \left( - \frac{55}{3} + 16\zeta_3 \right)
- \frac{16}{27}\,T_F^2 n_f^2 \ri]\,.
\end{align}
The anomalous dimension $\gamma^V$ can be determined up to three-loop order from the partial three-loop expression for the on-shell quark form factor in QCD. We have
\begin{align}
\g_0^V =&
-6 C_F
\,,
\nn\\
\g_1^V =&
C_F^2 \left( -3 + 4\pi^2 - 48\zeta_3 \right)
+ C_F C_A \left( - \frac{961}{27} - \frac{11\pi^2}{3} + 52\zeta_3 \right)
+ C_F T_F n_f \left( \frac{260}{27} + \frac{4\pi^2}{3} \right)
\,.
\end{align}

\bibliography{refs}

%%%%%%%%%%%%%%%%%%%%%%%%%%%%%%%%%%%%%%%%%%%%%%%%%%%%%%%%%%%%%
\bibliographystyle{JHEP}

\end{document}